\documentclass[twocolumn,pre,aps,floatfix,superscriptaddress,longbibliography]{revtex4-2}
\pdfoutput=1 %This is for the ArXiv submission only
\usepackage{amsmath,amssymb,eucal,graphicx,float,epstopdf}
\usepackage{epsfig,algorithm,algcompatible}
\usepackage[utf8]{inputenc}

\usepackage[colorlinks=true, urlcolor=blue, anchorcolor=blue, citecolor=blue,filecolor=blue,linkcolor=blue,menucolor=blue]{hyperref}

\usepackage{subfigure}

\def\Plus{\texttt{+}}

\begin{document}

\title{A Quantum Reservoir Computing Approach to Quantum Stock Movement Forecasting in Quantum-Invested Markets}

\author{Wendy Otieno}
\author{Alexandre Zagoskin}
\author{Alexander G. Balanov}
\author{Juan Totero Gongora}
\author{Sergey E. Savel'ev}
\affiliation{Department of Physics, Loughborough University, Loughborough LE11 3TU, UK}

\begin{abstract}
We present a quantum reservoir computing (QRC) framework based on a small-scale quantum system comprising at most six interacting qubits, designed for nonlinear financial time-series forecasting.  We apply the model to predict future daily closing trading volumes of 20 quantum-sector publicly traded companies over the period from April 11, 2020, to April 11, 2025, as well as minute-by-minute trading volumes during out-of-market hours on July 7, 2025. Our analysis identifies optimal reservoir parameters that yield stock trend (up/down) classification accuracies exceeding $86 \%$. Importantly, the QRC model is platform-agnostic and can be realized across diverse physical implementations of qubits, including superconducting circuits and trapped ions. These results demonstrate the expressive power and robustness of small-scale quantum reservoirs for modeling complex temporal correlations in financial data, highlighting their potential applicability to real-world forecasting tasks on near-term quantum hardware.
\end{abstract}

\maketitle

\section{Introduction}

The complex, non-linear \cite{liu2022quantum,li2024forecasting} and volatile nature of stock returns have made stock market prediction an arduous endeavor. Economical (GDP growth, interest rates, inflation) \cite{liu2022quantum, smithers2022economics, quah2006improving, buzan2024market, ebner2008institutions}, geopolitical \cite{liu2022quantum, buzan2024market, ebner2008institutions, davis2012politics} (government policies \cite{kara2011predicting}, wars, sanctions, trade agreements), societal \cite{liu2022quantum, buzan2024market, ebner2008institutions} (consumer behaviour, social movements, demographic shifts, technological adoption \cite{pastor2009technological}), environmental \cite{davis2012politics} (natural disasters, climate change policies), and psychological \cite{bollen2011twitter, loang2024psychological, bikhchandani2000herd, ooi2025behavioural} (herd behavior from investors, behavioral influences, spontaneous panic-selling) factors add complexities to stock price forecasting as it introduces a wide range of elements that are strenuous to integrate into predictive models. This makes it incredibly difficult to quantify a mathematical model that captures every nuances of the stock market. Despite this challenge, various aspects of the market's behaviour have been captured by mathematical models such as quantitative financial models (Black-Scholes option pricing and risk management \cite{black1973pricing}), stochastic processes (Brownian motions \cite{Bachelier1900, kanazawa2018kinetic, kanazawa2018derivation, garcin2022forecasting}, random walk models \cite{samuelson2016proof,mandelbrot1997variation, Walter2021random, taylor1982tests}), agent-based modelling \cite{thurner2011systemic,klimek2015bail, farmer2009economy} and decision-making strategies that involve the intersection of game theory and behavioural economics \cite{Lyons2021, Chauhan2024, Marmon2025}. These models act as approximations where they are beneficial in forecasting controlled scenarios. However, adjustments to these models are needed to account for the evolution of real-world conditions.

The Efficient Market Hypothesis (EMH) introduced by Eugene F. Fama \cite{fama1970efficient} defines a theoretical benchmark of perfect market efficiency. It claims that (1) past prices have no predictive information (weak form) (2) public information does not lead to consistent excess returns (semi-strong form) and (3) public and private information cannot generate consistent excess returns making forecasting trends impossible (strong form) \cite{fama1970efficient}. With this, EMH defines a theoretical state of perfect market efficiency in which the possibility of prediction is not feasible. In principle, perfect information market efficiency is an ideal state. In reality, real financial data can and do deviate from this perfect efficiency  \cite{de1985does,grabowski2019technology,gu2004increasing,silver2024agent,naseer2015efficient}. They exhibit nonlinear and dynamic structures that contradict EMH, as evidenced by well-documented anomalies such as volatility clustering \cite{cont2001empirical,cont2007volatility}, volume-driven spikes, momentum (predictable returns) \cite{moskowitz2012time, he2021social}, return autocorrelation \cite{campbell1998econometrics,mackinlay2007econometrics,holtes2024return} and short term predictability \cite{easley1992time} which draw market microstructure effects, revealing clear temporal memory in price and volume formation. These deviations are what predictive models such as Artificial Neural Networks (ANNs) and Quantum Machine Learning (QML) exploit. ANN and QML models can detect nonlinear patterns in financial data and learn its hidden structure by operating in high dimensional state spaces where the interwoven interactions between price, volume, returns and order flow exists. QML is advantageous in exploring extremely large, high-dimensional data spaces enabling them to learn complex correlations. Thus if QRC can demonstrate significant predictive power for future daily closing volumes (DCV) and market trends (up/down), this is an implication of the presence of nonlinear, exploitable structure in the DCV financial time series. Such structure contradict EMH and therefore QRC performances show proof of deviations from perfect market efficiency.

To tackle all complexities associated with the inherent nature of the stock market, advanced techniques such as classical machine learning, and now quantum computing, have been explored to achieve better accuracy in stock market price prediction. This entails employing various types of specialized  Artifical Neural Networks (ANNs) i.e. Long Short Term Memory (LSTM) networks, Recurrent Neural Networks (RNNs) and Self Organizing Fuzzy Neural Networks (SOFNNs), to find non-linear relationships, trends and patterns in stock price data. LSTM networks have proven to predict directional movement (binary up/down classifcation) of S$\&$P 500 index stocks from 1992 to 2015, which exhibited high volatility and short term reversal return profile, with an accuracy of $56\%$-$60\%$ \cite{fischer2018deep}. These networks outperform traditional memory-free machine learning classification models (i.e. Random Forests (RAF), Deep Neural Networks (DNN) and Logistic Regression (LOG) \cite{fischer2018deep}) by learning from past patterns after preserving sequential dependencies. Likewise, ANNs have also provided an accuracy of $75.54\%$ on directional movement of the Istanbul Stock Exchange (ISE) National Index surpassing the predicted accuracy of Support Vector Machines (SVM) by about $1.056\%$ when using 10 technical indicators as input features \cite{kara2011predicting}. Moreover, self-organizing fuzzy neural network have demonstrated the influence of twitter's public sentiments (collective mood states) on stock market movement in Dow Jones Industrial Average (DJIA) \cite{bollen2011twitter} by showing that mood dimensions significantly improves DJIA prediction accuracy (predicted accuracy of $87.6\%$) in DJIA's up/down closing values \cite{bollen2011twitter}. 

Apart from forecasting the stock market's directional trend, LSTM-RNN models - models beneficial in optimizing investment portfolios and risk management - have been shown to predict the daily closing price of Amazon (AMZN) with root mean squared error (RMSE) of 2.51 and mean absolute percentage error (MAPE) $1.84\%$ \cite{varadharajan2024stock}. These results were obtained by analyzing historical stock data to help capture hidden patterns and non-linear relationships and evaluating various hyperparameters to determine the hyperparameters' influence on the prediction accuracy \cite{varadharajan2024stock}. The ability to preserve sequential dependencies and learn from past patterns places ANNs above traditional models as the latter models reduces its ability to capture stock market trends over time by analyzing each input independently. Despite all these favourable results, ANNs still suffer from low convergence speed and high complexity in algorithm \cite{liu2022quantum}. 

Recently, quantum machine learning (QML) has been utilized for stock price prediction \cite{liu2022quantum,mourya2025contextual, srivastava2023potential}, option pricing \cite{stamatopoulos2020option, mourya2025contextual}, time series forecasting \cite{emmanoulopoulos2022quantum, orlandi2024enhancing} and many other financial applications such as fraud detection \cite{mironowicz2024applications}, portfolio optimization \cite{zhou2025quantum, cohen2020portfolio} and risk management \cite{zhou2025quantum}. Quantum computing overcomes the limitation of classical computing by applying two fundamental quantum principles: superposition and entanglement. Quantum Support Vector Machines (QSVM) combined with Quantum Annealing (QA) have shown to predict the stock market up/down movements with a maximum prediction accuracy of $58.94\%$ for companies such as Apple, Visa, Honeywell and Johnson $\&$ Johnson \cite{srivastava2023potential}. Future stock prices of Apple and Google, using quantum single-task learning QSTL and quantum multi-task learning QMTL, were predicted with an accuracy of $62.21\%$ (Apple - QSTL), $71.83\%$ (Apple - QMTL), $56.46\%$ (Google - QSTL) and $68.30\%$ (Google - QMTL) \cite{liu2022quantum}. Additionally, quantum neural networks (QNNs) composed of parametrized quantum circuits (PQCs) outperformed classical models in forecasting time series dealing with higher noise variations during effective training, suggesting the QNN's ablity to learn complex patterns \cite{emmanoulopoulos2022quantum}. Furthermore, combining fuzzy logics and wavelets to quantum annealing models (D-wave 2000Q) led to prediction accuracies as little as $80\%$ for specific markets with well processed data \cite{cohen2020portfolio}. All these contributions have paved the way in improving quantum models for accurate financial forecasting and trend (up/down) prediction. The ability of QRC to exploit subtle nonlinear dependencies and long range temporal correlations \cite{fujii2017harnessing, rosato2025study, kobayashi2024feedback, ivaki2025quantum} can be applied to financial data allowing it to detect weak inefficiencies in the data which violate strict EMH assumptions.

By leveraging the unique properties of entanglement and superposition, quantum computers can tackle complex financial data much faster than classical computers. However challenges still occur in limited quantum hardware due to early stage development. Consequently, data limitations, noise sensitivity, integration issues (combining quantum models to existing classical computing framework used in modern financial infrastructures) and algorithmic challenges (i.e. fine tuning and optimization requirements) arise. Quantum hardware realization for scalability and large-scale system is pivotal to unlocking the full potential of quantum computing as it enables fault-tolerant quantum computing (less susceptible to decoherence and noise) \cite{yoder2025tour,bourassa2021blueprint,butt2024measurement}, uses modular architectures (i.e quantum system on chip) to overcome physical limitations \cite{jnane2022multicore, monroe2014large, sang2025toward, li2024heterogeneous}, can support billions of quantum operators to solve real world problems \cite{ibm2025} (such as optimization, complex PDEs modelling,...)  faster with lower power consumption and more. All these are beneficial in tackling physical qubit constraints, fabrication challenges, noise sensitivity and control complexities of qubits. Some compelling development in quantum hardware realization for scalability and large-scale system involves a modular, scalable system on chip (QSoC) framework developed by researchers at MIT and MITRE \cite{li2024heterogeneous}, Cryo-CMOS multiplexing for Silicon Spin Qubits developed by the SCALLOP project \cite{bartee2025spin} and a framework involving hundreds of thousands to millions of physical qubits for practical quantum advantage developed by Microsoft and Collaborators \cite{beverland2022assessing}. The latter development highlights how architecture design choices that incorporate qubit size, speed and controllability can meet scaling requirements \cite{beverland2022assessing}. 

\begin{figure*}[t!]
\centering
\includegraphics[width = 16.75cm]{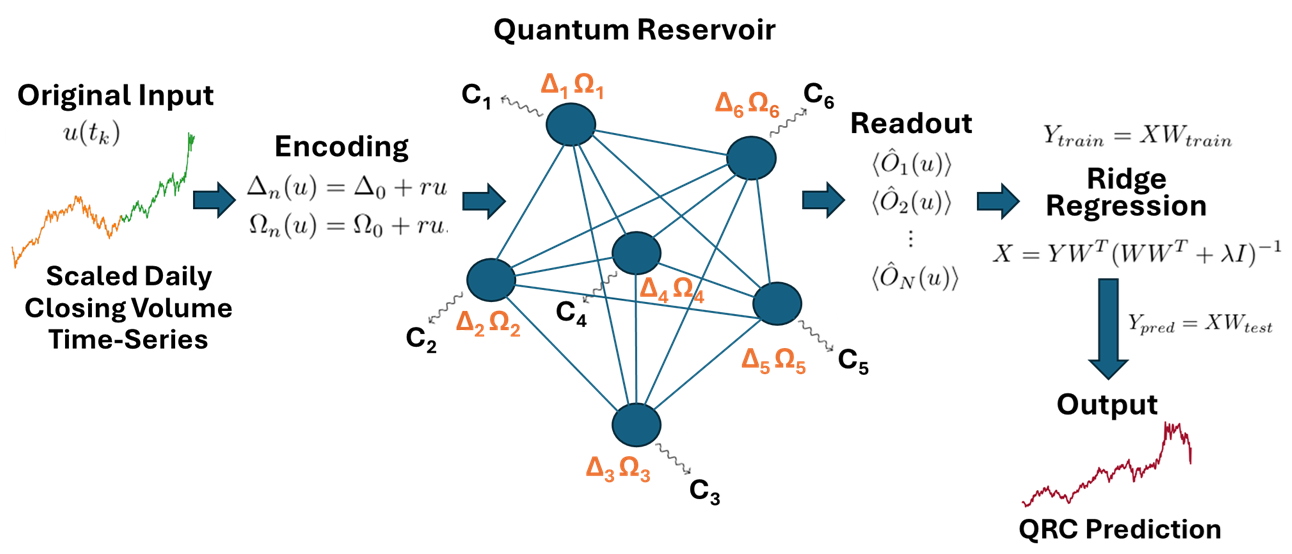}
\caption{The architecture of the QRC framework consists of an (a) input (b) encoding scheme (c) quantum reservoir (QR) (d) readout layer (e) linear regression and (f) output. A classical input time series is quantum encoded via Hamiltonian parameter encoding. The quantum reservoir consists of at most six qubits randomly connected and arranged in an all-to-all topology. The QR projects the quantum-encoded input to a higher dimensional feature space to capture temporal corelations. Readout layer generates a readout matrix containing the population inversion of each qubit populations. Linear Ridge regression determines the linear readout coefficients that relates the targeted input to the QR prediction. This framework models a D-Wave machine.}
\label{fig:QRCscheme}
\end{figure*}

In this paper, we introduce a scalable and minimalistic QRC framework in Section \ref{QRCframeworkqubits} which is employed to predict the future daily closing volumes of 20 quantum invested companies. This minimalistic system consists of a quantum reservoir comprised of up to six transmon superconducting qubits. The QR is scalable as newly added qubits easily couple with existing qubits via mutual connection with the optical waveguide. We implement non-autonomous prediction on the encoded input datasets in Section \ref{nonautonomouspred}, which requires good non-linearity and trace of memory for better predicted accuracy, and evaluate the performance of the QR by measuring the MSE, NMSE and RMSE. We predict future daily closing volume for each company using data that span 5 years, April 11 2020 - 2025, and minute-by-minute volumes of one day trading for July 7, 2025. We also evaluate the accuracy of the predicted stock market movement. We conclude in Section \ref{conclusionQRC} that our reservoir is able to predict stock market trends with high accuracy. This demonstrates how quantum reservoirs can improve accuracy and robustness in stock market predictive models. Thus, overcoming complexities and volatility in complex stock market data.

\section{Methods}
\label{QRCframeworkqubits}

The QRC architecture is composed of an input, encoding layer, quantum reservoir (QR), readout layer, linear regression and output (see Fig~\ref{fig:QRCscheme}). This is a model of a D-wave system. The QR is a quantum system consisting of $N$ individual qubits. The dynamics of the reservoir are governed by the Hamiltonian:
\begin{equation}
\hat{H}(u) = \sum_{n=1}^{N} \bigg[-\Delta_{n}(u) \hat{\sigma}_n^d + \frac{\Omega_{n}(u)}{2}  \hat{\sigma}_n^x \bigg] +  \sum_{m>n} \frac{V_{mn} (\hat{\sigma}_m^d  \otimes \hat{\sigma}_n^d)}{N-1}.
\label{HamiltonianEq}
\end{equation}
The first term in $\eqref{HamiltonianEq}$ describes the contribution of the individual qubits to the dynamics of the QR. Here $\Delta_{n}$ refers to the detuning of the $n$-th qubit while $\Omega_{n}$ gives the Rabi frequency. $\Omega_{n}$ and $\Delta_{n}$ are the Hamiltonian parameters in which the input time series $u$ is encoded. $\hat{\sigma}_n^d$ is an operator which projects onto qubit $n$'s excited state. It is written in terms of the Pauli operator $\hat{\sigma}_n^z$: $\hat{\sigma}_n^d = 0.5 (\hat{1} - \hat{\sigma}_n^z)$. $\hat{\sigma}_n^{x,y,z}$ are Pauli matrices describing the $n$-th qubit.

The second term in $\eqref{HamiltonianEq}$ accounts two-body interactions between various pairs of the qubits. In our model, the qubit-qubit interactions are all-to-all. That is, each qubit has a direct connection to all other qubits i.e. enabling interactions between any pair of qubits. This type of connectivity significantly enhances quantum algorithms by providing efficient entanglement distribution and fast information exchange. The parameter $V_{mn}$ represents the interaction between qubits $m$ and $n$. 

We encode the input time series $u$ using one of the following parameter encoding options:
\begin{align*} \label{hamiltoniandeltaohm}
\Delta_n(u) &= \Delta_{0} + ru, \\
\Omega_n(u) &= \Omega_{0}  + ru,
\end{align*}
where $\Delta_{n}$ and $\Omega_{n}$ are randomly chosen from a uniform distribution centered on $\Delta_{0}$ and $\Omega_{0}$ respectively while $r$ is a scaling parameter. The values of $\Delta_{n}$ introduces heterogeneity into the reservoir ensuring richer computational dynamics. $\Omega_{n}$ influences the dynamics of quantum state evolution (i.e. modulating decoherence or coherence, influencing entanglement dynamics, etc...). The interaction strength between qubits $V_{mn}$ are also chosen randomly around $V_0$. By encoding $u$ into the Hamiltonian parameters $\Delta_{n}$ and $\Omega_{n}$, it can be mapped onto a higher-dimensional feature space by the quantum reservoir. 

The QR dynamics are described by the time evolution of the density matrix $\rho$:
\begin{equation}
\frac{d \rho}{dt} = -\frac{i}{\hbar}[H(t),\rho(t)] + \mathcal{L}[\rho].
\end{equation}
$[H(t),\rho(t)]$ is a commutator of the Hamiltonian $H(t)$ and density $\rho(t)$ that ensures unitary quantum evolution. Dissipation is added to the system via Linbaldian operators $\mathcal{L}[\rho]$ which satisfy
\begin{equation}
\mathcal{L}[\rho] = \sum_{n} ( C_{n}\rho(t) C_{n}^{\Plus} - 0.5 [ C_{n}^{\Plus} C_{n},\rho(t)] ).
\label{dissiptationL}
\end{equation}
The first and last term in $\eqref{dissiptationL}$ characterizes the relaxation and dephasing process respectively \cite{andreev2021emergence} where $C_{n} = \sqrt{\gamma_n} \sigma_{n}^{\Plus}$ is the collapse operator and $\gamma_{n}$ is the qubit's strength of dissipative coupling. 

The reservoir receives the encoded time series $u$ and performs single qubit evolution from $T_{e} = 0$ to $T_{e} = 2\pi/\Omega_{0}$. That is, qubits start in their initial state at time $T_{e} = 0$ and evolve according to their Hamiltonian dynamics until time $T_{e} = 2\pi/\Omega_{0}$. The evolution takes advantage of detuning influences, rabi oscillations and inter-qubit interactions to generate a non-linear feature space. With the presence of heterogeneity, distinct qubits evolve at different rates leading to varying time-dependent representation. Qubits become entangled as they evolve forming a highly connected feature space. Hence, encoded $u$ is mapped from a linear space to a higher-dimensional feature space as qubit interactions introduce nonlinear dependencies which enhances feature extraction.

The feature space transformation of $u(t)$ is represented by the expectations
\begin{equation}
    \langle \hat{O}_n(u) \rangle = \langle \psi_0 | e^{i\hat{H}T_e} \hat{O}_n(u) e^{-i\hat{H}T_e} | \psi_0 \rangle
\end{equation}
with
\begin{equation}
|\psi_0 \rangle = \otimes^{N}_{n} a_n |0 \rangle + \sqrt{(1 - a_{n}^2)}e^{-i\phi_n} |1\rangle.
\end{equation}
These expectations are a set of observable operators $\{\hat{O}_n\}$ whose values construct a reservoir state 
\begin{equation}
W = [\langle \hat{O}_1(u) \rangle \langle \hat{O}_2(u) \rangle ... \langle \hat{O}_N(u) \rangle]^{T}.
\end{equation}
The reservoir state is then passed onto the readout layer for training purposes. 

Time is discretized as $t_k = k \Delta t$ where the time index and time step are $k$ and $\Delta t$ respectively. The input series $u(t)$ is discretized in $K$ total points i.e. $u(t_k)$ for $k \in [0,K]$. In our case, the qubit population $w_{n}(t_{k})$ is expressed as the expectation value of the observable operator $\langle \hat{O}_n(u(t_k)) \rangle$.

During the training process, the QR maps the encoded  input into high-dimensional phase space of the coupled qubits while the readout layer learns the relationship between the transformed encoded $u(t_k)$ and the target time series $y_{k}$. Temporal and non-linear features from a time series of daily closing volume are naturally extracted by the QR without adjusting weights explicitly. Thus, the reservoir remains fixed (it is not trained) and does not require backpropagation which is an essential feature of reservoir computing approach.

The QR generates readout, which is the set of values of the population inversion for each qubit populations $w_{n}(t_{k})$ for $n \in [1,N]$. This produces an $N \times N_{train}$ readout matrix 

\[
W_{train} = 
\begin{pmatrix}
 w_{1}(t_0) & w_1(t_1) & \cdots & w_1(t_{Ntrain}) \\
 w_{2}(t_0) & w_2(t_1) & \cdots & w_2(t_{Ntrain}) \\
 \vdots & \vdots & \vdots & \vdots \\
 w_{N}(t_0) & w_N(t_1) & \cdots & w_N(t_{Ntrain}) \\
\end{pmatrix}
\]
which is utilized to find linear readout coefficients $X$ that satisfies
\begin{equation}
Y_{train} = XW_{train}
\end{equation}
for a target time series $Y_{train}$. $X$ is determined using Ridge regression
\begin{equation}
X = YW^{T}(WW^{T} + \lambda I)^{-1}
\end{equation}
where we set the regression parameter $\lambda$ to $ 10^{-4}$. $\lambda$ is used to ensure the managing of overfitting, particularly in noisy or complex quantum states. It is the readout layer, where learning occurs, that is trained via ridge regression. Hence, the complexities of training is reduced to solely optimizing the readout layer.

After the linear readout coefficients $X$ are determined, we extract $W_{test}$ for the testing time-series
\[
W_{test} = 
\begin{pmatrix}
 w_{1}(t_{Ntrain + 1}) & w_1(t_{Ntrain + 2}) & \cdots & w_1(t_{K}) \\
 w_{2}(t_{Ntrain + 1}) & w_2(t_{Ntrain + 2}) & \cdots & w_2(t_{K}) \\
 \vdots & \vdots & \vdots & \vdots \\
 w_{N}(t_{Ntrain + 1}) & w_N(t_{Ntrain + 2}) & \cdots & w_N(t_{K}) \\
\end{pmatrix}
\]
and apply
\begin{equation}
Y_{pred} = XW_{test}
\end{equation}
to determine the testing prediction $Y_{pred}$ of the QRC framework.

\subsection{Performance Metrics}

Mean Squared Error (MSE), Normalized Mean Squared Error (NMSE) and Root Mean Squared Error (RMSE) are employed to quantify the framework's training and testing performance:
\begin{align}
    MSE[y(t_k), \hat{y}(t_k)] &= \sum_{k} |y(t_k) - \hat{y}(t_k)|^2, \\
    NMSE[y(t_k), \hat{y}(t_k)] &= \frac{MSE}{\sigma^2(y(t_k))},
\end{align}
and
\begin{equation}
    RMSE[y(t_k), \hat{y}(t_k)] = \sqrt{MSE}.
\end{equation}
Here, $y(t_k)$ is the targeted time series, $\hat{y}(t_k)$ is the QRC's predicted output, and $\sigma^2(y(t_k))$ is the variance of the targeted time series. The smaller the value of the MSE and NMSE, the better the QR's predictive performance (i.e. the prediction is closer to the actual value of the DCV). The absolute closeness between $y(t_k)$ and $\hat{y}(t_k)$ is characterized by RMSE.

We also include the Mean Absolute Percentage Error (MAPE) given by
\begin{equation}
MAPE = \frac{100}{N_{total}} \sum_{k} | \frac{y(t_k)-\hat{y}(t_k)}{y(t_k)} |.
\end{equation}

The MAPE shows the relative closeness (percentage) between $\hat{y}(t_k)$ and $y(t_k)$. Certain thresholds of MAPE values are set to categorize the fit between $y(t_k)$ and $\hat{y}(t_k)$. The fit is classified as excellent, good, reasonable and poor when the MAPE value falls in the range $0\%$ - $5\%$, $>5\%$ - $10\%$, $>10\%$ - $20\%$ and $>20\%$ respectively.

Lastly, we compute the direction accuracy (DA) for stock market trend prediction. The direction accuracy represents the correct prediction of the increase or decrease of the stock market data. The predicted trend is characterized by comparing the current predicted data point with the previous predicted data point. If $\hat{y}(t_{k}) - \hat{y}(t_{k-1}) > 0$, then we say that the stock market has increased (up prediction). If $\hat{y}(t_{k}) - \hat{y}(t_{k-1}) < 0$, the stock market has fallen (down prediction). If the actual trend matches with the predicted trend, then we characterize it as correct. We give the accuracy score as:

\begin{equation}
DA = \frac{number\;of\;correct\;trend\;predictions}{total\;number\;of\;trend\;predictions}.
\end{equation}

QRC performance is typically assessed by averaging multiple simulations to account for stochastic variability. In this studies, a fixed random seed is applied to the QRC framework to produce consistent, repeatable results rather than varying across multiple simulations. This choice still provides meaningful information about the model's predictive behaviour.

\section{Standardized Moments in Empirical Stock Data}

We evaluate the QRC performance of 20 companies: Rigetti Computing, Inc (RGTI), Quantum Computing Inc (QUBT), Quantum Corp (QMCO), QuantumScape Corporation (QS), IonQ Inc (IONQ), Quantum eMotion Corp (QNC.V), Defiance Quantum ETF (QTUM), Arqit Quantum Inc (ARQQ), Quantum-SI Incorporated (QSI), Microsoft (MSFT), D-Wave Quantum Inc (QBTS), SEALQ Corp (LAES), Zapata AI (ZPTA), IBM (IBM), NVIDIA (NVDA), Form Factor Inc (FORM), Honeywell (HON), Google (GOOG), Amazon (AMZN) and Intel (INTC). 

The quantum-invested companies can be split into three markets:
\begin{enumerate}
\item Micro/Small-cap markets - QMCO, RGTI, QS, IONQ, QTUM, QSI, QBTS, LAES, ZPTA, FORM, ARQQ.
\item Large-cap markets - IBM, HON, INTC.
\item Mega-cap markets - GOOG, AMZN, NVDA, QUBT, QNC.V, MSFT.
\end{enumerate}

To quantify how far the closing volume data of (a) April 11 2020 - 2025 regular trading hours and (b) July 7, 2025 out of trading hours deviates from a normal distribution, the Gaussian moments
\begin{equation*}
\Gamma_{n}^{Gaussian} =
\begin{cases}
    (n - 1)!! & \text{$n$ is even} \\
    0        & \text{$n$ is odd},    
\end{cases}
\end{equation*}
are compared to the empirical moments
\begin{equation*}
    \Gamma_{n} = \frac{\langle (y - \mu)^n   \rangle}{\langle (y - \mu)^\frac{n}{2}\rangle},
\end{equation*}
where
\begin{equation*}
    \langle (y - \mu)^n   \rangle = \frac{1}{K}  \sum_{k=1}^{K} (y(t_k)-\mu)^n, \hspace{0.3cm} \mu = \frac{1}{K} \sum_{k=1}^{K} y(t_k)
\end{equation*}
\cite{de2022investigation,de2022ai}. Financial returns are known to have fat tails, volatility clustering, non-stationarity and skewness. The standardised moment ratios (SMR) $R_{n} = \Gamma_{n}^{Gaussian}/\Gamma_{n}$ aids in highlighting these deviations. 

\begin{figure*}[t!]
    \centering
    \includegraphics[width=5.75cm]{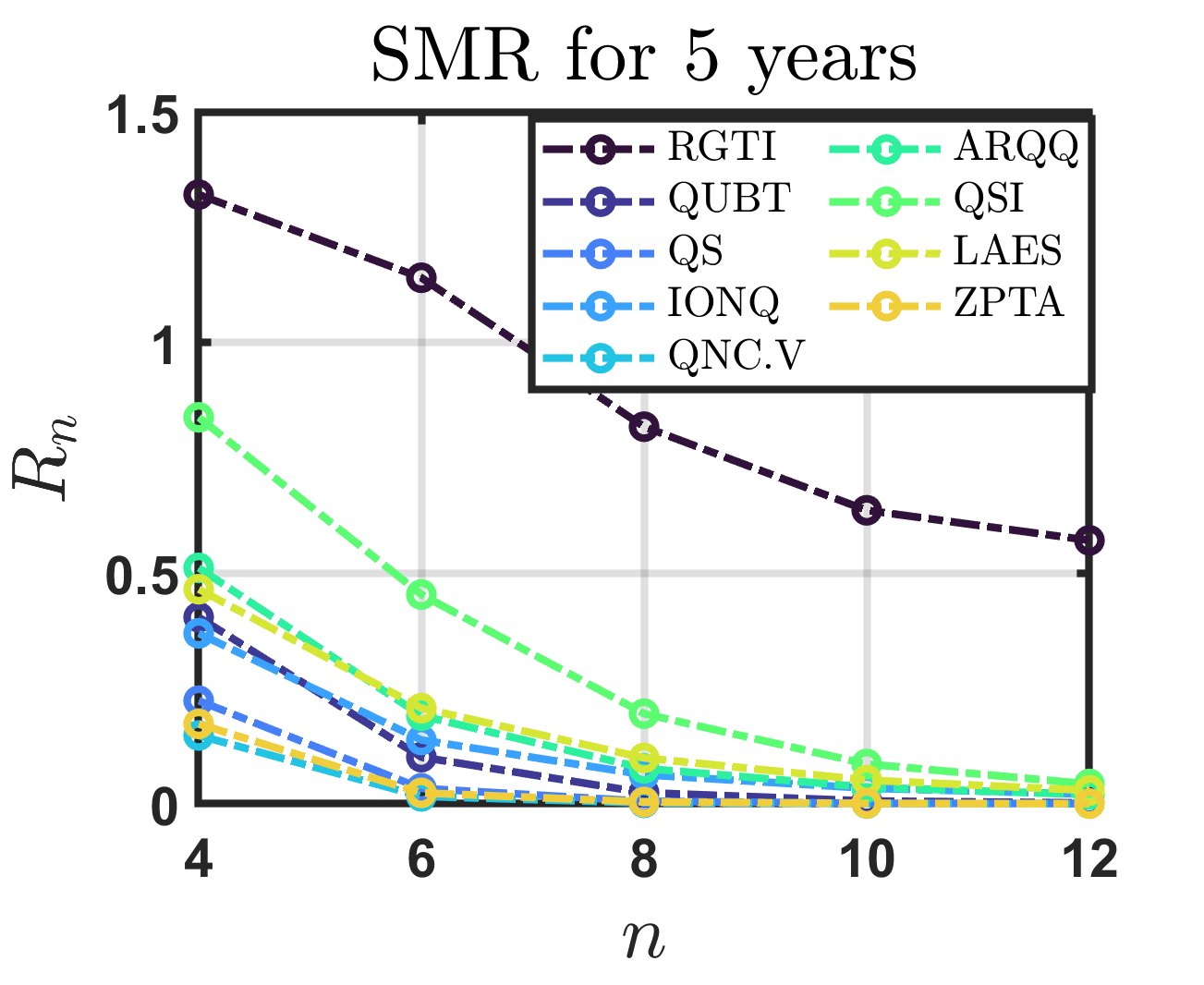}
    \includegraphics[width=5.75cm]{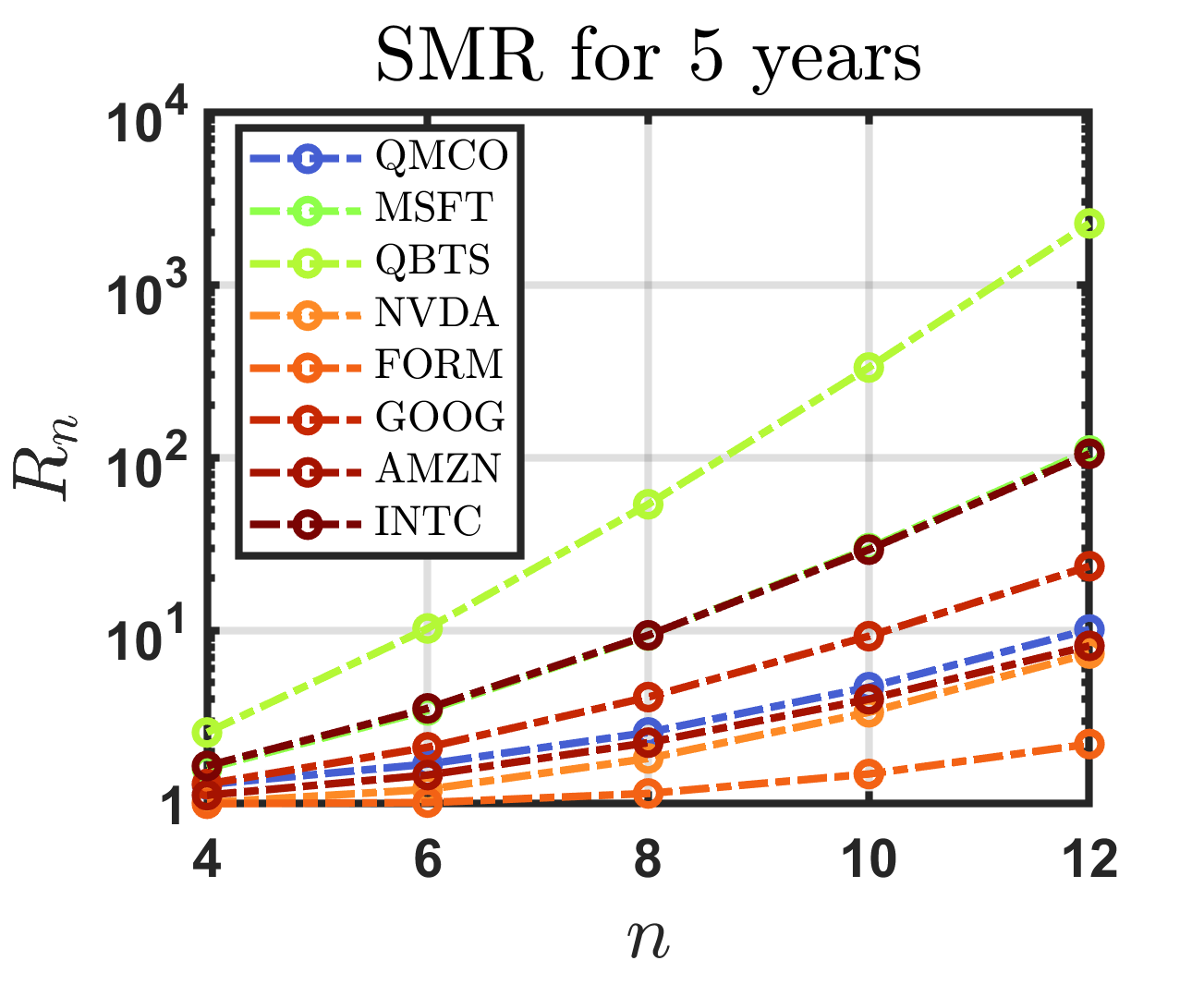}
    \includegraphics[width=5.75cm]{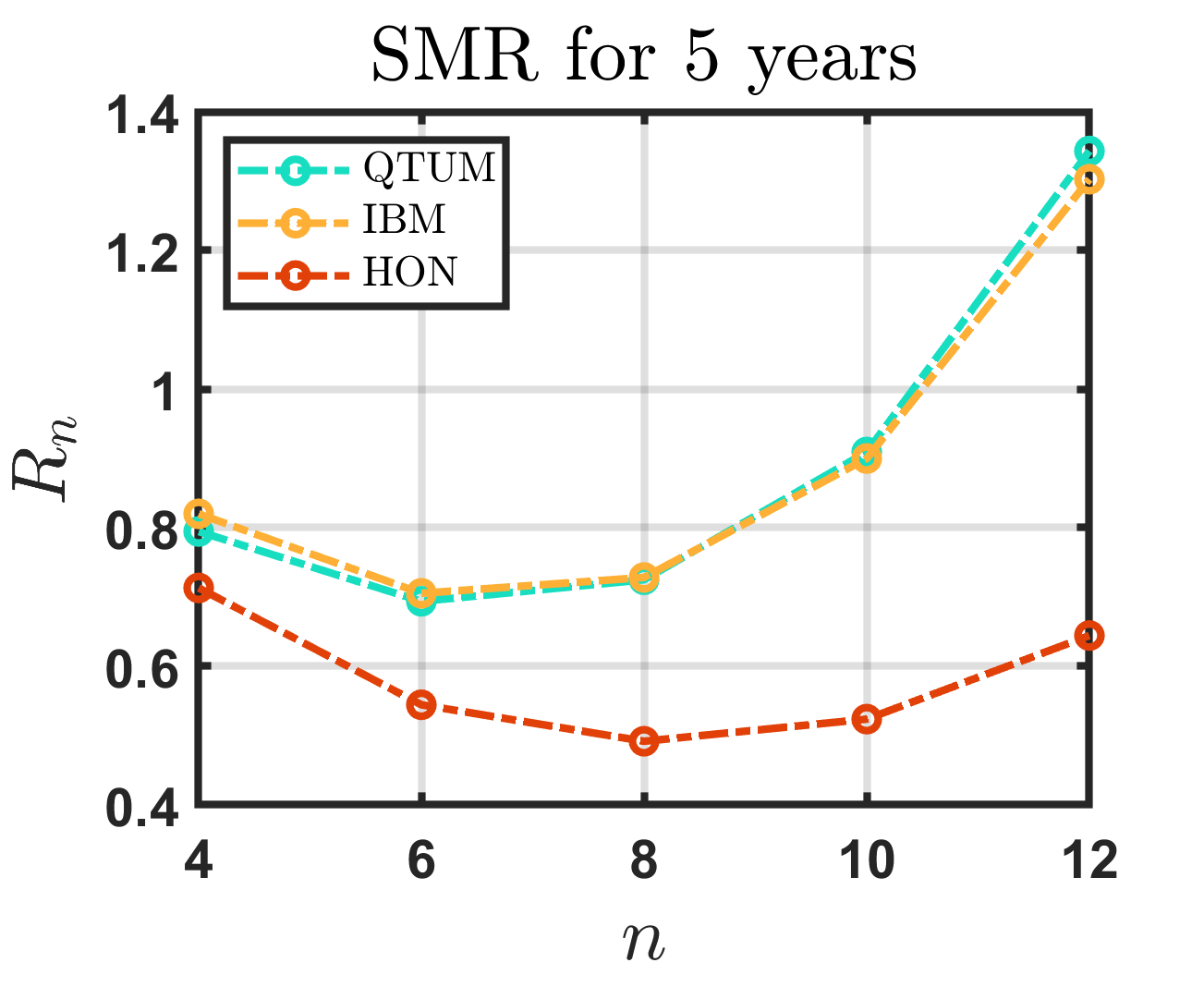}
    \includegraphics[width=5.75cm]{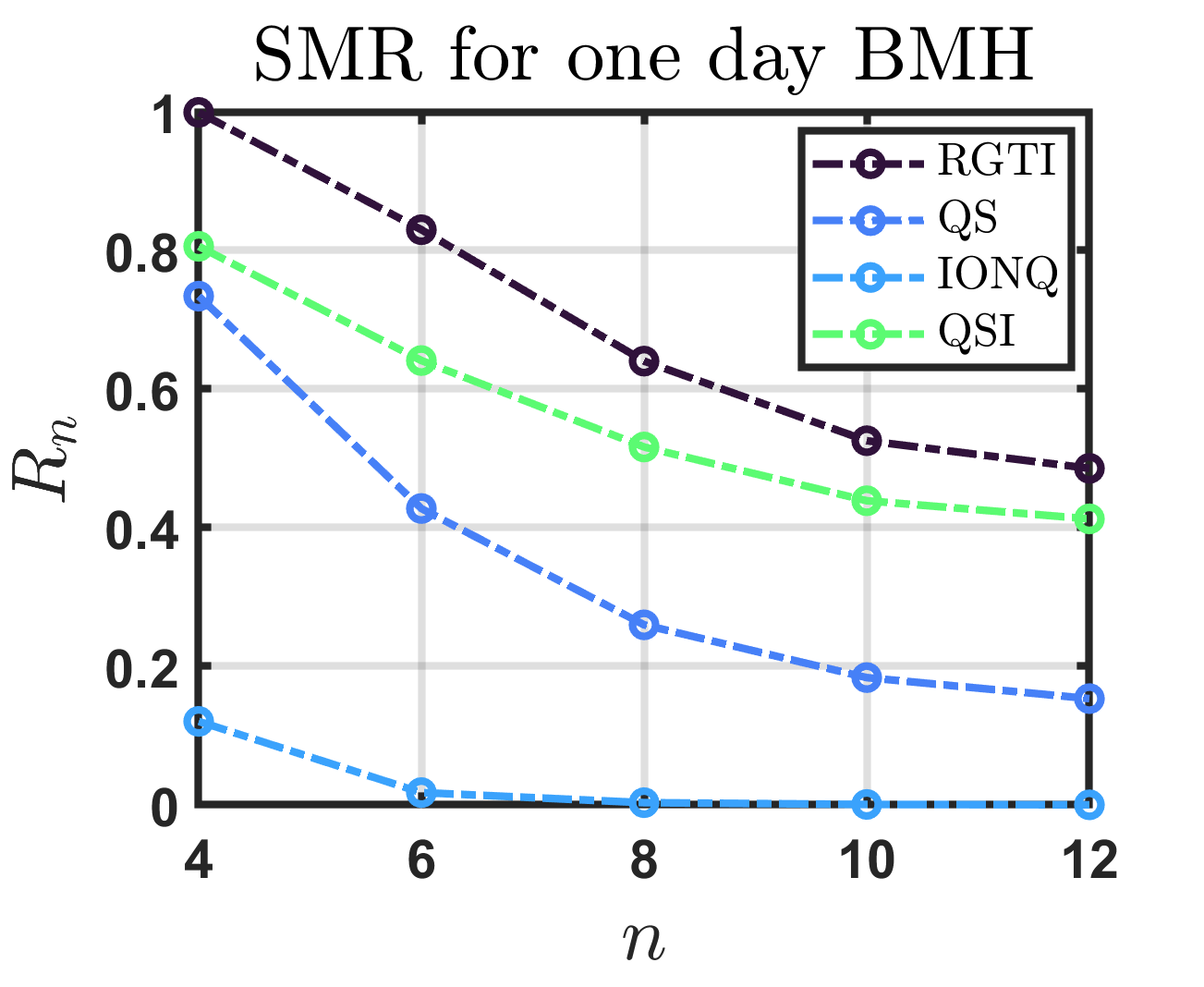}
    \includegraphics[width=5.75cm]{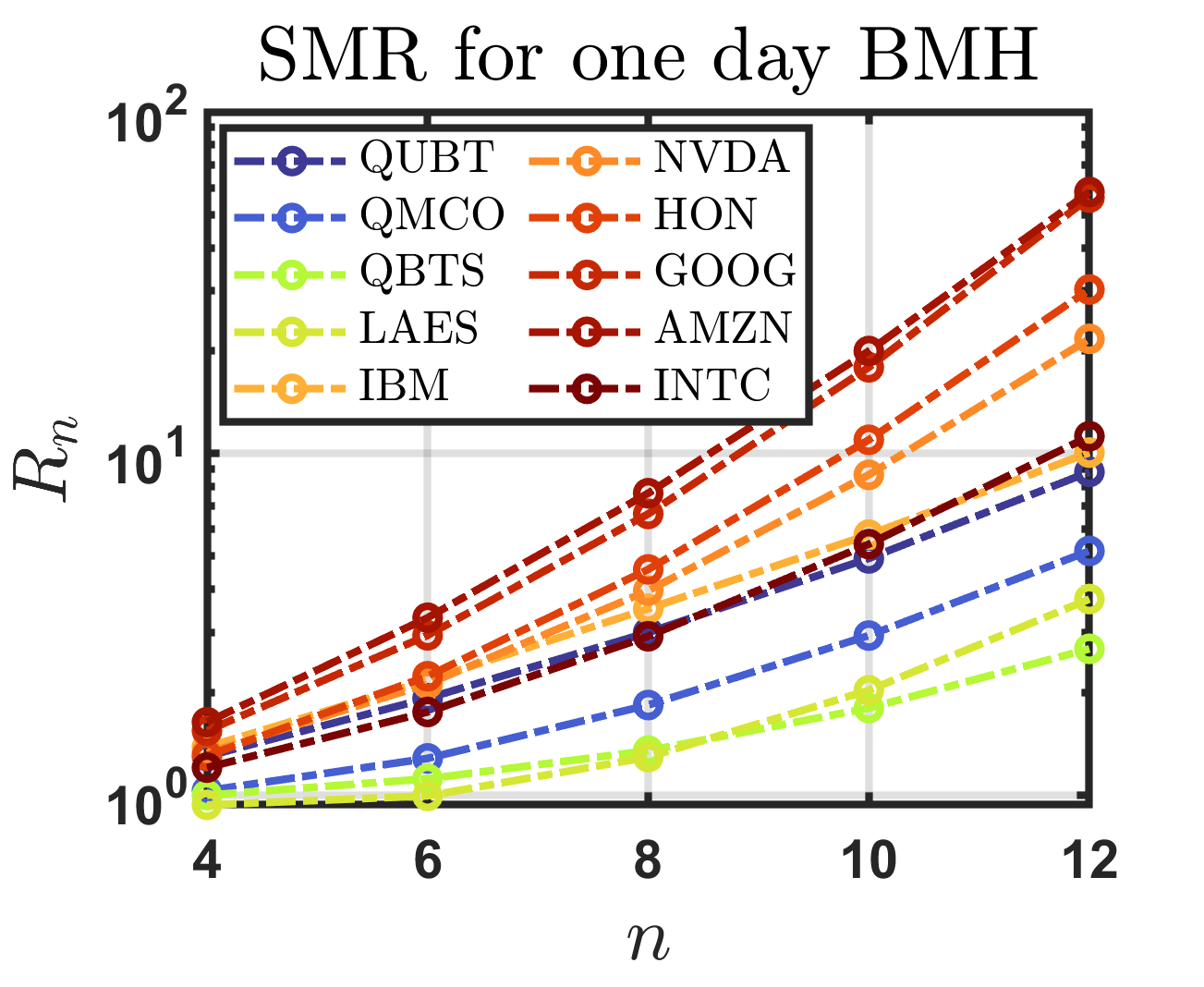}
    \includegraphics[width=5.75cm]{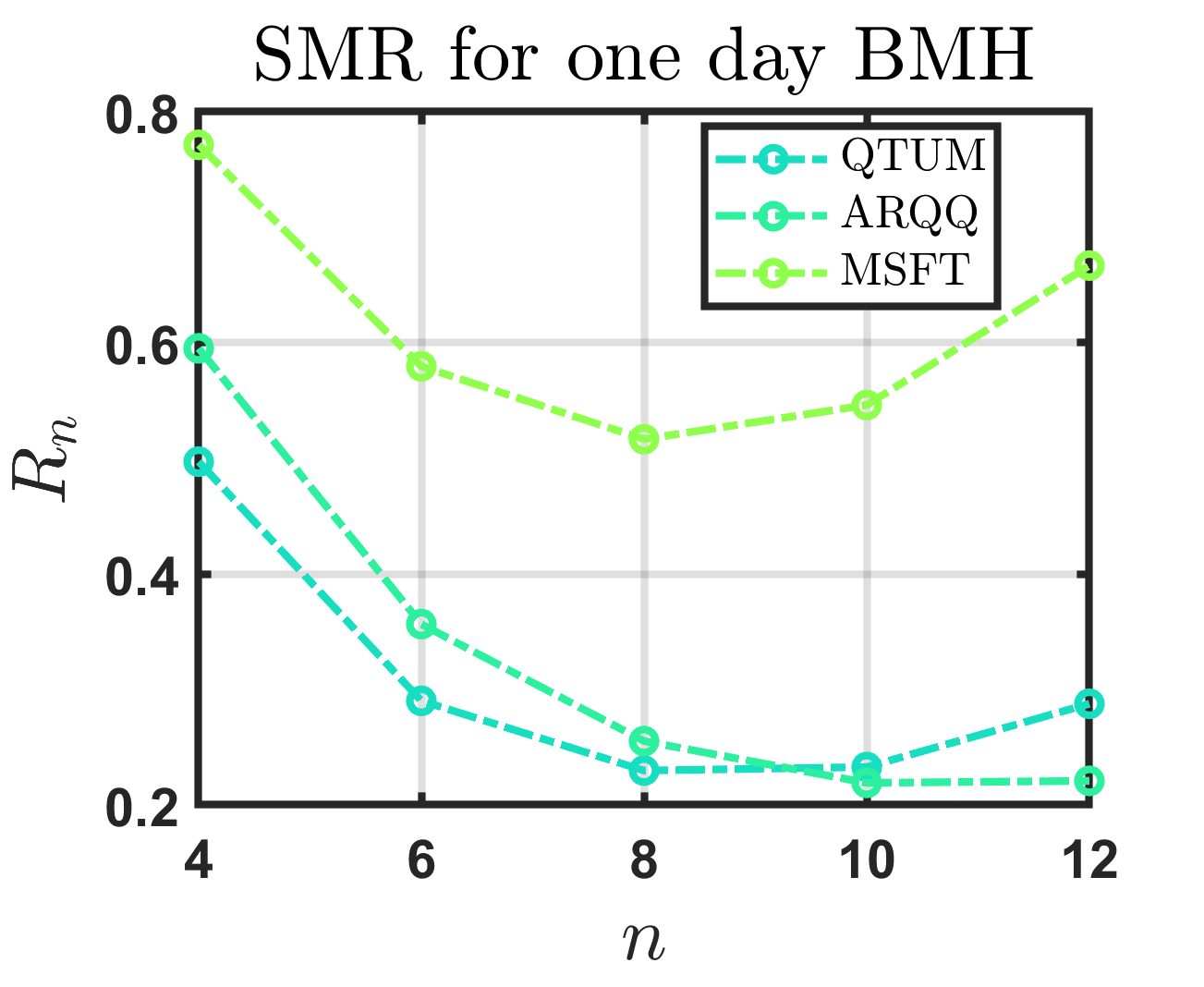}
    \includegraphics[width=5.75cm]{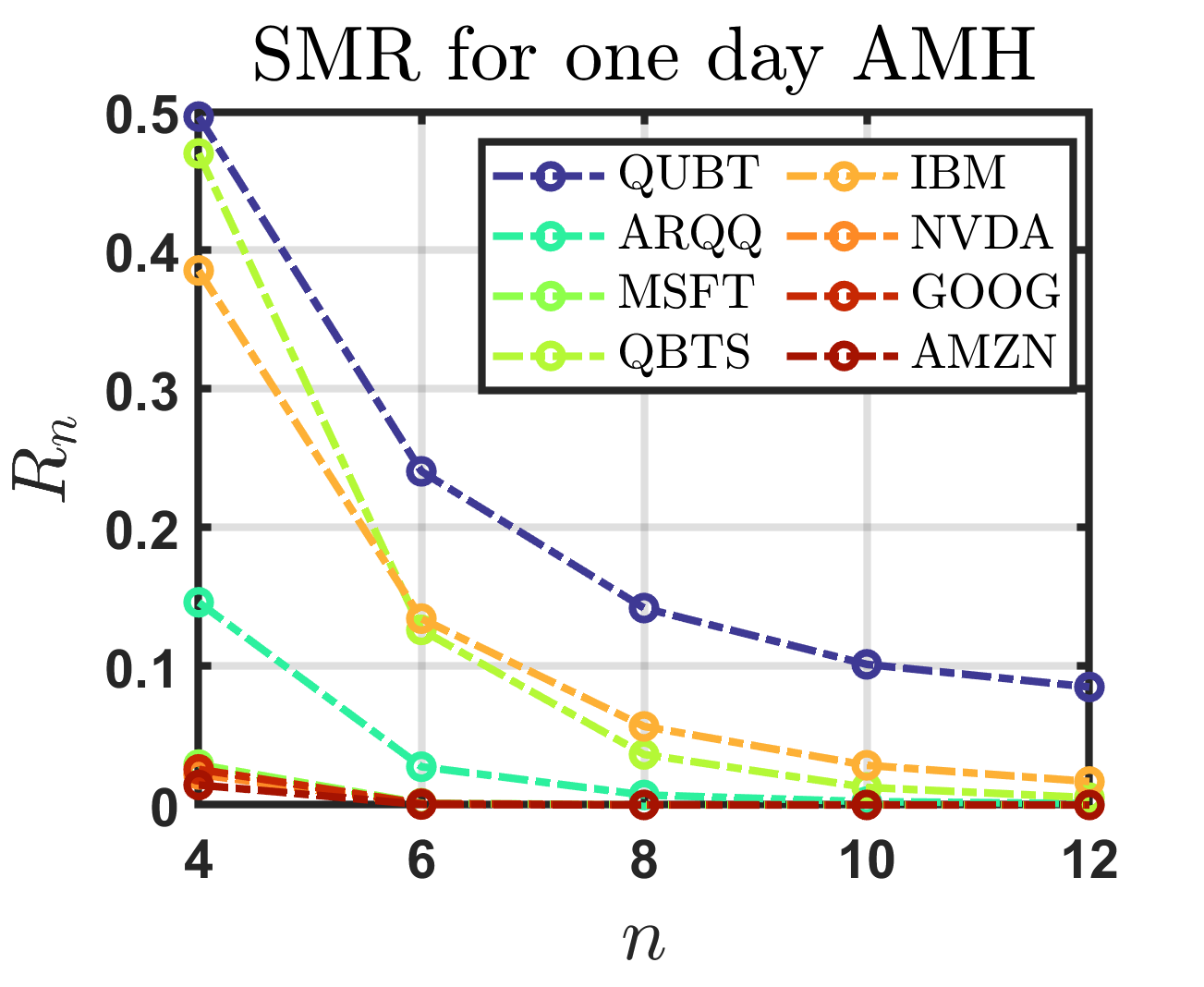}
    \includegraphics[width=5.75cm]{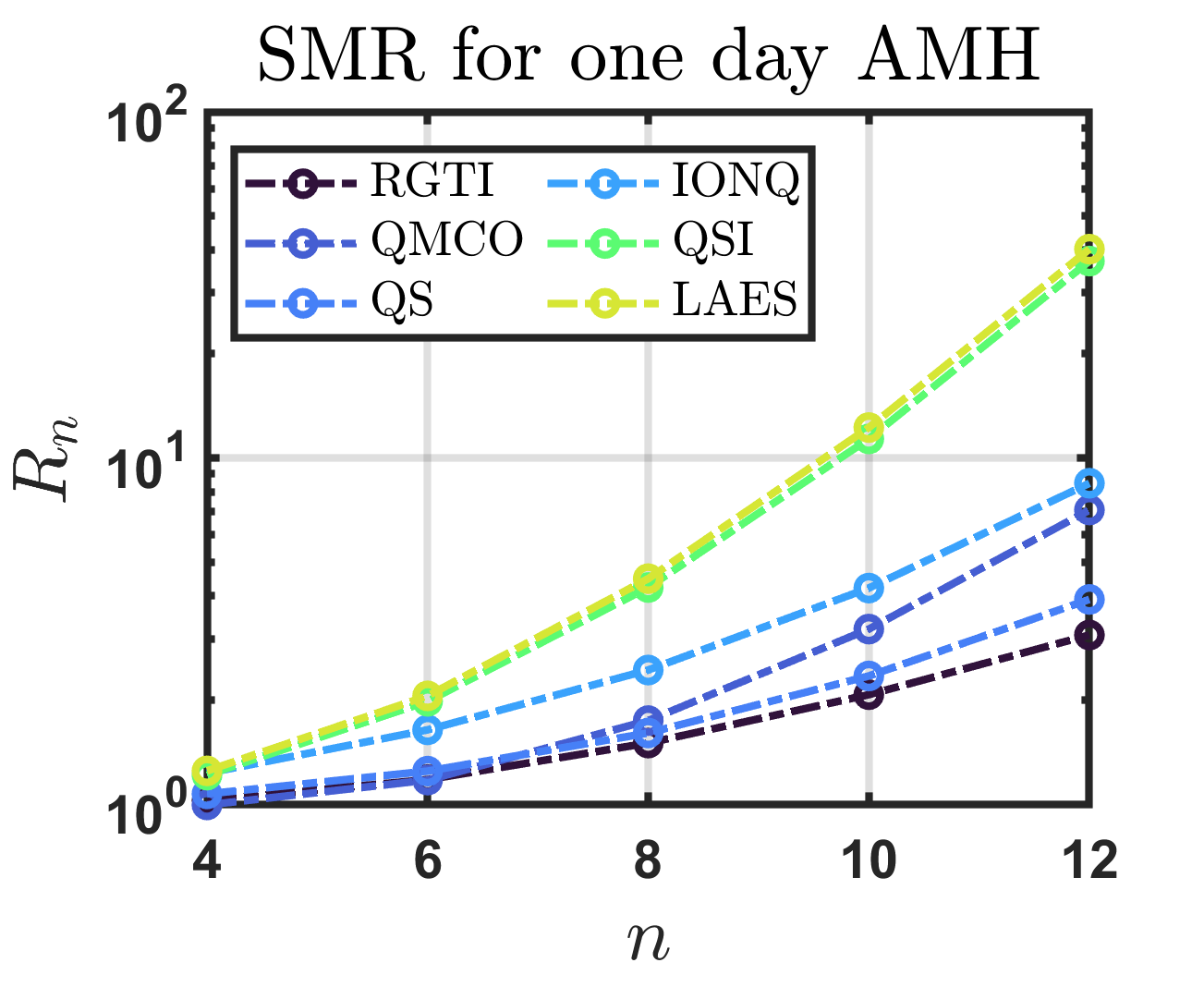}
    \includegraphics[width=5.75cm]{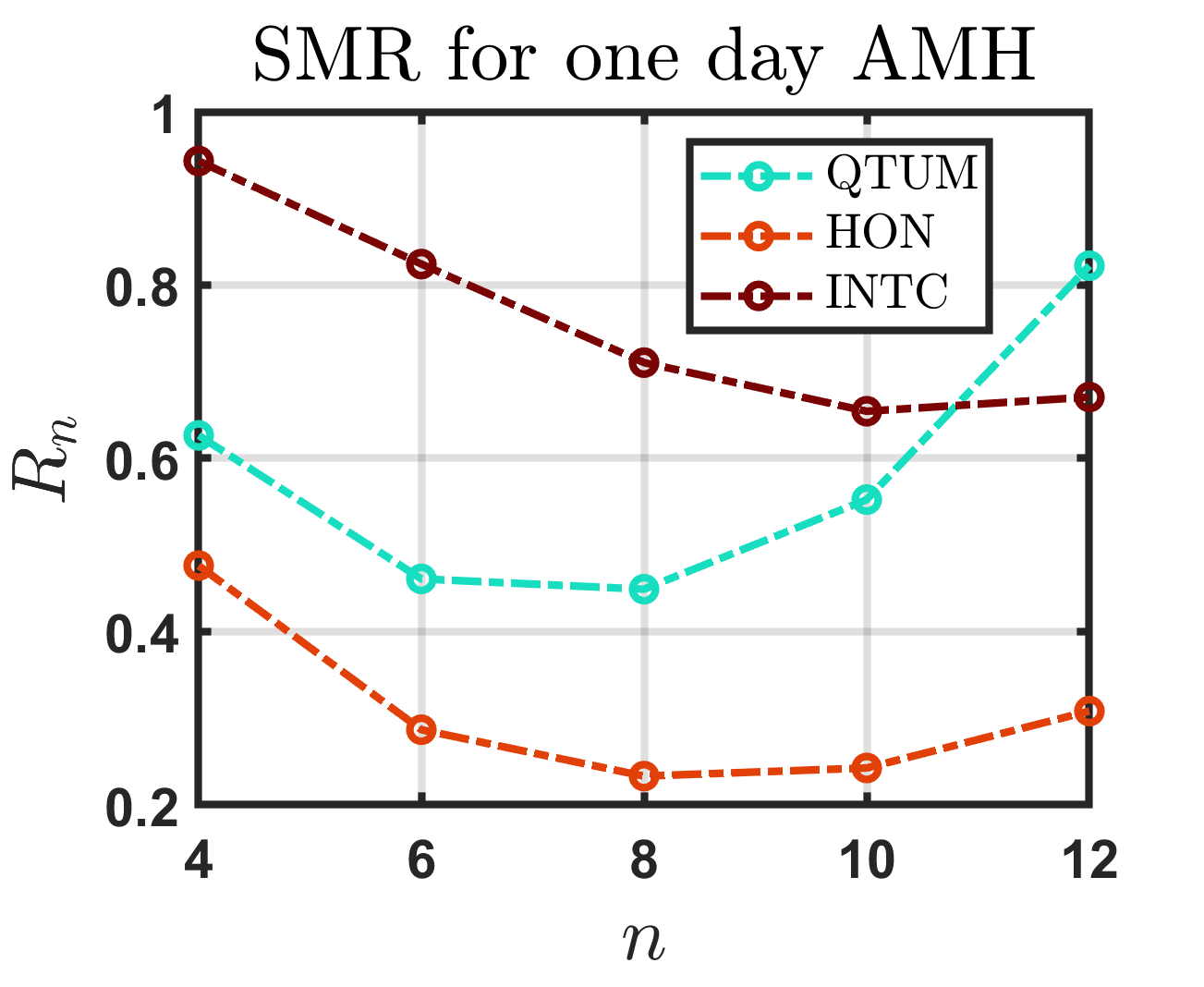}
    \caption{Standardised Moment Ratios for (a) April 11, 2020 - 2025 in-market hours (b) July 7, 2025 pre-market hours and (c) July 7, 2025 after-market hours. During the 5 year period, nine companies exhibit heavy tails (see top row -- left plot) while eight companies experience lighter tails (see top row -- middle plot). Three companies QTUM, IBM and HON have mixed tails. For pre-market hours (BMH), most companies exhibit lighter tails (see second row -- middle plot) indicating stable volume and high liquidity. Numerous companies during after-market hours (AMH) experience heavy tails suggesting low liquidity and high volatility.}
    \label{STOCKCORR}
\end{figure*}

Fig~\ref{STOCKCORR} demonstrates the dependence of SMR on moments $n$ for April 11 2020 - 2025 (top row) and July 7, 2025 (second and third row). During the 5 year period, micro/small-cap markets RGTI, QS, IONQ, ARQQ, QSI, LAES and ZPTA exhibit heavy tails ($R_n$ decreases with $n$) along with mega-cap markets QUBT and QNC.V. Heavy tails are typically observed in micro/small cap markets due to high volatility, thin order books and irregular liquidity. This generates irregular volume dynamics with severe and frequent spikes in trading activity. Market-wide shocks, episodic liquidity withdrawal and also institutional order-flow clustering account for the heavy tails observed in mega-cap markets QUBT and QNC.V. The Pearson correlation between $R_n$ for all moment orders ($n = 4,6,8,10,12$) ranges from 0.7940 to 0.9799. This indicates a monotonic decay pattern that is broadly similar, for these nine companies, differing in the mid-order moments $(n = 4,6)$. The companies, as a whole, exhibit partial systematic co-movement with an idiosyncratic structure in conditions that are normal or moderately extreme. However, when the correlations are restricted to higher moment orders $(n = 8,10,12)$, the minimum correlation rises sharply to 0.9799. This suggests that the companies share strong co-movement in the extreme tails, as the companies have nearly identical extreme tail structure, and they are likely to experience severe shocks in tandem (i.e. crash together in rare events). 

Mega-cap markets MSFT, NVDA, GOOG, AMZN and large-cap market INTC experience lighter tails as $R_n$ increases with $n$. This is expected as large/mega-cap markets have stable volume and high liquidity. We also observe lighter tails in small-cap markets QMCO, QBTS and FORM. For the lighter-tail companies, the minimum high order correlation (0.9799) is slightly higher than the correlation for all moments (0.9654). The $R_n$ decay pattern is highly similar across companies indicating strong systematic co-movement in the whole moment distribution. This reflects a more stable, more liquid and less jump-driven market behaviour with few idiosyncratic shocks (consistent risk structure). In addition to that, the high order correlation value shows that the companies are independent on the extreme tail convergence, unlike heavy tail companies, as the companies share a extreme tail mechanism that is common.

Non-monotonic behaviour between SMR and moments $n$ is seen for QTUM (small-cap), IBM (large-cap) and HON (large-cap). QTUM and IBM have strong systematic co-movement in the whole moment distribution with a correlation of 0.9990. This means they have similar liquid regimes, volatility bursts and trading patterns leading to an aligned body, mid-tail and tail behaviour. They form a sub-cluster inside the non-monotonic group as their non-monotonic patterns occur in similar places. HON is an outlier in the group with a correlation value of 0.3774 (with QTUM) and 0.4190 (with IBM). HON experiences idiosyncratic mid-order behaviour. Suprisingly, when considering higher order moments, all companies share a systematic extreme tail and are likely to experience rare, severe shocks simultaneously. 

\begin{table*}[t!]
    \centering

    \begin{tabular}{|c|c|c|c|c|c|c|c|c|c|c|c|c|c|c|c|c|c|c|c|c|}
        \hline 
        Company & RGTI & QUBT & QMCO & QS & IONQ & QNC.V & QTUM & ARQQ & QSI & MSFT  \\ \hline
         Number of Training Points & 598 & 754 & 754 & 701 & 643 & 753 & 754 & 607 & 663 & 754 \\ \hline
         Number of Testing Points & 400 & 503 & 503 & 468 & 430 & 502 & 503 & 405 & 443 & 503 \\ \hline
         Minimum DCV & 0.375 & 0.421 & 2.5099 & 3.470 & 3.099 & 0.0299 & 23.709 & 3.7899 & 0.6330 & 158.423 \\ \hline
         Maximum DCV & 20.0 & 25.68 & 189.0 & 131.67 & 51.07 & 1.7799 & 86.2317 & 951.5 & 22.4099 & 464.854 \\ \hline
    \end{tabular}
    \caption{Number of training point, testing points, minimum daily closing volume and maximum daily closing volume for the companies Rigetti Computing, Inc (RGTI), Quantum Computing Inc (QUBT), Quantum Corp (QMCO), QuantumScape Corporation (QS), IonQ Inc (IONQ), Quantum eMotion Corp (QNC.V), Defiance Quantum ETF (QTUM), Arqit Quantum Inc (ARQQ), Quantum-SI Incorporated (QSI), Microsoft (MSFT).}
    \label{table1data}
\end{table*}

\begin{table*}[t!]
    \centering
    \begin{tabular}{|c|c|c|c|c|c|c|c|c|c|c|c|c|c|c|c|c|c|c|c|c|}
     \hline 
        Company & QBTS & LAES & ZPTA & IBM & NVDA & FORM & HON & GOOG & AMZN & INTC \\ \hline
         Number of Training Points & 652 & 283 & 155 & 754 & 754 & 754 & 754 & 754 & 754 & 754 \\ \hline
         Number of Testing Points & 435 & 190 & 104 & 503 & 503 & 503 & 503 & 503 & 503 & 503 \\ \hline
         Minimum DCV & 0.41299 & 0.342 & 0.0003 & 83.268 & 6.713 & 18.190 & 110.8916 & 60.5303 & 81.8199 & 18.1299 \\ \hline
         Maximum DCV & 12.3999 & 21.96999 & 5.6999 & 264.7399 & 149.416 & 62.220 & 234.7436 & 207.4736 & 242.0599 & 62.0833 \\ \hline
    \end{tabular}
    \caption{Number of training point, testing points, minimum daily closing volume and maximum daily closing volume for the companies D-Wave Quantum Inc (QBTS), SEALQ Corp (LAES), Zapata AI (ZPTA), IBM (IBM), NVIDIA (NVDA), Form Factor Inc (FORM), Honeywell (HON), Google (GOOG), Amazon (AMZN) and Intel (INTC).}
    \label{table2data}
\end{table*}

In pre-market hours for July 7, 2025 volume data (see Fig~\ref{STOCKCORR} second row), heavy tail companies (RGTI, QS, IONQ, QSI) show strong $R_n$ shape similarity with a minumum correlation of 0.8602 across all $n$ and almost perfectly aligned extreme tails with minimum correlation value of 0.9931. Lighter tail companies (QUBT, QMCO, QBTS, LAES, IBM, NVDA, HON, GOOG, AMZN, INTC) have a highly consistent and systematic risk structure with minimum correlation of 0.9750 across all moments and 0.9915 in the higher order moments. The non-monotonic companies (QTUM, ARQQ, MSFT) show 0.7590 correlation in all moments indicating substantial mid-order idiosyncrasy. The higher order correlation -0.6188 observed for ARQQ and MSFT reveals that the extreme tail irregularities move in opposite directions. This means no common pre-market tail risk mechanism is shared. 

In after-market hours (see Fig~\ref{STOCKCORR} last row), the minimum correlation for all moments and higher moments are 0.9414 and 0.9672 respectively for heavy tail companies QUBT, ARQQ, MSFT, QBTS, IBM, NVDA, GOOG and AMZN. This indicates strong shape similarity and closely aligned extreme tails. Lighter tail companies RGTI, QMCO, QS, IONQ, QSI and LAES also have a stable and systematic risk structure as minimum correlations 0.9715 and 0.9878 are observed in all $n$ and higher order $n$ respectively. Computing the correlation of QTUM with the other non-monotonic companies HON and INTC we get negative minimum correlations -0.1532 in all $n$ and -0.4859 in higher $n$. We see no similarity across full moment distribution and extreme-tail behaviour move in opposite direction. 

RGTI, QS, QSI and IONQ exhibit heavy tails and lighter tails in pre-market hours and after-market hours respectively. Some companies can exhibit such behaviour in pre-market hours due to thin order books, irregular liquidity and overnight information shocks and in after-market hours due to stable liquidity and low volatility. This shows that tail behaviour is session-dependent (risky in one session and stable in another) with pre-market behaviour dominated by idiosyncratic jumps and after-market by systematic predictable flows.

\section{Results and discussion}
\subsection{Five Year Stock Market Prediction}
\label{nonautonomouspred}

Future values of the input-time series $u(t_k)$, which is the Daily Closing Volume (DCV) data of each company, independently, over the period April 11, 2020 - 2025 (a 5-year span), are predicted via non-autonomous prediction.  

\begin{figure*}[htp!]
    \centering
    \includegraphics[width=7.25cm]{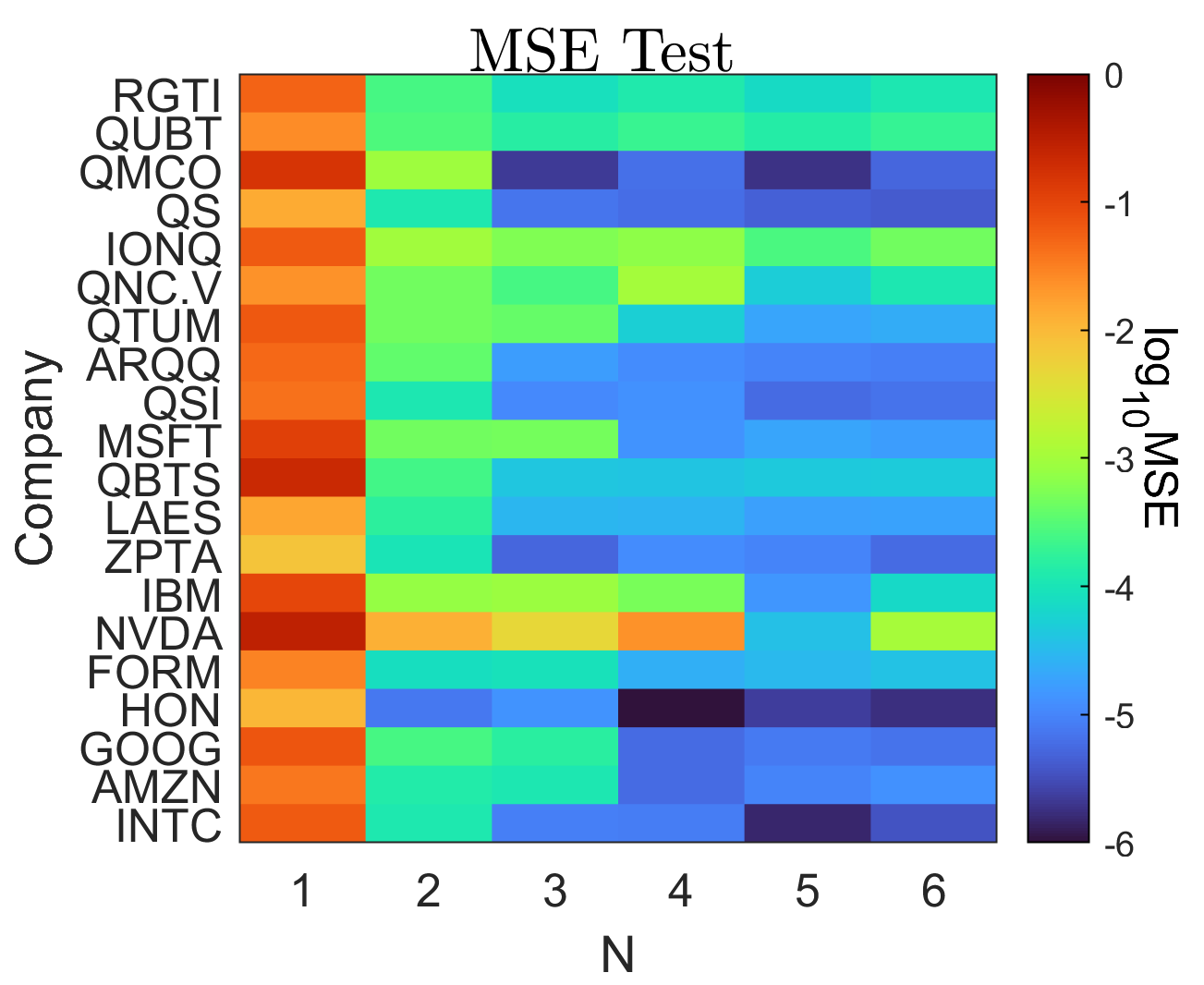}
    \includegraphics[width=7.25cm]{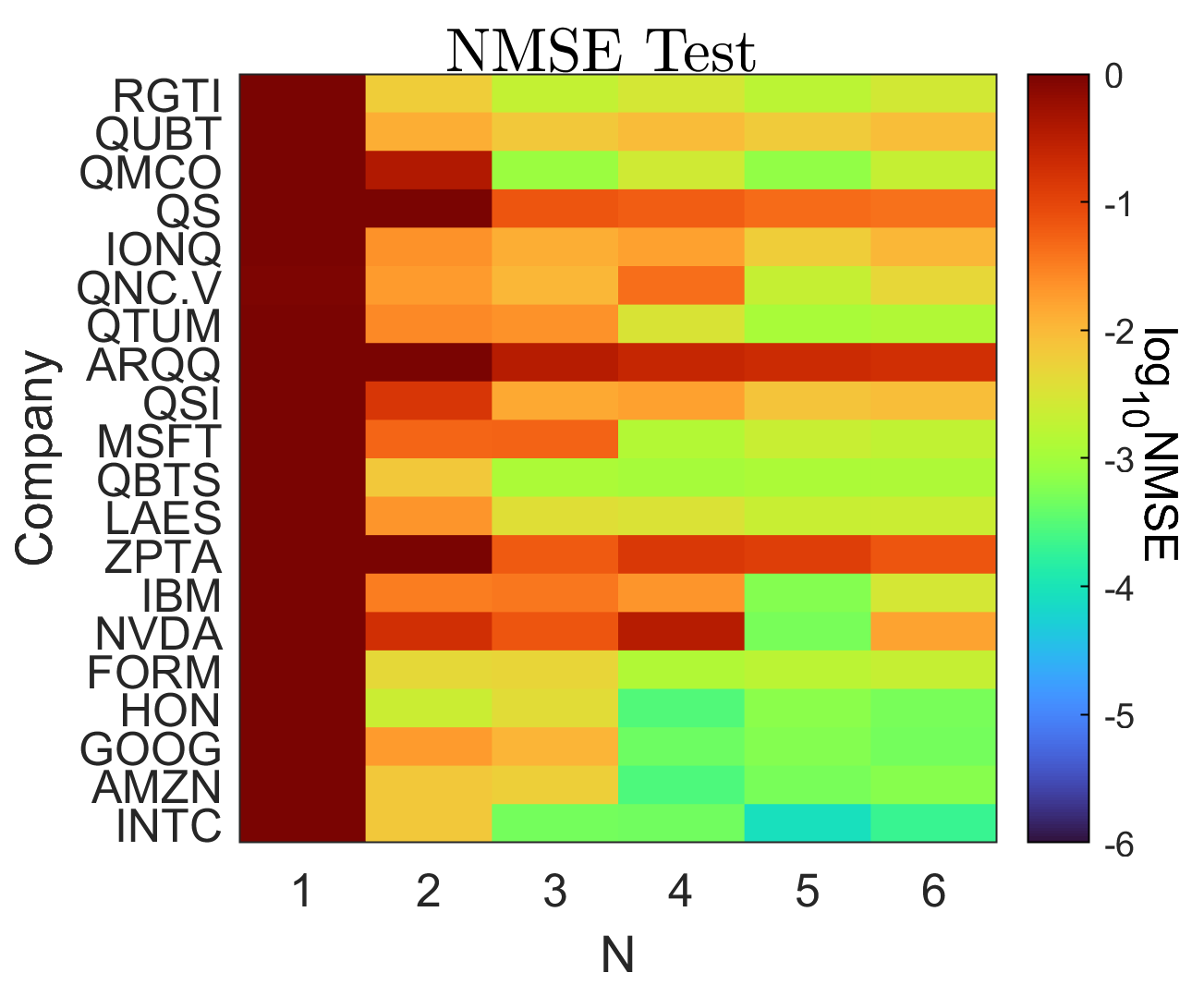}
    \includegraphics[width=7.25cm]{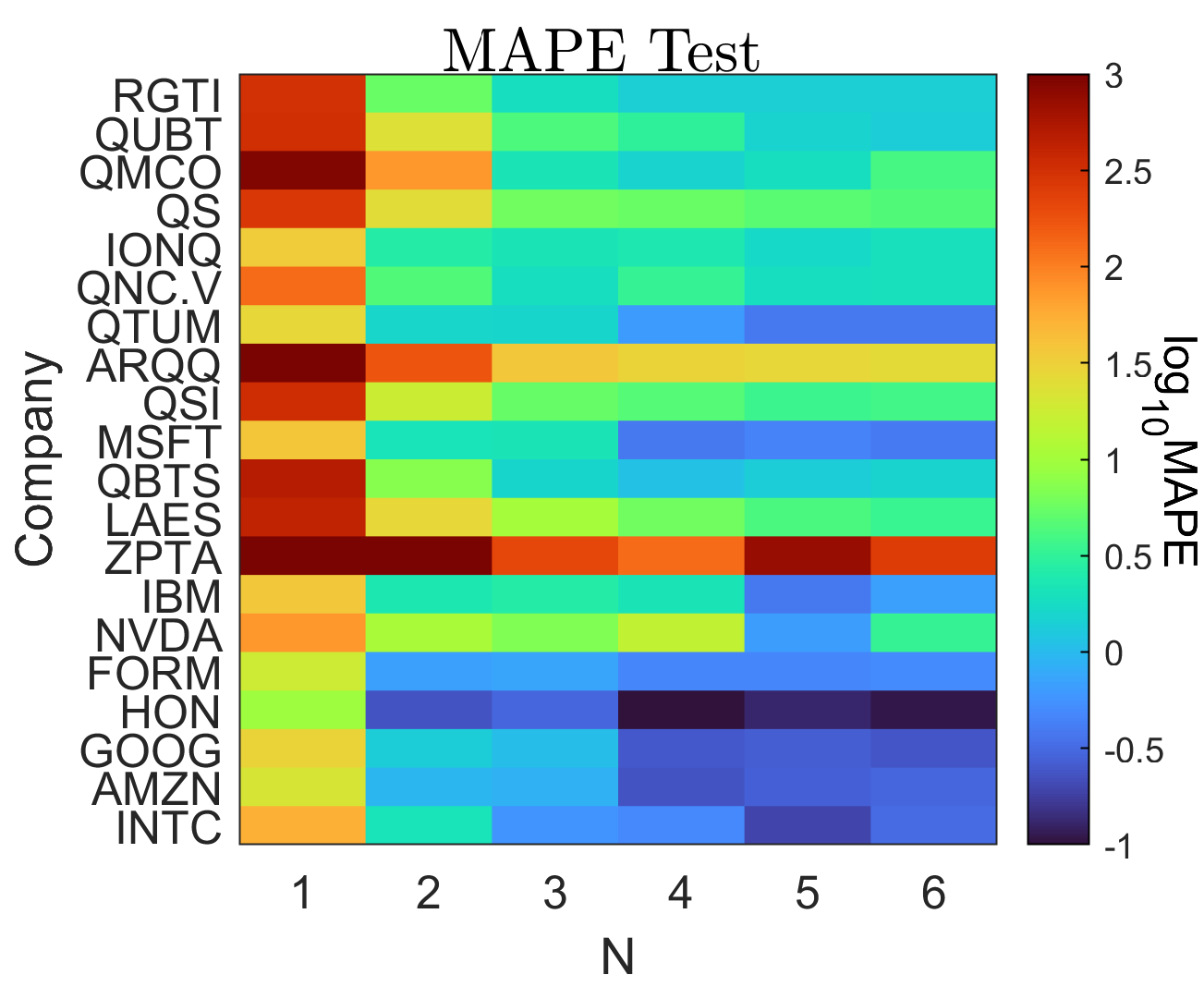}
    \includegraphics[width=7.25cm]{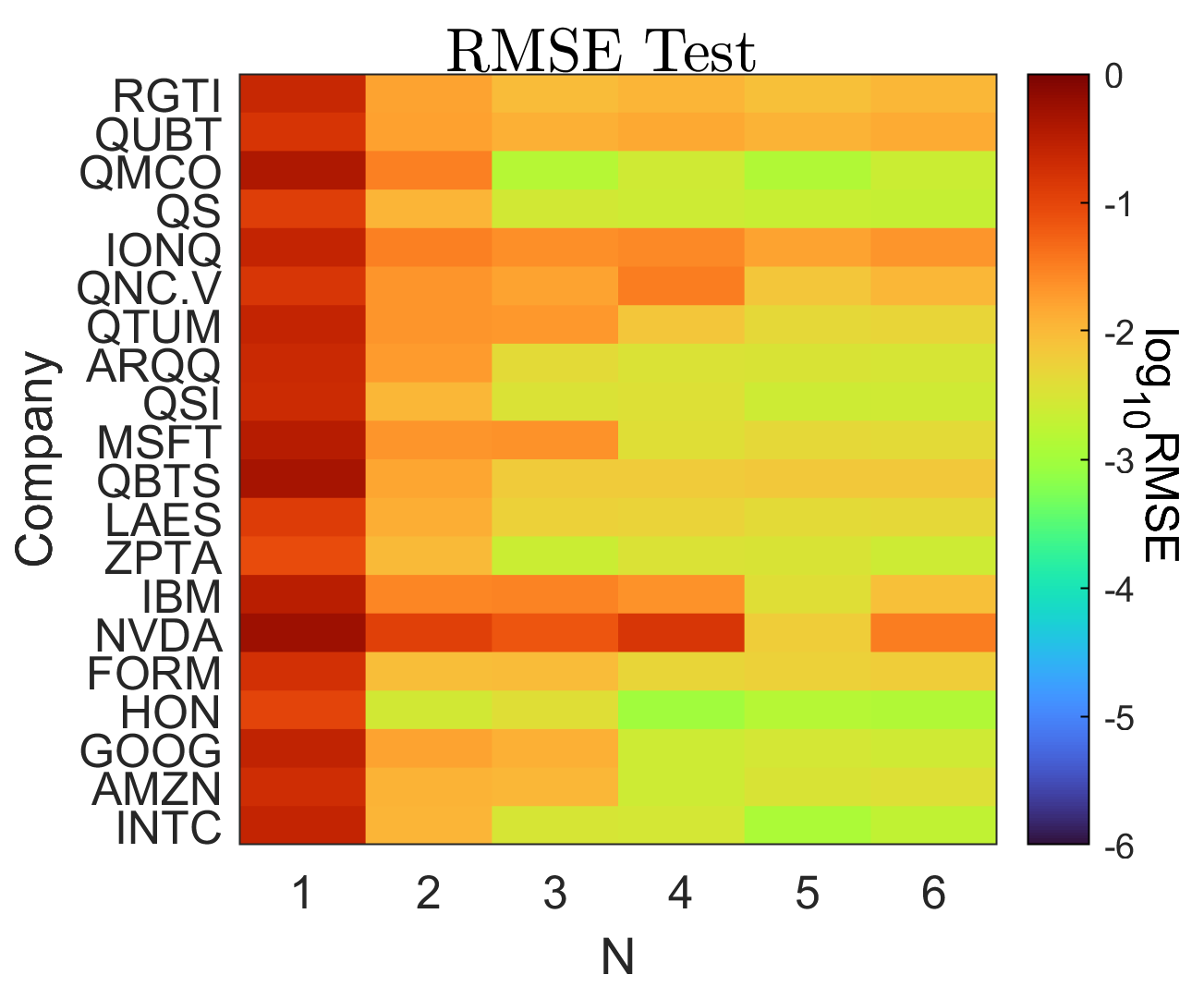}
    \caption{2D Colormap showcasing the dependency of MSE Test (top left), NMSE Test (top right), MAPE Test (bottom left) and RMSE Test (bottom right) on the number of qubits $N$ for the 20 quantum-invested companies. The MSE Test values range from $9.9301 \times 10^{-7}$ - $0.2879$. MSE Test, NMSE Test, RMSE Test and MAPE Test peak at $N = 1$ and slightly/drastically drop depending on the company at $N \geq 2$. At N = 5, eleven companies (RGTI, QUBT, QMCO, IONQ, QNC.V, QTUM, QSI, LAES, IBM, NVDA, INTC) experience low MSE, NMSE and RMSE.}
    \label{MSETestResults}
\end{figure*}

For each company, $60\%$ and $40\%$ of the data are set as the training and testing dataset respectively. Tables \ref{table1data} and \ref{table2data} show the number of training and testing points for each company along with the minimum and maximum DCV over the 5 year span. Initially, we divide the DCV by the maximum value so that the initial timeseries $u(t_k)$ ranges between $0$ and $1$ for a better performance from the QRC. We use the QR parameters $\Delta_0 = 8$ (mean detuning), $\Omega_0 = 6$ (mean rabi frequency), $T_{e} = \pi$ , $N_{steps} = 3000$ (number of time steps in the evolution) and $c_{\gamma} = 1 \times 10^{-8}$ (decoherence rate) to evaluate how the number of qubits $N \in [1,6]$ affects the stock market accuracy of the QRC framework. To incorporate memory into the QRC framework, we add delay embeddings by setting $\delta = 6$.

Fig~\ref{MSETestResults} shows the dependency of MSE Test, NMSE Test, MAPE Test and RMSE Test on the number of qubits $N$ as a 2D colormap. $N = 1$ yields the highest MSE Test, for each company, ranging from 0.0076 to 0.2879. HON has the lowest error, MSE test = $ 9.9301 \times 10^{-7}$, at $N = 4$. Overall, $N = 5$ yields the lowest error for 11 companies (RGTI, QUBT, QMCO, IONQ, QNC.V, QTUM, QSI, LAES, IBM, NVDA, INTC). MSFT, QBTS, FORM, HON, GOOG and AMZN exhibit the least MSE test at $N = 4$, ZPTA at $N = 3$ and QS and ARQQ at $N = 6$. The MSE Test ranges from $9.9301 \times 10^{-7}$ - $0.2879$. It is a non-monotonic function of $N$.

Similar to MSE Test, NMSE Test and RMSE Test have the same companies with the least errors at $N = 5$ (RGTI, QUBT, QMCO, IONQ, QNC.V, QTUM, QSI, LAES, IBM, NVDA, INTC), $N = 4$ (MSFT, QBTS, FORM, HON, GOOG, AMZN), $N = 6$ (QS, ARQQ) and $N = 3$ (ZPTA). RMSE Test ranges from $9.9650 \times 10^{-4}$ - $1.5028 \times 10^{-1}$ for $N \geq 2$ indicating extremely accurate fit between the QRC testing prediction and the actual targeted testing time-series. When $N = 1$, RMSE Test varies from $8.7245 \times 10^{-2}$ - $5.3659 \times 10^{-1}$ meaning the prediction for companies such as NVDA (0.5369), QBTS (0.4672) and QMCO (0.3969) e.t.c are quite poor. The ten companies with the best MAPE Test for $N \geq 2$ are IONQ, QNC.V, QTUM, MSFT, IBM, FORM, HON, GOOG, AMZN and INTC with an error that is less than $5\%$ (see Fig~\ref{MSETestResults} left bottom plot). HON has a highly accurate MAPE Test value ranging from $0.0922\%$ - $0.2975\%$. ZPTA and ARQQ have poor MAPE Test values ($>20\%$) for all $N$.

\begin{figure}[htp!]
    \centering
    \includegraphics[width=\linewidth]{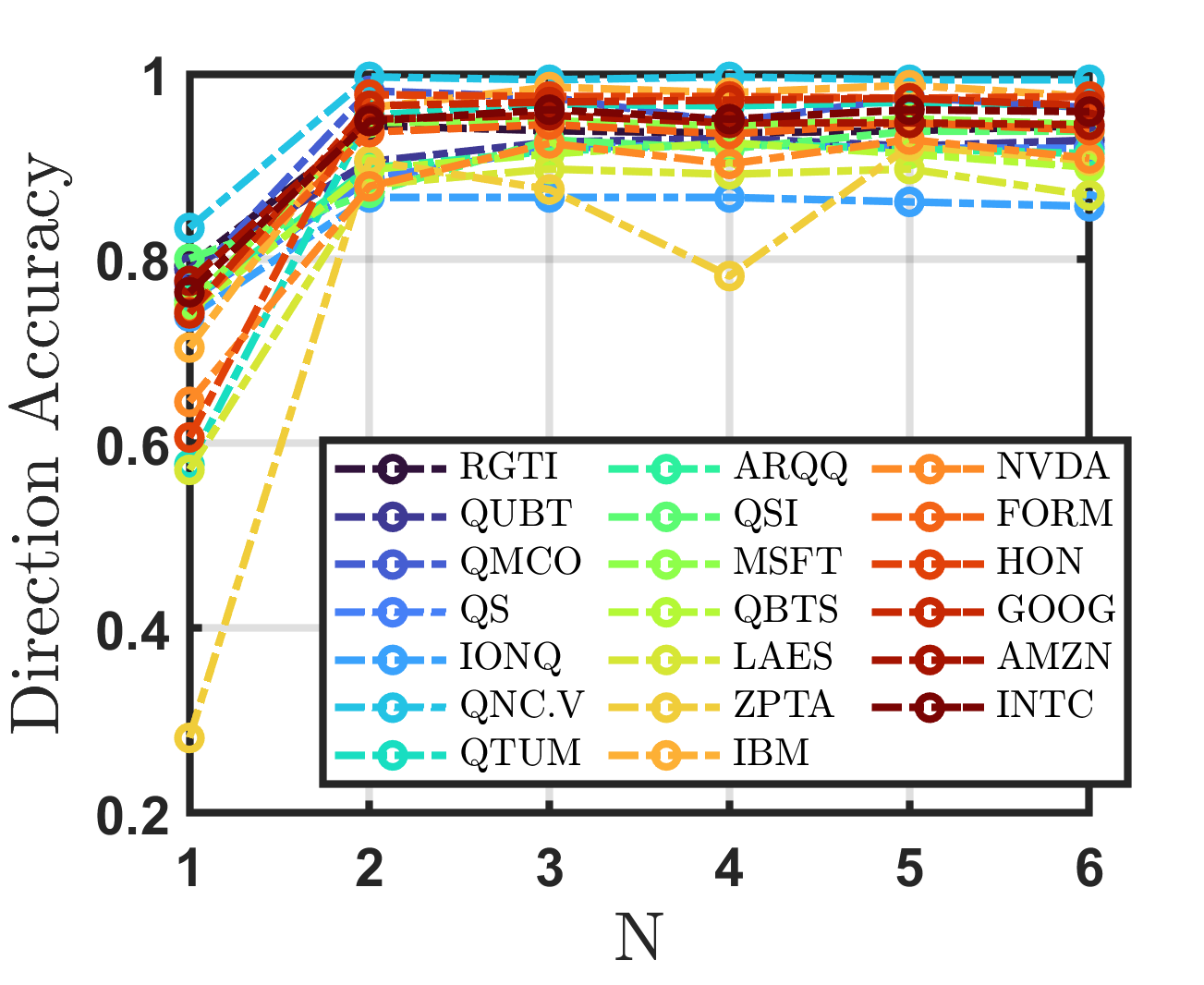}
    \caption{The direction accuracy is a non-monotonic function of the number of qubits $N$. For nineteen companies, the accuracy is greater than $85\%$ when $N \geq 2$. ZPTA has the lowest accuracy at $N = 1$ and $N = 4$ with DA = 0.2813 and 0.7813 respectively.}
    \label{DAvsNcompanies}
\end{figure}

Fig~\ref{DAvsNcompanies} shows the direction accuracy (DA) of each company for $N = 1$ - $6$ qubits and $\delta = 6$. Surprisingly, one qubit is more than enough for the QRC to sufficiently predict the direction trend (up/down) in daily closing volumes with a DA $> 70\%$ for 15 companies and DA $> 77\%$ for 6 companies (RGTI, QUBT, QMCO, QNC.V, QSI, AMZN). Overall, the DA for QNC.V is the highest at $N = 2$ and $N = 4$ (DA = $0.9969$) indicating the non-monotonic dependency of DA on $N$. Nineteen companies have an accuracy greater than $85\%$ for $N \geq 2$. ZPTA yields the lowest DA at $N = 1$ (0.2813) and $N = 4$ (0.7813).

In conclusion, $N \geq 2$ is more than enough for the HPE-QRC (Hamiltonian Parameter Encoded QRC) framework to predict the up/down trend in DCV with high accuracy. To have high DA and low MSE, NMSE and RMSE in the QRC output, the best choice is $N \geq 3$. 

\subsubsection{Benchmarking MLP, ESN, QIESN and HPE-QRC}
The performance of the HPE-QRC is compared across various machine learning ML models: Echo State Network (ESN), Quantum Inspired Echo State Network (QIESN) and  Multilayer Perceptron (MLP). MLP and ESN are purely classical models while QIESN incorporates quantum-inspired feature maps to a classical ESN to enrich the reservoir's dynamics. We choose the optimal number of qubits, $N = 5$, for the HPE-QRC framework since all companies experience low MSE, NMSE, MAPE and RMSE values yielding the best predictive performance.

\begin{figure}[b!]
    \centering
    \includegraphics[width=8cm]{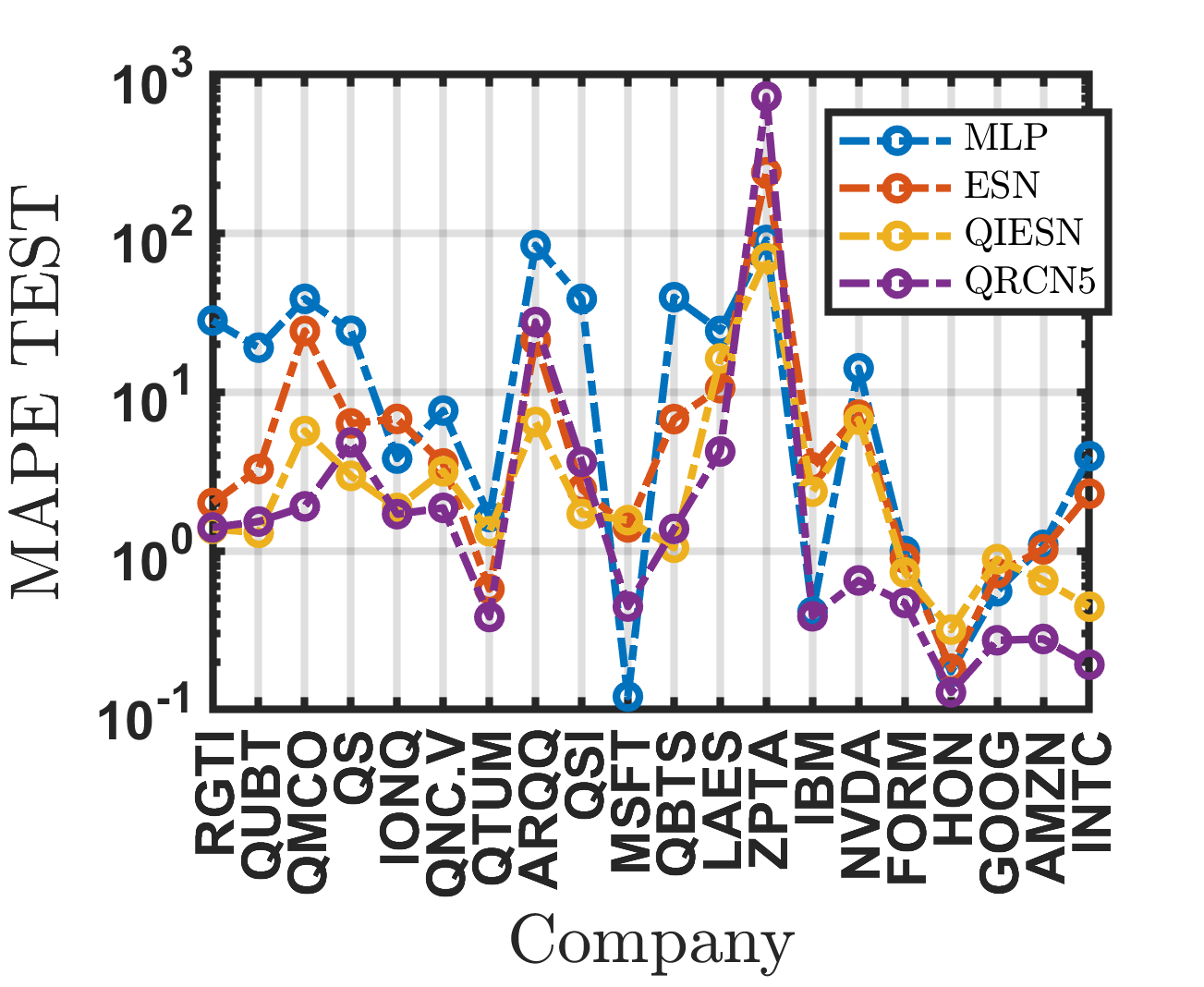}
    \includegraphics[width=8cm]{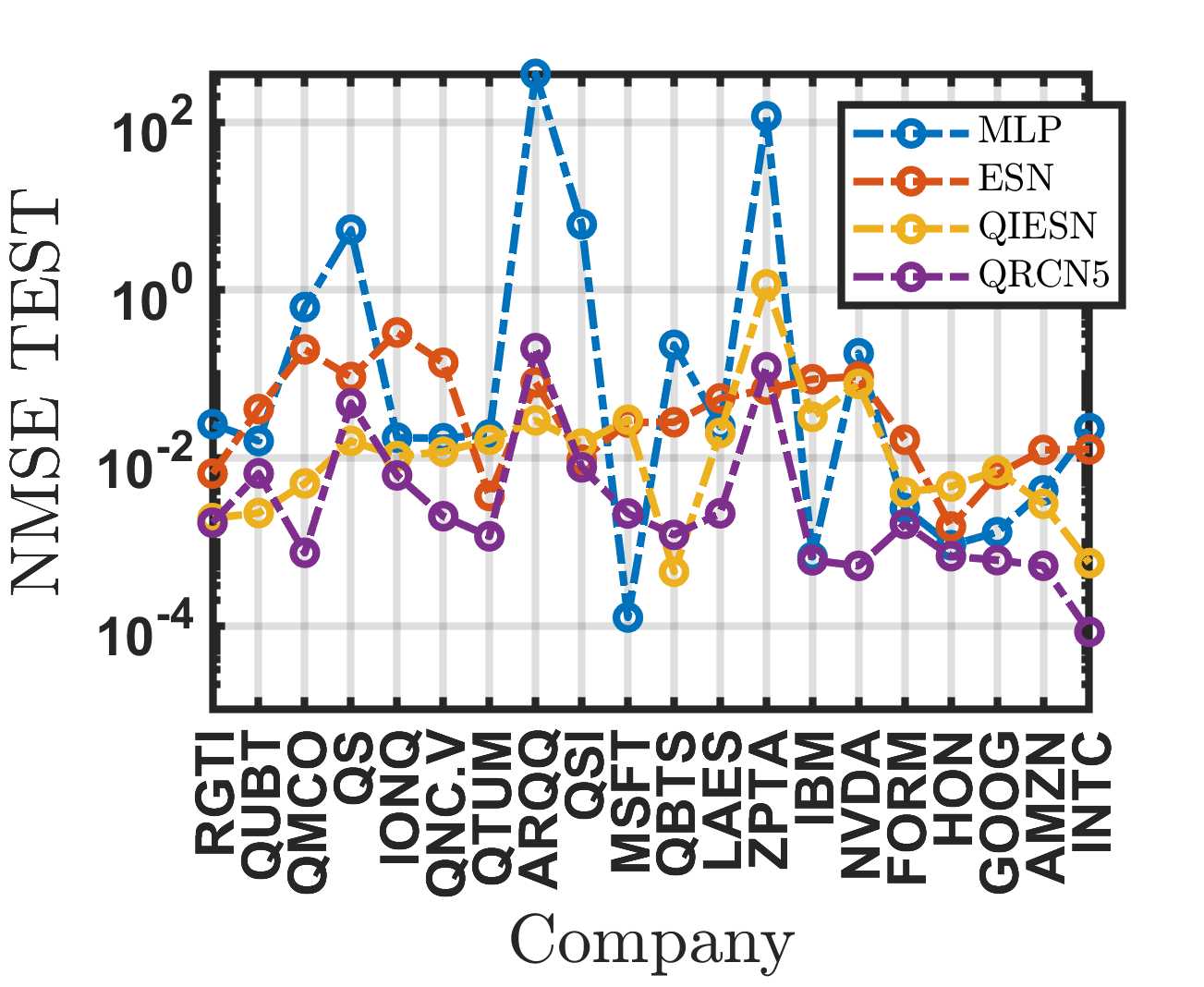}
    \caption{MAPE TEST (top) and NMSE Test (bottom) of MLP (blue), ESN (orange), QIESN (yellow) and HPE-QRC (purple) when HPE-QRC has five qubits. HPE-QRC outperforms MLP, ESN and QIESN in 12 companies for MAPE Test and 14 companies for NMSE Test.}
    \label{NMSEMAPEvsNcompaniescompare}
\end{figure}

Each model is implemented for the purpose of DCV prediction over a 5 year span with the following parameters:
\begin{enumerate}
\item MLP: Structure - Dense(156) $\rightarrow$ Dense(136) $\rightarrow$ Dense(1), Activation function = ReLU, Optimizer = Adam, Epochs = 60, Batch Size = 16, Input = $60\%$ of DCV dataset, (Training Set, Validation Set) = ($50\%$ Input, $50\%$ Input).
\item ESN: Number of Reservoirs =  400, Spectral Radius = 0.95, Leak rate = 0.8.
\item QIESN: Number of Reservoirs = 400, Spectral Radius = 0.95, Leak rate = 0.8, Sparsity = 0.1, Random State = 42, Quantum Feature Map = Sin(x).

\item HPE-QRC: $N = 5$, $\Delta_0 = 8$, $\Omega_0 = 6$, $c_{\gamma} = 1 \times 10^{-8}$, $\delta = 6$.
\end{enumerate}
We compare the MAPE Test, NMSE Test and DA results to see the competitive nature of the HPE-QRC framework with the notable techniques.  

Fig~\ref{NMSEMAPEvsNcompaniescompare} (top) compares the MAPE Test for all four models when HPE-QRC has five qubits in the QR. HPE-QRC outperforms MLP, ESN and QIESN in 12 companies (QMCO, IONQ, QNC.V, QTUM, LAES, IBM, NVDA, FORM, HON, GOOG, AMZN, INTC) as it predicts DCV values numerically close to the targeted dataset for these companies. Solely comparing HPE-QRC with MLP, ESN and QIESN independently, HPE-QRC dominates in 18, 17 and 13 respectively. Overall QIESN and HPE-QRC outperform MLP and ESN. MAPE Test has the worst performance at $N = 1$ indicating one qubit does not focus on the quantitative behaviour of the targeted dataset.

HPE-QRC also outperforms MLP, ESN and QIESN in 14 companies when considering NMSE Test (see Fig~\ref{NMSEMAPEvsNcompaniescompare} (bottom)). The same behaviour is observed for MSE Test and RMSE Test. At $N = 1$, HPE-QRC has its worst performance across all companies.

\begin{figure}[htp!]
    \centering
    \includegraphics[width=8.5cm]{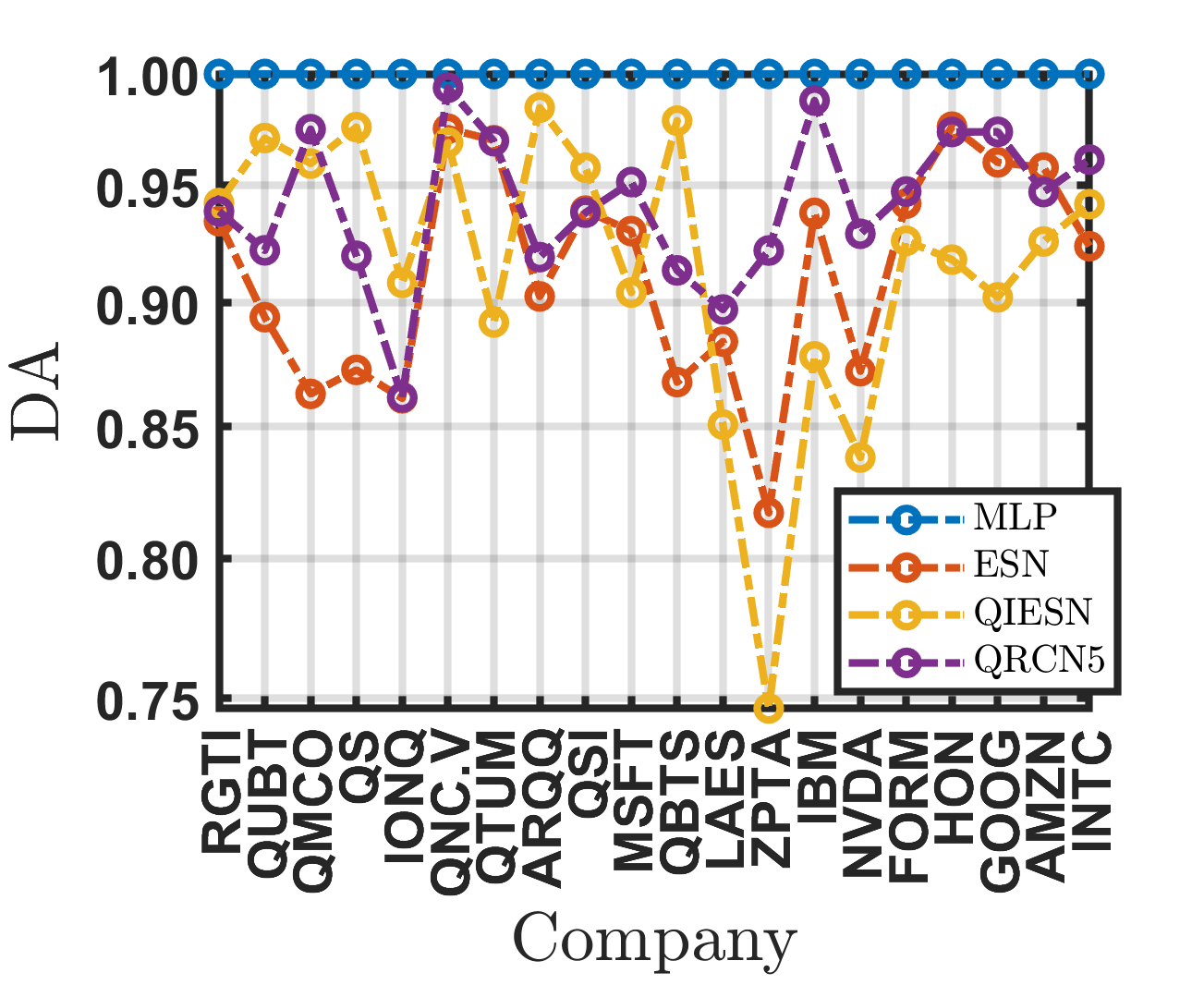}
    \caption{HPE-QRC outperforms ESN and QIESN in 16 and 13 companies respectively for $N = 5$ where DA $> 85 \%$. MLP performs better than ESN, QIESN and HPE-QRC for all companies.}
    \label{DAvsNcompaniescompare}
\end{figure}

Fig~\ref{DAvsNcompaniescompare} shows the direction accuracy of MLP, ESN, QIESN and HPE-QRC. Overall, MLP captures the direction trend compared to ESN, QIESN and HPE-QRC by having the highest DA for all companies. However, when comparing HPE-QRC with QIESN and HPE-QRC with ESN, HPE-QRC has a higher DA in thirteen and sixteen companies respectively. 

\subsection{One Day Stock Market Prediction}

In this section, we evaluate the performance of the QRC on one day stock market prediction. We take the volume dataset of July 7, 2025 and divide the dataset into three sections: Pre-market hours (4:00 - 9:29 a.m. EST), In-market hours (9:30 a.m. - 4:00 p.m. EST) and After-market hours (4:01 - 8:00 p.m. EST). The volume data is sampled at one-minute interval for each company. The dataset for the pre-market and after-market hours are further split into training points ($60\%$) and testing points ($40\%$). We investigate the QRC's performance on forecasting future values of pre-market and after-market hours using one-step non-autonomous prediction. 

The motivation in evaluating the QRC's performance on out of hours stock prediction (pre-market and after-market hours) stems from assessing the QRC's capabilities in capturing the high volatility and low liquidity in out of hours stock. The heightened volatility and reduced liquidity can be attributed to news-driven sporadic price jumps which frequently coincide with major information releases (ex. technological milestones, funding announcements, market reports and forecasts, partnerships, etc) that are quite often announced outside regular trading hours. By capturing the nuanced dynamics of out-of-market hours, the QRC can anticipate transitional behaviour and potential spillover effects which unfold in regular trading hours.

In general, we evaluate the effects of $\Delta_0$, $\Omega_0$ and $N$ on the overall performance of the QRC for one day stock trend prediction. We vary $\Delta_0 \in [1,10]$, $\Omega_0 \in [1,10]$ and $N \in [1,6]$ for all companies with the exception of QNC.V, FORM and ZPTA which lack pre-market and after-market volume data.

\setlength{\tabcolsep}{4pt}
 \begin{table*}[t!]
    \centering
    \begin{tabular}{|c|c|c|c|c|c|c|c|}
     \hline 
        Company & Highest DA & $(N,\Delta_0,\Omega_0)$ & Lowest MSE & Lowest NMSE & Lowest MAPE & Lowest RMSE \\ \hline
         RGTI & 0.8737 & $(5,1,4$-$10)$, $(5,2,8$-$10)$  & $4.1689 \times 10^{-6}$ & $0.2368$ & 0.1622 & 0.0020 \\  \hline
         QUBT & 0.9167 & $(5,1,5$-$10)$, $(5,2,8$-$10)$ & $2.5398 \times 10^{-6}$ & $0.1651$ & 0.1239 & $1.5937 \times 10^{-3}$ \\  \hline
         QMCO & 1 & $(3,3,4)$, $(3,5,7)$ & $8.2296 \times 10^{-6}$ & $1.0771$ & 0.2598 & $2.8687 \times 10^{-3}$ \\ 
         {}  & {} & $(3,7,10)$, $(5,1,2)$  & {} & {} & {} & {} \\ 
         {}  & {} & $(5,2,4$-$6)$, $(5,3,7$-$10)$ & {} & {} & {} & {} \\ 
         {} & {} & $(5,4,10)$ & {} & {} & {} & {} \\ \hline
         QS  & 0.9831 & $(5,1,2$-$7)$, $(5,2,4$-$10)$  & $5.6813 \times 10^{-7}$ & 0.0674 & 0.0647 & $7.5374 \times 10^{-4}$ \\ 
         {}  & {} & $(5,3,6$-$10)$, $(5,4,8$-$10)$  & {} & {} & {} & {} \\ 
         {} & {} & $(6,1,1$-$10)$, $(6,2,5$-$10)$ & {} & {} & {} & {} \\ 
         {}  & {} & $(6,3,8$-$10)$ & {} & {} & {} & {} \\ \hline
         ARQQ  & 1 & (6,2,2), (6,4,5) & $7.0777 \times 10^{-6}$ & 0.4161 & 0.2269 & $2.6604 \times 10^{-3}$ \\ 
         {}  & {} & $(6,5,6$-$7)$, $(6,6,[7,9])$ & {} & {} & {} & {} \\ 
         {}  & {} & $(6,7,8)$, $(6,8,9$-$10)$ & {} & {} & {} & {} \\ \hline
         QSI  & 0.8889 & (1,1,2), (1,2,4) & $1.2970 \times 10^{-6}$ & 0.0456 & 0.0986 & $1.1388 \times 10^{-3}$ \\ 
         {}  & {} & $(1,3,6)$, $(1,4,8)$ & {} & {} & {} & {} \\
         {}  & {} & $(1,5,9$-$10)$, $(5,1,8$-$10)$ & {} & {} & {} & {} \\ \hline
         QBTS  & 0.8864 & $(5,1,7$-$10)$, $(6,2,4$-$10)$ & $1.9799 \times 10^{-6}$ & 0.2024 & 0.1255 & $1.4071 \times 10^{-3}$ \\ 
         {} & {} & $(6,1,6$-$10)$, $(6,3,6$-$10)$ & {} & {} & {} & {} \\ 
         {}  & {} & $(6,4,8$-$10)$, $(6,5,10)$ & {} & {} & {} & {} \\ \hline
         LAES  & 0.8684 & $(4,2,[3,4])$, $(4,3,5$-$7)$ & $4.0388 \times 10^{-6}$ & 0.2951 & 0.1357 & $2.0097 \times 10^{-3}$ \\
         {}  & {} & $(4,4,7$-$10)$, $(4,5,9$-$10)$ & {} & {} & {} & {} \\
         {}  & {} & $(4,6,10)$, $(5,1,1)$, $(5,3,4)$ & {} & {} & {} & {} \\
         {}  & {} & $(5,4,6)$, $(5,5,7)$, $(5,6,9)$ & {} & {} & {} & {} \\ 
         {}  & {} & $(5,7,10)$, $(6,3,3)$, $(6,4,4)$ & {} & {} & {} & {} \\
         {} & {} & $(6,8,9)$, $(6,9,10)$  & {} & {} & {} & {} \\
         \hline 
         HON  & 1 & (2,1,1), (5,3,2), $(5,3,5)$ & $9.2616 \times 10^{-6}$ & 0.9922 & 0.2672 & $3.0433 \times 10^{-3}$ \\
         {}  & {} & $(5,4,7)$,$(5,5,9)$, $(5,7,5)$  & {} & {} & {} & {} \\
         {}  & {} & $(5,8,6)$ & {} & {} & {} & {} \\ \hline 
    \end{tabular}
    \caption{The highest directional accuracy (DA) for the companies RGTI, QUBT, QS, ARQQ, QSI, QBTS, LAES and HON occur in multiple parameters $(N,\Delta_0,\Omega_0)$ where $N = 5$, $N = 5$, $N = [3,5]$, $N = [5,6]$, $N = 6$, $N = [1,5]$, $N = [5,6]$, $N = [4,5,6]$ and $N = [2,5]$ respectively. The parameters which yield the lowest MSE Test, NMSE Test and RMSE Test values occur at $(5,1,10)$, $(5,1,10)$, $(5,4,10)$, $(5,1,7)$, $(6,6,9)$, $(5,1,10)$, $(5,1,10)$, $(4,5,10)$ and $(5,3,5)$ for RGTI, QUBT, QS, ARQQ, QSI, QBTS, LAES and HON respectively with a MAPE Test that is less than $0.27\%$.}
    \label{table3data}
\end{table*}

\subsubsection{Pre-Market Hours}

Optimal parameters $(N,\Delta_0,\Omega_0)$ exist for nine companies (RGTI - 37, QUBT - 57, QMCO - 80, QS - 436, ARQQ - 140, QSI - 133, QBTS - 48, LAES - 22 and HON - 7) that yield a DA that is greater than $86\%$. The number of optimal parameters that exhibit the highest DA are dependent on the company (see Table~\ref{table3data}). For example QUBT has it highest DA (DA $= 0.9167$) in nine parameters $(N,\Delta_0,\Omega_0) = (5,1,5$-$10)$ and $(5,2,8$-$10)$ while QS has a high DA of $0.9831$ in 41 parameters i.e. $(N,\Delta_0,\Omega_0) = (5,1,2$-$7)$, $(5,2,4$-$10)$ and many more. QMCO, ARQQ and HON have multiple parameters with $DA = 1$ since the number of data points are extremely small (less than thirty points) making it easier for the parameters to enhance the QRC's ability to capture trend patterns.   

The lowest values for MSE Test, NMSE Test, MAPE Test and RMSE Test occur at $(5,1,10)$, $(5,1,10)$, $(5,4,10)$, $(5,1,7)$, $(6,6,9)$, $(5,1,10)$, $(5,1,10)$, $(4,5,10)$ and $(5,3,5)$ for RGTI, QUBT, QS, ARQQ, QSI, QBTS, LAES and HON respectively. This shows that parameters including $N = 5$ lead to low MSE, NMSE, MAPE and RMSE Test with a high DA. Also, the MAPE TEST in these parameters are less than $0.27\%$ indicating an excellent fit. 

To find optimal parameters that include all 17 companies, we would need to set the requirement DA $ > 0.68$ since MSFT has a maximum DA value of $0.6833$ in six parameters. In fact the maximum DA value for RGTI, QUBT, QMCO, QS, IONQ, QTUM, ARQQ, QSI, MSFT, QBTS, LAES, IBM, NVDA, HON, GOOG, AMZN and INTC are $0.8737$, $0.9167$, $1$, $0.9831$, $0.8544$, $0.8571$, $1$, $0.8889$, $0.6883$, $0.8864$, $0.8684$, $0.8372$, $0.7217$, $1$, $0.8539$, $0.72$ and $0.7949$ respectively. This shows that the QRC's predictive performance for pre-hours market data is good since parameters exists for each company that give satisfactory DA results for MSFT (DA $ = 0.6833$), good results for NVDA, AMZN and INTC ($0.71 < $ DA $< 0.8$) and great results for the remaining 13 companies (DA $> 0.83$). 

\begin{figure}[htp!]
    \centering
    \includegraphics[width=\linewidth]{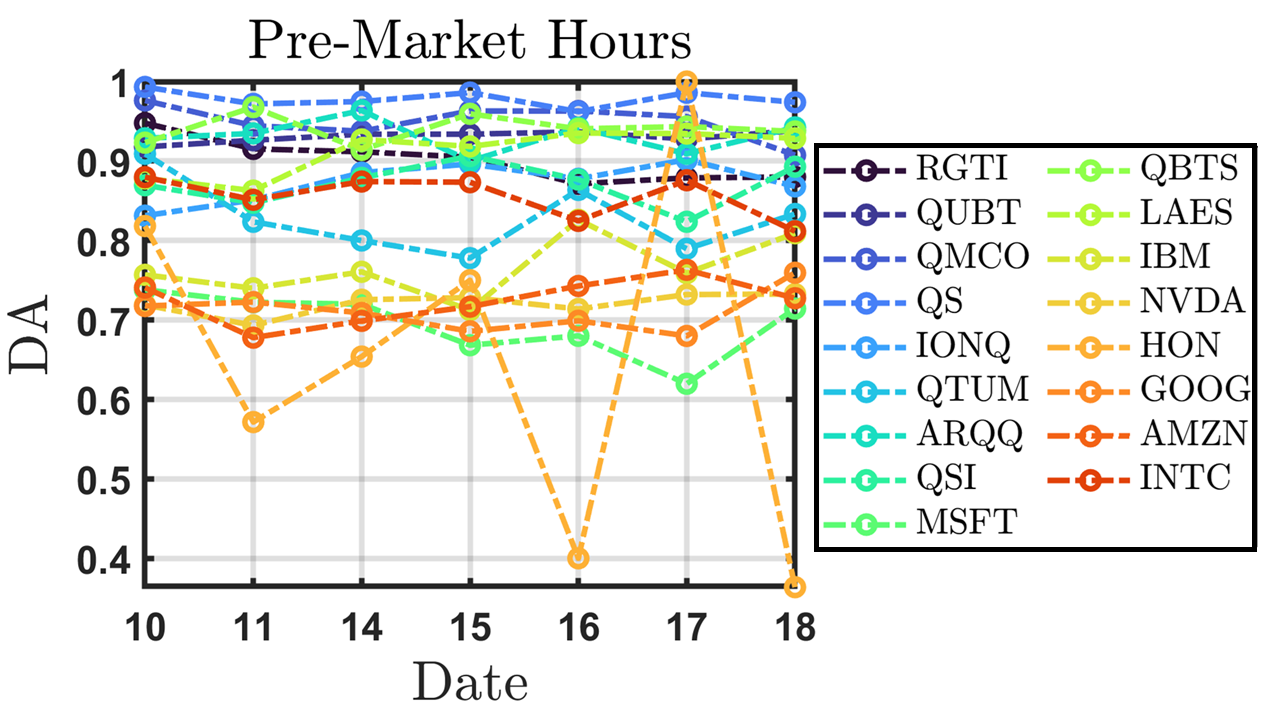}
    \caption{Direction Accuracy (DA) for 17 companies using July 7, 2025 dataset as training and predicting up/down trends for July 10-11 and July 14-18. Using the parameters $(N,\Delta_0,\Omega_0) = (5,1,10)$ we find a DA accuracy greater than $69\%$ for thirteen companies.}
    \label{BMHfutureprediction}
\end{figure}

\setlength{\tabcolsep}{3pt}
 \begin{table*}[t!]
    \centering
    \begin{tabular}{|c|c|c|c|c|c|c|c|}
     \hline 
        Company & Highest DA & $(N,\Delta_0,\Omega_0)$ & Lowest MSE & Lowest NMSE & Lowest MAPE & Lowest RMSE \\ \hline
         QUBT & 0.8615 & $(6,5,[2,6])$, $(6,6,7)$, $(6,7,9)$& $2.9809 \times 10^{-7}$ & $1.3323$ & 0.0466 & $5.4598 \times 10^{-4}$ \\  
         {} & {} & $(6,8,10)$ & {} & {} & {} & {} \\  \hline
         QMCO & 0.8889 & $(2,3,1)$, $(3,4,3)$, $(3,5,4)$ & $3.0688 \times 10^{-5}$ & $1.3521$ & 0.4484 & 0.0055 \\ 
         {}  & {} & $(3,6,5)$, $(3,7,6)$, $(3,8,7$-$8)$  & {} & {} & {} & {} \\ 
         {}  & {} & $(3,9,8$-$9)$, $(3,10,[6,9,10])$, $(4,9,3)$ & {} & {} & {} & {} \\ 
         {} & {} & $(6,4,2), (6,5,2), (6,10,4)$ & {} & {} & {} & {} \\ \hline
         QS  & 0.8889 & $(6,3,6)$, $(6,4,8)$  & $1.9828 \times 10^{-6}$ & 1.5254 & 0.1173 & 0.0014 \\ \hline
         QBTS  & 0.8696 & $(6,2,4$-$6)$, $(6,3,7$-$10)$ & $7.8058 \times 10^{-7}$ & 0.8165 & 0.0769 & $8.8351 \times 10^{-4}$ \\ 
         {} & {} & $(6,4,9$-$10)$& {} & {} & {} & {} \\ \hline
         HON  & 1 & $(1,6,4$-$10)$, $(2,1,2$-$10)$ & $9.4326 \times 10^{-7}$ & 0.9867 & 0.0848 & $9.7122 \times 10^{-4}$ \\
         {}  & {} & $(3,7,[3,6$-$10])$, $(4,6,[2,6$-$10])$ & {} & {} & {} & {} \\ 
         {}  & {} & $(5,8,[1,3,7$-$10])$, $(6,10,[2,4,6$-$7])$, ... & {} & {} & {} & {} \\\hline 
    \end{tabular}
    \caption{QUBT, QMCO, QS, QBTS and HON have the highest DA in multiple parameters $(N,\Delta_0,\Omega_0)$ where $N = 6$, $N = [2,3,4,6]$, $N = 6$, $N = 6$, $N = 6$ and $N = 1$-$6$ respectively. HON has 346 parameters which exhibit the highest DA at DA = 1. We show thirty seven parameters in this table for HON. The lowest MSE Test, NMSE Test and RMSE Test values occur at $(6,7,9)$, $(3,8,8)$, $(6,4,8)$, $(6,4,9)$, and $(5,3,4)$ for QUBT, QMCO, QS, QBTS and HON respectively with a MAPE Test that is less than $0.5\%$.}
    \label{table4data}
\end{table*}

In most cases, low DA (for example DA $< 0.20$) means that the framework's output is a reflection of the desired output. This suggests that certain parameters $(N,\Delta_0,\Omega_0)$ cause the QR to generate inverted output relative to the desired stock market data while others align with the orientation of the desired output. Therefore, the reservoir dynamics robustly encode directional features and one can utilize polarity inversion as a tool for predictive alignment. 

Additionally, we train the QRC using the dataset from July 7, 2025 for each company independently and see if it can predict future stock trend patterns for July 10-11 and July 14-18. We choose parameters $(N,\Delta_0,\Omega_0) = (5,1,10)$ to implement this predictive task. Fig~\ref{BMHfutureprediction} shows the DA for July 10-11 and July 14-18 using July 7 as the training dataset. We find that the stock trend prediction for these parameters are greater than $77\%$ overall for eleven companies (RGTI, QUBT, QMCO, QS, IONQ, QTUM, ARQQ, QSI, QBTS, LAES and INTC)
indicating that despite being trained on only one day, it is enough for the QRC to demonstrate great predictive performance across future pre-market days. Adding MSFT, NVDA, GOOG, IMB and AMZN leads to DA $>61\%$ across all dates. The predictive performance of HON is poor on July 11, 16 and 18 with DA = $0.574$, $0.4$ and $0.3636$ respectively. However, for July 10 and 17, the DA is high as DA = 0.8181 and 1 respectively.

Surprisingly, as the days increased, the DA did not monotonically decrease. In sixteen companies, July 18 has a DA $ > 70\%$. Maintaining a high accuracy over time without retraining indicates that the inductive bias of the QRC is great. It is able to extract and retain meaningful features just by using limited data indicating strong model generalization. This suggests a form of temporal invariance where the QRC can consistently model cyclical liquidity behaviour that rear across Pre-market sessions. 

\subsubsection{After-Market Hours}

Table~\ref{table4data} shows the optimal parameters with the highest DA for QUBT, QMCO, QS, QBTS and HON. DA $> 0.86$ is observed exclusively in these companies for after-market DCV data. The lowest value of MSE Test, NMSE Test and RMSE Test for QUBT, QMCO, QS, QBTS and HON are seen in the parameters $(6,7,9)$, $(3,8,8)$, $(6,4,8)$, $(6,4,9)$, and $(5,3,4)$ respectively. The MAPE values for these parameters are less than $0.5\%$. 

The order of magnitude of MSE Test is less than -5 and the NMSE Test ranges from 0.8165 - 1.5254. Because after-market hours have little trading volume, the variance shrinks in quiet market conditions making NMSE appear large relative to the MSE.

Table~\ref{table3data} and Table~\ref{table4data} show that the direction accuracy in pre-market hours dataset are better than after-market hours. This is because the response to previous close news is much more settled before pre-market hours begins due to the extended time to process and evaluate the information without active trading pressure. When pre-market hours open, the market responds with more direction and intentionality which causes the increase of volume and reliability. This results to smoother trajectories being formed by pre-market data in the quantum reservoir state space. Thus, the QRC's ability to predict input patterns that it can detect and learn from efficiently is enhanced.

\begin{figure}[htp!]
    \centering
    \includegraphics[width=\linewidth]{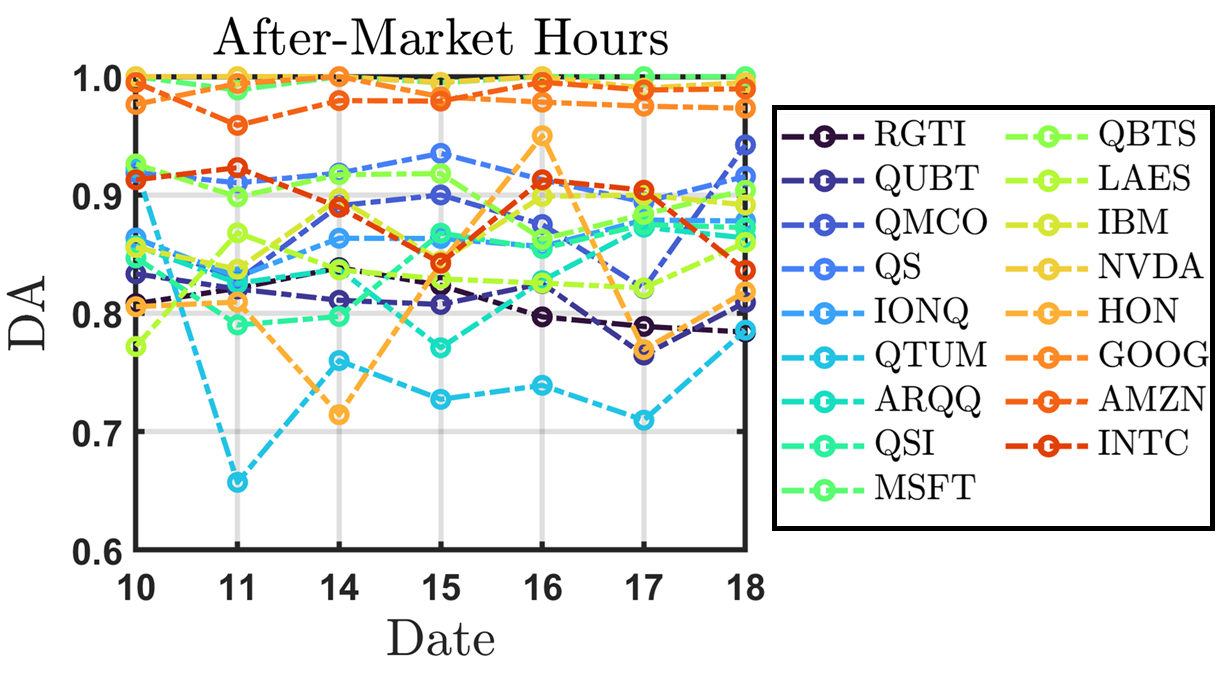}
    \caption{Using the parameters $(N,\Delta_0,\Omega_0) = (5,1,10)$, we find a DA accuracy greater than $76\%$ for 15 companies when predicting After-Market hours for July 10 - 11 and July 14 - 18. The After-Market hours dataset for July 7, 2025 is used as the training dataset.}
    \label{AMHfutureprediction}
\end{figure}

Fig~\ref{AMHfutureprediction} shows the DA for July 10 - 11 and July 14 - 18 for After-Market hours. Again, the QRC is trained on After-Market hours for July 7, 2025. We again choose the parameter $(N,\Delta_0,\Omega_0) = (5,1,10)$ for all 17 companies. DA $> 76 \%$ for 15 companies (RGTI, QUBT, QMCO, QS, IONQ, ARQQ, QSI, MSFT, QBTS, LAES, IBM, NVDA, GOOG, AMZN and INTC) and DA $> 65 \%$ when HON and QTUM are included. HON and QTUM under performs on July 11 and 14 respectively. Overall, the high DA is maintained for further days, with DA $> 78\%$ on July 18, suggesting that the QRC framework can capture latent structure within low volumes and high volatility of After-Market hours behaviour. This offers insight into the nonlinear dependencies which conventional models often do not consider. 

The last two investigations involve: (1) predicting After-Market hours data for July 10 - 11 and July 14 - 18 using Pre-Market hours dataset for July 7, 2025 as input and (2) predicting Pre-Market hours data for July 10 - 11 and July 14 - 18 using After-Market hours dataset for July 7, 2025 as input. For the first case, when After hours data is sparse (limited or messy), Pre-Market data insights can help decode After-Market gaps as Pre-Market trader moves based on overnight news or expectations often hint the market's behaviour later. Therefore, if the QRC model sees similar patterns over time i.e. stock rises pre-market and dips after-market, it can make predictive model based on that knowledge even on unseen days. We again choose parameters $(N,\Delta_0,\Omega_0) = (5,1,10)$. 

\begin{figure}[t!]
    \centering
    \includegraphics[width=\linewidth]{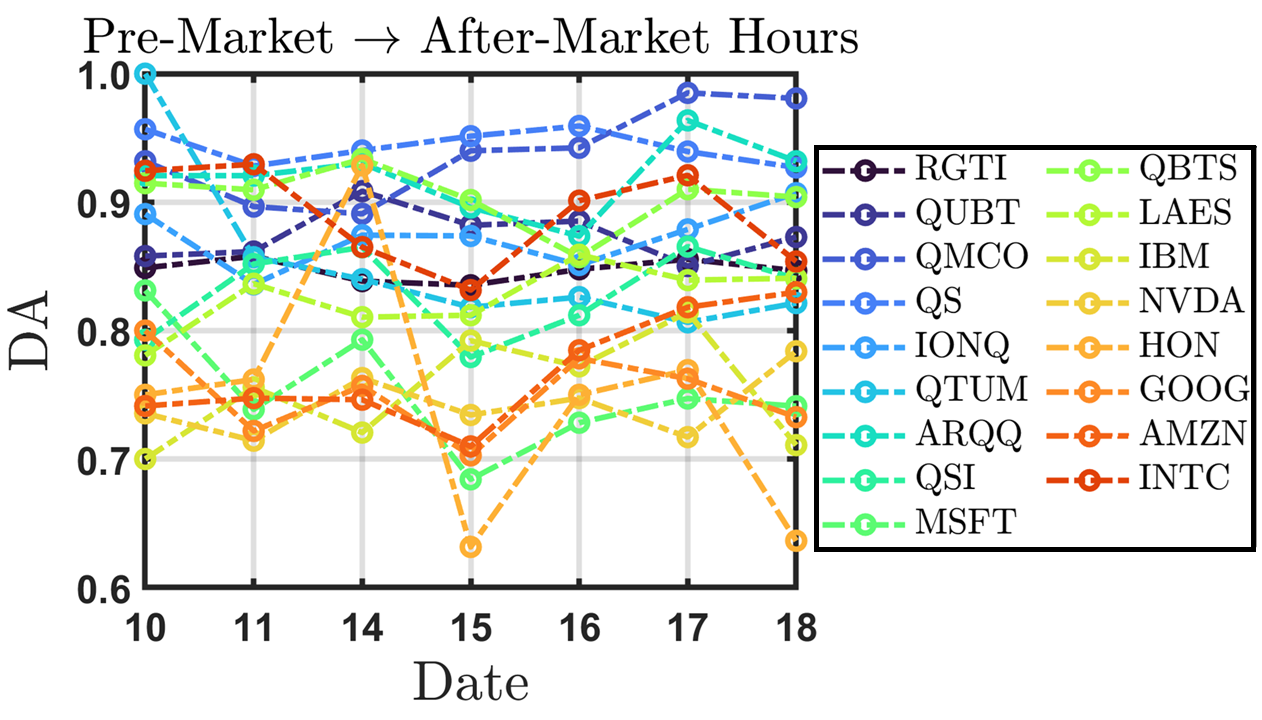}
    \caption{When predicting After-Market hours data for July 10 - 11 and July 14 - 18, a DA accuracy greater than $63 \%$ is observed for $(N,\Delta_0,\Omega_0) = (5,1,10)$ when the Pre-Market hours dataset for July 7, 2025 is the training dataset.}
    \label{BMHtoAMHfutureprediction}
\end{figure}

\begin{figure}[b!]
    \centering
    \includegraphics[width=\linewidth]{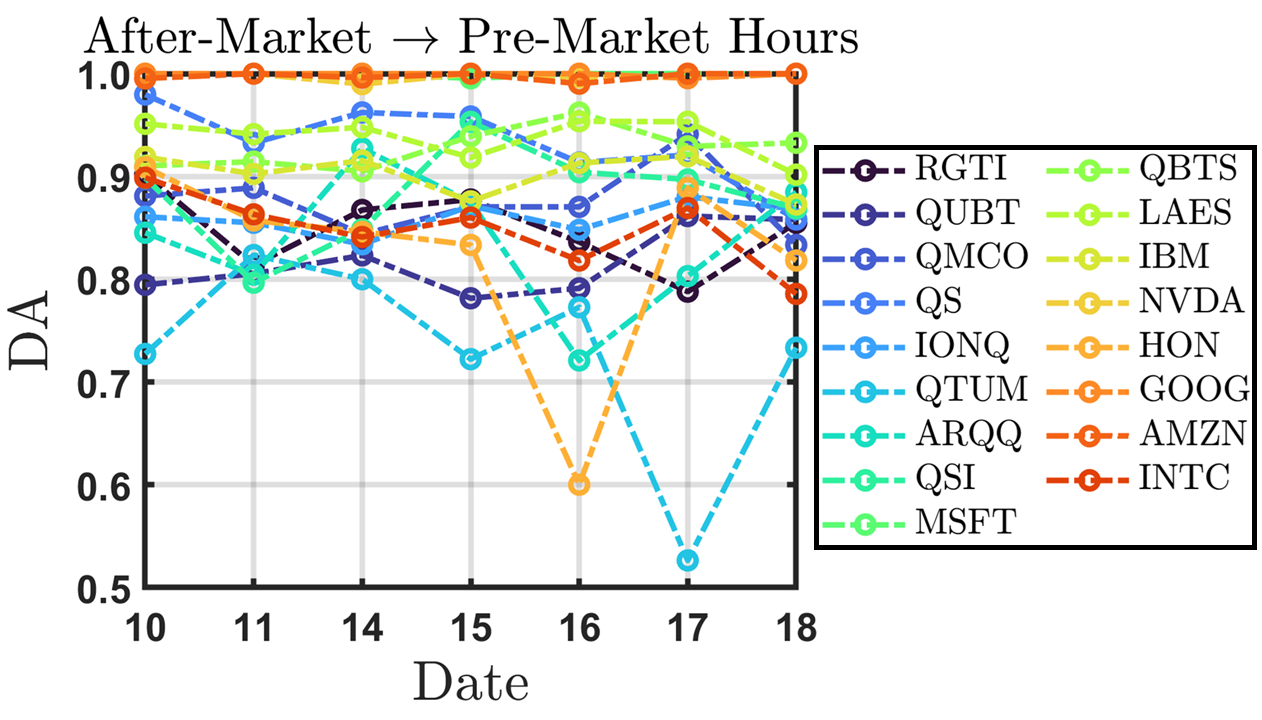}
    \caption{A DA accuracy greater than $52\%$ is observed when predicting After-Market hours data for July 10 - 11 and July 14 - 18 using Pre-Market hours dataset for July 7, 2025 as input with $(N,\Delta_0,\Omega_0) = (5,1,10)$.}
    \label{AMHtoBMHfutureprediction}
\end{figure}

For the first task, the up/down predictions are over $63\%$ for all companies and maintained for further days (see Fig~\ref{BMHtoAMHfutureprediction}). RGTI, QUBT, QMCO, QS, IONQ, QTUM, ARQQ, QBTS and INTC have a DA $> 80 \%$ which is maintained from July 10 - July 18. The QRC captures trend patterns in pre-market data for July 7 that is essential in predicting after-market hours data from July 10 - July 18.

The up/down predictions for the second task are over $52\%$. This is due to QTUM's and HON's performance on July 17 (DA = 0.5263) and July 16 (DA = 0.6) respectively. Removing HON and QTUM leads to an accuracy greater than $70 \%$ maintained for further days. Therefore using the after-market data for July 7, the QRC captures trend patterns very well for pre-market data of QMCO, QS, IONQ, MSFT, QBTS, LAES, IBM, NVDA, GOOG and AMZN with DA $> 83\%$. This is dominant in MSFT, NVDA, GOOG and AMZN where $DA > 98.98 \%$.

\section{Conclusion}
\label{conclusionQRC}

We present a quantum reservoir, within a QRC framework, consisting of up to six two-level qubits connected via an all-to-all topology. We demonstrate that this QRC framework is capable of predicting future closing daily volumes and up/down stock trends of 20 quantum-invested companies. We start by characterizing the tail behaviour in daily closing volume data by computing the standardized moment ratios (SMR) and Pearson correlation between companies. For the 5 year period (April 11 2020 - 2025), heavy-tail and lighter-tail companies experience strong co-movement in the extreme tails alongside non-monotonic companies QTUM and IBM. The non-monotonic company HON exhibits idiosyncratic mid-order behaviour. Pre-market tail behaviour for July 7, 2025 varies sharply across heavy-tail, lighter-tail, and non-monotonic tail companies. Heavy and lighter tail companies experience systematic extreme tails co-movement while non-monotonic companies exhibit idiosyncratic distortions. After-market behaviour have systematic co-movement in the heavy and lighter tail companies while non-monotonic companies remain entirely idiosyncratic. Despite heavy tails or lighter tails, the HPE-QRC is capable of great predictive performance.

The HPE-QRC performance is enhanced by applying non-linear post processing delay embeddings to the readouts. To incorporate memory into the framework, we add delay embeddings. Adding delay embeddings have demonstrated significant improvements in the QRC performance  \cite{zhu2024practical,mccaul2025minimalquantumreservoirshamiltonian}. We find that the Hamiltonian parameters $\Delta_0 = 8$ and $\Omega_0 = 6$, delay $\delta = 6$ along with $60\%$ of the initial data points as training points, enables low MSE Train, MSE Test, MAPE Test, NMSE Test and RMSE Test values, depending on the number of qubits $N$ used, with minimal training. For most companies, low NMSE, MSE and RMSE can occur at $N = 2$, $3$, $4$, $5$ and $6$ depending on the initial stock market time series. This is for a 5 year span: April 11 2020 - 2025.

We additionally capture the effects of $\Delta_0$, $\Omega_0$ and $N$ on the overall performance of the QRC  for Pre-Market and After-Marker hours trend prediction. We find optimal parameters $(N,\Delta_0,\Omega_0)$ which yield an accuracy of over $86\%$. We also show that training the QRC on the dataset of one date (i.e. July 7, 2025  pre-market or after-market hours) is more than enough to predict the directional accuracy of future days. 

\section*{DECLARATIONS}

\subsection*{Competing Interests}
The authors declare no competing interests.

\subsection*{Availability of data and materials}
To support reproducibility, all dataset, codes and trained model parameters will be made publicly available upon publication.

\subsection*{Funding}

The authors acknowledge funding from the European Union and the European Innovation Council via the Horizon Europe Projects QRC-4-ESP (Grant Agreement No. 101129663 and QUEST (Grant Agreement No. 101156088). We also would like to acknowledge funding from the UK Research and Innovation (UKRI) Horizon Europe guarantee schemes for the projects QRC-4-ESP (Grant No. 101129663 and QUEST (Grant No. 101156088) and the UK Engineering and Physical Sciences Research Council (ESPRC) under the grant No. EP/WO28344/1.

\subsection*{Author's contributions}

AZ conceptualised and supervised the work. SS conceptualised the standardized moments in empirical stock data and inclusion of Efficient Market Hypothesis. AB and JG provided and supervised the development of initial codes. AZ, SS, AB, and WO interpreted the results and contributed to the writing and editing of the paper. JG provided insights into adding delay embeddings. WO gathered the financial data, conducted the numerical simulations, and contributed to the writing of the first draft. 

\bibliography{main}

\begin{thebibliography}{71}%
\makeatletter
\providecommand \@ifxundefined [1]{%
 \@ifx{#1\undefined}
}%
\providecommand \@ifnum [1]{%
 \ifnum #1\expandafter \@firstoftwo
 \else \expandafter \@secondoftwo
 \fi
}%
\providecommand \@ifx [1]{%
 \ifx #1\expandafter \@firstoftwo
 \else \expandafter \@secondoftwo
 \fi
}%
\providecommand \natexlab [1]{#1}%
\providecommand \enquote  [1]{``#1''}%
\providecommand \bibnamefont  [1]{#1}%
\providecommand \bibfnamefont [1]{#1}%
\providecommand \citenamefont [1]{#1}%
\providecommand \href@noop [0]{\@secondoftwo}%
\providecommand \href [0]{\begingroup \@sanitize@url \@href}%
\providecommand \@href[1]{\@@startlink{#1}\@@href}%
\providecommand \@@href[1]{\endgroup#1\@@endlink}%
\providecommand \@sanitize@url [0]{\catcode `\\12\catcode `\$12\catcode
  `\&12\catcode `\#12\catcode `\^12\catcode `\_12\catcode `\%12\relax}%
\providecommand \@@startlink[1]{}%
\providecommand \@@endlink[0]{}%
\providecommand \url  [0]{\begingroup\@sanitize@url \@url }%
\providecommand \@url [1]{\endgroup\@href {#1}{\urlprefix }}%
\providecommand \urlprefix  [0]{URL }%
\providecommand \Eprint [0]{\href }%
\providecommand \doibase [0]{https://doi.org/}%
\providecommand \selectlanguage [0]{\@gobble}%
\providecommand \bibinfo  [0]{\@secondoftwo}%
\providecommand \bibfield  [0]{\@secondoftwo}%
\providecommand \translation [1]{[#1]}%
\providecommand \BibitemOpen [0]{}%
\providecommand \bibitemStop [0]{}%
\providecommand \bibitemNoStop [0]{.\EOS\space}%
\providecommand \EOS [0]{\spacefactor3000\relax}%
\providecommand \BibitemShut  [1]{\csname bibitem#1\endcsname}%
\let\auto@bib@innerbib\@empty
%</preamble>
\bibitem [{\citenamefont {Liu}\ and\ \citenamefont
  {Ma}(2022)}]{liu2022quantum}%
  \BibitemOpen
  \bibfield  {author} {\bibinfo {author} {\bibfnamefont {G.}~\bibnamefont
  {Liu}}\ and\ \bibinfo {author} {\bibfnamefont {W.}~\bibnamefont {Ma}},\
  }\bibfield  {title} {\bibinfo {title} {A quantum artificial neural network
  for stock closing price prediction},\ }\href@noop {} {\bibfield  {journal}
  {\bibinfo  {journal} {Information Sciences}\ }\textbf {\bibinfo {volume}
  {598}},\ \bibinfo {pages} {75} (\bibinfo {year} {2022})}\BibitemShut
  {NoStop}%
\bibitem [{\citenamefont {Li}\ \emph {et~al.}(2024{\natexlab{a}})\citenamefont
  {Li}, \citenamefont {Kamaruddin}, \citenamefont {Yuhaniz},\ and\
  \citenamefont {Al-Jaifi}}]{li2024forecasting}%
  \BibitemOpen
  \bibfield  {author} {\bibinfo {author} {\bibfnamefont {Q.}~\bibnamefont
  {Li}}, \bibinfo {author} {\bibfnamefont {N.}~\bibnamefont {Kamaruddin}},
  \bibinfo {author} {\bibfnamefont {S.~S.}\ \bibnamefont {Yuhaniz}},\ and\
  \bibinfo {author} {\bibfnamefont {H.~A.~A.}\ \bibnamefont {Al-Jaifi}},\
  }\bibfield  {title} {\bibinfo {title} {Forecasting stock prices changes using
  long-short term memory neural network with symbolic genetic programming},\
  }\href@noop {} {\bibfield  {journal} {\bibinfo  {journal} {Scientific
  reports}\ }\textbf {\bibinfo {volume} {14}},\ \bibinfo {pages} {422}
  (\bibinfo {year} {2024}{\natexlab{a}})}\BibitemShut {NoStop}%
\bibitem [{\citenamefont {Smithers}(2022)}]{smithers2022economics}%
  \BibitemOpen
  \bibfield  {author} {\bibinfo {author} {\bibfnamefont {A.}~\bibnamefont
  {Smithers}},\ }\href@noop {} {\emph {\bibinfo {title} {The Economics of the
  Stock Market}}}\ (\bibinfo  {publisher} {Oxford University Press},\ \bibinfo
  {year} {2022})\BibitemShut {NoStop}%
\bibitem [{\citenamefont {Quah}(2006)}]{quah2006improving}%
  \BibitemOpen
  \bibfield  {author} {\bibinfo {author} {\bibfnamefont {T.-S.}\ \bibnamefont
  {Quah}},\ }\bibfield  {title} {\bibinfo {title} {Improving returns on stock
  investment through neural network selection},\ }in\ \href@noop {} {\emph
  {\bibinfo {booktitle} {Artificial Neural Networks in Finance and
  Manufacturing}}}\ (\bibinfo  {publisher} {IGI Global},\ \bibinfo {year}
  {2006})\ pp.\ \bibinfo {pages} {152--164}\BibitemShut {NoStop}%
\bibitem [{\citenamefont {Buzan}\ and\ \citenamefont
  {Falkner}(2024)}]{buzan2024market}%
  \BibitemOpen
  \bibfield  {author} {\bibinfo {author} {\bibfnamefont {B.}~\bibnamefont
  {Buzan}}\ and\ \bibinfo {author} {\bibfnamefont {R.}~\bibnamefont
  {Falkner}},\ }\href@noop {} {\emph {\bibinfo {title} {The Market in Global
  International Society: An English School Approach to International Political
  Economy}}}\ (\bibinfo  {publisher} {Oxford University Press},\ \bibinfo
  {year} {2024})\BibitemShut {NoStop}%
\bibitem [{\citenamefont {Ebner}\ and\ \citenamefont
  {Beck}(2008)}]{ebner2008institutions}%
  \BibitemOpen
  \bibfield  {author} {\bibinfo {author} {\bibfnamefont {A.}~\bibnamefont
  {Ebner}}\ and\ \bibinfo {author} {\bibfnamefont {N.}~\bibnamefont {Beck}},\
  }\href@noop {} {\emph {\bibinfo {title} {The institutions of the market:
  organizations, social systems, and governance}}}\ (\bibinfo  {publisher}
  {Oxford University Press, USA},\ \bibinfo {year} {2008})\BibitemShut
  {NoStop}%
\bibitem [{\citenamefont {Davis}(2012)}]{davis2012politics}%
  \BibitemOpen
  \bibfield  {author} {\bibinfo {author} {\bibfnamefont {G.~F.}\ \bibnamefont
  {Davis}},\ }\bibfield  {title} {\bibinfo {title} {Politics and financial
  markets},\ }\href@noop {} {\bibfield  {journal} {\bibinfo  {journal} {The
  Oxford handbook of the sociology of finance}\ ,\ \bibinfo {pages} {33}}
  (\bibinfo {year} {2012})}\BibitemShut {NoStop}%
\bibitem [{\citenamefont {Kara}\ \emph {et~al.}(2011)\citenamefont {Kara},
  \citenamefont {Boyacioglu},\ and\ \citenamefont
  {Baykan}}]{kara2011predicting}%
  \BibitemOpen
  \bibfield  {author} {\bibinfo {author} {\bibfnamefont {Y.}~\bibnamefont
  {Kara}}, \bibinfo {author} {\bibfnamefont {M.~A.}\ \bibnamefont
  {Boyacioglu}},\ and\ \bibinfo {author} {\bibfnamefont {{\"O}.~K.}\
  \bibnamefont {Baykan}},\ }\bibfield  {title} {\bibinfo {title} {Predicting
  direction of stock price index movement using artificial neural networks and
  support vector machines: The sample of the istanbul stock exchange},\
  }\href@noop {} {\bibfield  {journal} {\bibinfo  {journal} {Expert systems
  with Applications}\ }\textbf {\bibinfo {volume} {38}},\ \bibinfo {pages}
  {5311} (\bibinfo {year} {2011})}\BibitemShut {NoStop}%
\bibitem [{\citenamefont {P{\'a}stor}\ and\ \citenamefont
  {Veronesi}(2009)}]{pastor2009technological}%
  \BibitemOpen
  \bibfield  {author} {\bibinfo {author} {\bibfnamefont {L.}~\bibnamefont
  {P{\'a}stor}}\ and\ \bibinfo {author} {\bibfnamefont {P.}~\bibnamefont
  {Veronesi}},\ }\bibfield  {title} {\bibinfo {title} {Technological
  revolutions and stock prices},\ }\href@noop {} {\bibfield  {journal}
  {\bibinfo  {journal} {American Economic Review}\ }\textbf {\bibinfo {volume}
  {99}},\ \bibinfo {pages} {1451} (\bibinfo {year} {2009})}\BibitemShut
  {NoStop}%
\bibitem [{\citenamefont {Bollen}\ \emph {et~al.}(2011)\citenamefont {Bollen},
  \citenamefont {Mao},\ and\ \citenamefont {Zeng}}]{bollen2011twitter}%
  \BibitemOpen
  \bibfield  {author} {\bibinfo {author} {\bibfnamefont {J.}~\bibnamefont
  {Bollen}}, \bibinfo {author} {\bibfnamefont {H.}~\bibnamefont {Mao}},\ and\
  \bibinfo {author} {\bibfnamefont {X.}~\bibnamefont {Zeng}},\ }\bibfield
  {title} {\bibinfo {title} {Twitter mood predicts the stock market},\
  }\href@noop {} {\bibfield  {journal} {\bibinfo  {journal} {Journal of
  computational science}\ }\textbf {\bibinfo {volume} {2}},\ \bibinfo {pages}
  {1} (\bibinfo {year} {2011})}\BibitemShut {NoStop}%
\bibitem [{\citenamefont {Loang}(2024)}]{loang2024psychological}%
  \BibitemOpen
  \bibfield  {author} {\bibinfo {author} {\bibfnamefont {O.}~\bibnamefont
  {Loang}},\ }\href {https://books.google.co.uk/books?id=XQg1EQAAQBAJ} {\emph
  {\bibinfo {title} {Psychological Drivers of Herding and Market
  Overreaction}}},\ Advances in Marketing, Customer Relationship Management,
  and E-Services\ (\bibinfo  {publisher} {IGI Global Scientific Publishing},\
  \bibinfo {year} {2024})\BibitemShut {NoStop}%
\bibitem [{\citenamefont {Bikhchandani}\ and\ \citenamefont
  {Sharma}(2000)}]{bikhchandani2000herd}%
  \BibitemOpen
  \bibfield  {author} {\bibinfo {author} {\bibfnamefont {S.}~\bibnamefont
  {Bikhchandani}}\ and\ \bibinfo {author} {\bibfnamefont {S.}~\bibnamefont
  {Sharma}},\ }\bibfield  {title} {\bibinfo {title} {Herd behavior in financial
  markets},\ }\href@noop {} {\bibfield  {journal} {\bibinfo  {journal} {IMF
  Staff papers}\ }\textbf {\bibinfo {volume} {47}},\ \bibinfo {pages} {279}
  (\bibinfo {year} {2000})}\BibitemShut {NoStop}%
\bibitem [{\citenamefont {Ooi}(2025)}]{ooi2025behavioural}%
  \BibitemOpen
  \bibfield  {author} {\bibinfo {author} {\bibfnamefont {K.~L.}\ \bibnamefont
  {Ooi}},\ }\bibfield  {title} {\bibinfo {title} {Behavioural finance and
  financial crises},\ }in\ \href@noop {} {\emph {\bibinfo {booktitle}
  {Demystifying Behavioral Finance: Foundational Theories to Contemporary
  Applications and Future Directions}}}\ (\bibinfo  {publisher} {Springer},\
  \bibinfo {year} {2025})\ pp.\ \bibinfo {pages} {79--93}\BibitemShut {NoStop}%
\bibitem [{\citenamefont {Black}\ and\ \citenamefont
  {Scholes}(1973)}]{black1973pricing}%
  \BibitemOpen
  \bibfield  {author} {\bibinfo {author} {\bibfnamefont {F.}~\bibnamefont
  {Black}}\ and\ \bibinfo {author} {\bibfnamefont {M.}~\bibnamefont
  {Scholes}},\ }\bibfield  {title} {\bibinfo {title} {The pricing of options
  and corporate liabilities},\ }\href@noop {} {\bibfield  {journal} {\bibinfo
  {journal} {Journal of political economy}\ }\textbf {\bibinfo {volume} {81}},\
  \bibinfo {pages} {637} (\bibinfo {year} {1973})}\BibitemShut {NoStop}%
\bibitem [{\citenamefont {Bachelier}(1900)}]{Bachelier1900}%
  \BibitemOpen
  \bibfield  {author} {\bibinfo {author} {\bibfnamefont {L.}~\bibnamefont
  {Bachelier}},\ }\bibfield  {title} {\bibinfo {title} {Théorie de la
  spéculation},\ }\href {https://doi.org/10.24033/asens.476} {\bibfield
  {journal} {\bibinfo  {journal} {Annales scientifiques de l'École Normale
  Supérieure}\ }\bibinfo {series} {3},\ \textbf {\bibinfo {volume} {17}},\
  \bibinfo {pages} {21} (\bibinfo {year} {1900})}\BibitemShut {NoStop}%
\bibitem [{\citenamefont {Kanazawa}\ \emph
  {et~al.}(2018{\natexlab{a}})\citenamefont {Kanazawa}, \citenamefont
  {Sueshige}, \citenamefont {Takayasu},\ and\ \citenamefont
  {Takayasu}}]{kanazawa2018kinetic}%
  \BibitemOpen
  \bibfield  {author} {\bibinfo {author} {\bibfnamefont {K.}~\bibnamefont
  {Kanazawa}}, \bibinfo {author} {\bibfnamefont {T.}~\bibnamefont {Sueshige}},
  \bibinfo {author} {\bibfnamefont {H.}~\bibnamefont {Takayasu}},\ and\
  \bibinfo {author} {\bibfnamefont {M.}~\bibnamefont {Takayasu}},\ }\bibfield
  {title} {\bibinfo {title} {Kinetic theory for financial brownian motion from
  microscopic dynamics},\ }\href@noop {} {\bibfield  {journal} {\bibinfo
  {journal} {Physical review E}\ }\textbf {\bibinfo {volume} {98}},\ \bibinfo
  {pages} {052317} (\bibinfo {year} {2018}{\natexlab{a}})}\BibitemShut
  {NoStop}%
\bibitem [{\citenamefont {Kanazawa}\ \emph
  {et~al.}(2018{\natexlab{b}})\citenamefont {Kanazawa}, \citenamefont
  {Sueshige}, \citenamefont {Takayasu},\ and\ \citenamefont
  {Takayasu}}]{kanazawa2018derivation}%
  \BibitemOpen
  \bibfield  {author} {\bibinfo {author} {\bibfnamefont {K.}~\bibnamefont
  {Kanazawa}}, \bibinfo {author} {\bibfnamefont {T.}~\bibnamefont {Sueshige}},
  \bibinfo {author} {\bibfnamefont {H.}~\bibnamefont {Takayasu}},\ and\
  \bibinfo {author} {\bibfnamefont {M.}~\bibnamefont {Takayasu}},\ }\bibfield
  {title} {\bibinfo {title} {Derivation of the boltzmann equation for financial
  brownian motion: Direct observation of the collective motion of
  high-frequency traders},\ }\href@noop {} {\bibfield  {journal} {\bibinfo
  {journal} {Physical review letters}\ }\textbf {\bibinfo {volume} {120}},\
  \bibinfo {pages} {138301} (\bibinfo {year} {2018}{\natexlab{b}})}\BibitemShut
  {NoStop}%
\bibitem [{\citenamefont {Garcin}(2022)}]{garcin2022forecasting}%
  \BibitemOpen
  \bibfield  {author} {\bibinfo {author} {\bibfnamefont {M.}~\bibnamefont
  {Garcin}},\ }\bibfield  {title} {\bibinfo {title} {Forecasting with
  fractional brownian motion: a financial perspective},\ }\href@noop {}
  {\bibfield  {journal} {\bibinfo  {journal} {Quantitative finance}\ }\textbf
  {\bibinfo {volume} {22}},\ \bibinfo {pages} {1495} (\bibinfo {year}
  {2022})}\BibitemShut {NoStop}%
\bibitem [{\citenamefont {Samuelson}(2016)}]{samuelson2016proof}%
  \BibitemOpen
  \bibfield  {author} {\bibinfo {author} {\bibfnamefont {P.~A.}\ \bibnamefont
  {Samuelson}},\ }\bibfield  {title} {\bibinfo {title} {Proof that properly
  anticipated prices fluctuate randomly},\ }in\ \href@noop {} {\emph {\bibinfo
  {booktitle} {The world scientific handbook of futures markets}}}\ (\bibinfo
  {publisher} {World Scientific},\ \bibinfo {year} {2016})\ pp.\ \bibinfo
  {pages} {25--38}\BibitemShut {NoStop}%
\bibitem [{\citenamefont {Mandelbrot}(1997)}]{mandelbrot1997variation}%
  \BibitemOpen
  \bibfield  {author} {\bibinfo {author} {\bibfnamefont {B.~B.}\ \bibnamefont
  {Mandelbrot}},\ }\href@noop {} {\emph {\bibinfo {title} {The variation of
  certain speculative prices}}}\ (\bibinfo  {publisher} {Springer},\ \bibinfo
  {year} {1997})\BibitemShut {NoStop}%
\bibitem [{\citenamefont {Walter}(2021)}]{Walter2021random}%
  \BibitemOpen
  \bibfield  {author} {\bibinfo {author} {\bibfnamefont {C.~P.}\ \bibnamefont
  {Walter}},\ }\bibfield  {title} {\bibinfo {title} {The random walk model in
  finance: A new taxonomy},\ }\bibfield  {journal} {\bibinfo  {journal} {SSRN
  Electronic Journal}\ }\href {https://doi.org/10.2139/ssrn.3908441}
  {10.2139/ssrn.3908441} (\bibinfo {year} {2021})\BibitemShut {NoStop}%
\bibitem [{\citenamefont {Taylor}(1982)}]{taylor1982tests}%
  \BibitemOpen
  \bibfield  {author} {\bibinfo {author} {\bibfnamefont {S.~J.}\ \bibnamefont
  {Taylor}},\ }\bibfield  {title} {\bibinfo {title} {Tests of the random walk
  hypothesis against a price-trend hypothesis},\ }\href@noop {} {\bibfield
  {journal} {\bibinfo  {journal} {Journal of Financial and Quantitative
  Analysis}\ }\textbf {\bibinfo {volume} {17}},\ \bibinfo {pages} {37}
  (\bibinfo {year} {1982})}\BibitemShut {NoStop}%
\bibitem [{\citenamefont {Thurner}(2011)}]{thurner2011systemic}%
  \BibitemOpen
  \bibfield  {author} {\bibinfo {author} {\bibfnamefont {S.}~\bibnamefont
  {Thurner}},\ }\bibfield  {title} {\bibinfo {title} {Systemic financial risk:
  agent based models to understand the leverage cycle on national scales and
  its consequences},\ }\href@noop {} {\bibfield  {journal} {\bibinfo  {journal}
  {IfP/fgS Working Paper}\ }\textbf {\bibinfo {volume} {14}} (\bibinfo {year}
  {2011})}\BibitemShut {NoStop}%
\bibitem [{\citenamefont {Klimek}\ \emph {et~al.}(2015)\citenamefont {Klimek},
  \citenamefont {Poledna}, \citenamefont {Farmer},\ and\ \citenamefont
  {Thurner}}]{klimek2015bail}%
  \BibitemOpen
  \bibfield  {author} {\bibinfo {author} {\bibfnamefont {P.}~\bibnamefont
  {Klimek}}, \bibinfo {author} {\bibfnamefont {S.}~\bibnamefont {Poledna}},
  \bibinfo {author} {\bibfnamefont {J.~D.}\ \bibnamefont {Farmer}},\ and\
  \bibinfo {author} {\bibfnamefont {S.}~\bibnamefont {Thurner}},\ }\bibfield
  {title} {\bibinfo {title} {To bail-out or to bail-in? answers from an
  agent-based model},\ }\href@noop {} {\bibfield  {journal} {\bibinfo
  {journal} {Journal of Economic Dynamics and Control}\ }\textbf {\bibinfo
  {volume} {50}},\ \bibinfo {pages} {144} (\bibinfo {year} {2015})}\BibitemShut
  {NoStop}%
\bibitem [{\citenamefont {Farmer}\ and\ \citenamefont
  {Foley}(2009)}]{farmer2009economy}%
  \BibitemOpen
  \bibfield  {author} {\bibinfo {author} {\bibfnamefont {J.~D.}\ \bibnamefont
  {Farmer}}\ and\ \bibinfo {author} {\bibfnamefont {D.}~\bibnamefont {Foley}},\
  }\bibfield  {title} {\bibinfo {title} {The economy needs agent-based
  modelling},\ }\href@noop {} {\bibfield  {journal} {\bibinfo  {journal}
  {Nature}\ }\textbf {\bibinfo {volume} {460}},\ \bibinfo {pages} {685}
  (\bibinfo {year} {2009})}\BibitemShut {NoStop}%
\bibitem [{\citenamefont {Lyons}\ and\ \citenamefont
  {Kass-Hanna}(2021)}]{Lyons2021}%
  \BibitemOpen
  \bibfield  {author} {\bibinfo {author} {\bibfnamefont {A.~C.}\ \bibnamefont
  {Lyons}}\ and\ \bibinfo {author} {\bibfnamefont {J.}~\bibnamefont
  {Kass-Hanna}},\ }\bibfield  {title} {\bibinfo {title} {Behavioral economics
  and financial decision making},\ }\bibfield  {journal} {\bibinfo  {journal}
  {SSRN Electronic Journal}\ }\href {https://doi.org/10.2139/ssrn.3887605}
  {10.2139/ssrn.3887605} (\bibinfo {year} {2021})\BibitemShut {NoStop}%
\bibitem [{\citenamefont {Chauhan}(2024)}]{Chauhan2024}%
  \BibitemOpen
  \bibfield  {author} {\bibinfo {author} {\bibfnamefont {A.}~\bibnamefont
  {Chauhan}},\ }\bibfield  {title} {\bibinfo {title} {The intersection of game
  theory and behavioral economics: A theoretical approach},\ }\bibfield
  {journal} {\bibinfo  {journal} {Advance}\ }\href
  {https://doi.org/10.22541/au.173627313.33277076/v1}
  {10.22541/au.173627313.33277076/v1} (\bibinfo {year} {2024})\BibitemShut
  {NoStop}%
\bibitem [{\citenamefont {Marmon}(2025)}]{Marmon2025}%
  \BibitemOpen
  \bibfield  {author} {\bibinfo {author} {\bibfnamefont {S.}~\bibnamefont
  {Marmon}},\ }\bibfield  {title} {\bibinfo {title} {Game theory applications
  in modern business strategy: Optimizing competitive decision-making in
  uncertain markets},\ }\bibfield  {journal} {\bibinfo  {journal} {SSRN
  Electronic Journal}\ }\href {https://doi.org/10.2139/ssrn.5141929}
  {10.2139/ssrn.5141929} (\bibinfo {year} {2025})\BibitemShut {NoStop}%
\bibitem [{\citenamefont {Fama}(1970)}]{fama1970efficient}%
  \BibitemOpen
  \bibfield  {author} {\bibinfo {author} {\bibfnamefont {E.~F.}\ \bibnamefont
  {Fama}},\ }\bibfield  {title} {\bibinfo {title} {Efficient capital markets: A
  review of theory and empirical work},\ }\href@noop {} {\bibfield  {journal}
  {\bibinfo  {journal} {The journal of Finance}\ }\textbf {\bibinfo {volume}
  {25}},\ \bibinfo {pages} {383} (\bibinfo {year} {1970})}\BibitemShut
  {NoStop}%
\bibitem [{\citenamefont {De~Bondt}\ and\ \citenamefont
  {Thaler}(1985)}]{de1985does}%
  \BibitemOpen
  \bibfield  {author} {\bibinfo {author} {\bibfnamefont {W.~F.}\ \bibnamefont
  {De~Bondt}}\ and\ \bibinfo {author} {\bibfnamefont {R.}~\bibnamefont
  {Thaler}},\ }\bibfield  {title} {\bibinfo {title} {Does the stock market
  overreact?},\ }\href@noop {} {\bibfield  {journal} {\bibinfo  {journal} {The
  Journal of finance}\ }\textbf {\bibinfo {volume} {40}},\ \bibinfo {pages}
  {793} (\bibinfo {year} {1985})}\BibitemShut {NoStop}%
\bibitem [{\citenamefont {Grabowski}(2019)}]{grabowski2019technology}%
  \BibitemOpen
  \bibfield  {author} {\bibinfo {author} {\bibfnamefont {D.}~\bibnamefont
  {Grabowski}},\ }\bibfield  {title} {\bibinfo {title} {Technology, adaptation
  and the efficient market hypothesis},\ }\href@noop {} {\bibfield  {journal}
  {\bibinfo  {journal} {Adaptation and the Efficient Market Hypothesis
  (December 4, 2019)}\ } (\bibinfo {year} {2019})}\BibitemShut {NoStop}%
\bibitem [{\citenamefont {Gu}(2004)}]{gu2004increasing}%
  \BibitemOpen
  \bibfield  {author} {\bibinfo {author} {\bibfnamefont {A.~Y.}\ \bibnamefont
  {Gu}},\ }\bibfield  {title} {\bibinfo {title} {Increasing market efficiency:
  Evidence from the nasdaq},\ }\href@noop {} {\bibfield  {journal} {\bibinfo
  {journal} {American Business Review}\ }\textbf {\bibinfo {volume} {22}},\
  \bibinfo {pages} {20} (\bibinfo {year} {2004})}\BibitemShut {NoStop}%
\bibitem [{\citenamefont {Silver}(2024)}]{silver2024agent}%
  \BibitemOpen
  \bibfield  {author} {\bibinfo {author} {\bibfnamefont {S.~D.}\ \bibnamefont
  {Silver}},\ }\bibfield  {title} {\bibinfo {title} {Agent expectations and
  news sentiment in the dynamics of price in a financial market},\ }\href@noop
  {} {\bibfield  {journal} {\bibinfo  {journal} {Review of Behavioral Finance}\
  }\textbf {\bibinfo {volume} {16}},\ \bibinfo {pages} {836} (\bibinfo {year}
  {2024})}\BibitemShut {NoStop}%
\bibitem [{\citenamefont {Naseer}\ and\ \citenamefont
  {Bin~Tariq}(2015)}]{naseer2015efficient}%
  \BibitemOpen
  \bibfield  {author} {\bibinfo {author} {\bibfnamefont {M.}~\bibnamefont
  {Naseer}}\ and\ \bibinfo {author} {\bibfnamefont {D.~Y.}\ \bibnamefont
  {Bin~Tariq}},\ }\bibfield  {title} {\bibinfo {title} {The efficient market
  hypothesis: A critical review of the literature},\ }\href@noop {} {\bibfield
  {journal} {\bibinfo  {journal} {The IUP journal of financial risk
  management}\ }\textbf {\bibinfo {volume} {12}},\ \bibinfo {pages} {48}
  (\bibinfo {year} {2015})}\BibitemShut {NoStop}%
\bibitem [{\citenamefont {Cont}(2001)}]{cont2001empirical}%
  \BibitemOpen
  \bibfield  {author} {\bibinfo {author} {\bibfnamefont {R.}~\bibnamefont
  {Cont}},\ }\bibfield  {title} {\bibinfo {title} {Empirical properties of
  asset returns: stylized facts and statistical issues},\ }\href@noop {}
  {\bibfield  {journal} {\bibinfo  {journal} {Quantitative finance}\ }\textbf
  {\bibinfo {volume} {1}},\ \bibinfo {pages} {223} (\bibinfo {year}
  {2001})}\BibitemShut {NoStop}%
\bibitem [{\citenamefont {Cont}(2007)}]{cont2007volatility}%
  \BibitemOpen
  \bibfield  {author} {\bibinfo {author} {\bibfnamefont {R.}~\bibnamefont
  {Cont}},\ }\bibfield  {title} {\bibinfo {title} {Volatility clustering in
  financial markets: empirical facts and agent-based models},\ }in\ \href@noop
  {} {\emph {\bibinfo {booktitle} {Long memory in economics}}}\ (\bibinfo
  {publisher} {Springer},\ \bibinfo {year} {2007})\ pp.\ \bibinfo {pages}
  {289--309}\BibitemShut {NoStop}%
\bibitem [{\citenamefont {Moskowitz}\ \emph {et~al.}(2012)\citenamefont
  {Moskowitz}, \citenamefont {Ooi},\ and\ \citenamefont
  {Pedersen}}]{moskowitz2012time}%
  \BibitemOpen
  \bibfield  {author} {\bibinfo {author} {\bibfnamefont {T.~J.}\ \bibnamefont
  {Moskowitz}}, \bibinfo {author} {\bibfnamefont {Y.~H.}\ \bibnamefont {Ooi}},\
  and\ \bibinfo {author} {\bibfnamefont {L.~H.}\ \bibnamefont {Pedersen}},\
  }\bibfield  {title} {\bibinfo {title} {Time series momentum},\ }\href@noop {}
  {\bibfield  {journal} {\bibinfo  {journal} {Journal of financial economics}\
  }\textbf {\bibinfo {volume} {104}},\ \bibinfo {pages} {228} (\bibinfo {year}
  {2012})}\BibitemShut {NoStop}%
\bibitem [{\citenamefont {He}\ \emph {et~al.}(2021)\citenamefont {He},
  \citenamefont {Li}, \citenamefont {Santi},\ and\ \citenamefont
  {Shi}}]{he2021social}%
  \BibitemOpen
  \bibfield  {author} {\bibinfo {author} {\bibfnamefont {X.}~\bibnamefont
  {He}}, \bibinfo {author} {\bibfnamefont {K.}~\bibnamefont {Li}}, \bibinfo
  {author} {\bibfnamefont {C.}~\bibnamefont {Santi}},\ and\ \bibinfo {author}
  {\bibfnamefont {L.}~\bibnamefont {Shi}},\ }\bibfield  {title} {\bibinfo
  {title} {Social interaction, stochastic volatility, and momentum},\
  }\href@noop {} {\bibfield  {journal} {\bibinfo  {journal} {Stochastic
  Volatility, and Momentum (April 19, 2021)}\ } (\bibinfo {year}
  {2021})}\BibitemShut {NoStop}%
\bibitem [{\citenamefont {Campbell}\ \emph {et~al.}(1998)\citenamefont
  {Campbell}, \citenamefont {Lo}, \citenamefont {MacKinlay},\ and\
  \citenamefont {Whitelaw}}]{campbell1998econometrics}%
  \BibitemOpen
  \bibfield  {author} {\bibinfo {author} {\bibfnamefont {J.~Y.}\ \bibnamefont
  {Campbell}}, \bibinfo {author} {\bibfnamefont {A.~W.}\ \bibnamefont {Lo}},
  \bibinfo {author} {\bibfnamefont {A.~C.}\ \bibnamefont {MacKinlay}},\ and\
  \bibinfo {author} {\bibfnamefont {R.~F.}\ \bibnamefont {Whitelaw}},\
  }\bibfield  {title} {\bibinfo {title} {The econometrics of financial
  markets},\ }\href@noop {} {\bibfield  {journal} {\bibinfo  {journal}
  {Macroeconomic Dynamics}\ }\textbf {\bibinfo {volume} {2}},\ \bibinfo {pages}
  {559} (\bibinfo {year} {1998})}\BibitemShut {NoStop}%
\bibitem [{\citenamefont {MacKinlay}\ \emph {et~al.}(2007)\citenamefont
  {MacKinlay}, \citenamefont {Lo},\ and\ \citenamefont
  {Campbell}}]{mackinlay2007econometrics}%
  \BibitemOpen
  \bibfield  {author} {\bibinfo {author} {\bibfnamefont {A.~C.}\ \bibnamefont
  {MacKinlay}}, \bibinfo {author} {\bibfnamefont {A.~W.}\ \bibnamefont {Lo}},\
  and\ \bibinfo {author} {\bibfnamefont {J.~Y.}\ \bibnamefont {Campbell}},\
  }\href@noop {} {\bibinfo {title} {The econometrics of financial markets}}
  (\bibinfo {year} {2007})\BibitemShut {NoStop}%
\bibitem [{\citenamefont {Holtes}(2024)}]{holtes2024return}%
  \BibitemOpen
  \bibfield  {author} {\bibinfo {author} {\bibfnamefont {G.}~\bibnamefont
  {Holtes}},\ }\bibfield  {title} {\bibinfo {title} {Return volatility
  estimates: A review and practical analysis},\ }\href@noop {} {\bibfield
  {journal} {\bibinfo  {journal} {Available at SSRN 4693434}\ } (\bibinfo
  {year} {2024})}\BibitemShut {NoStop}%
\bibitem [{\citenamefont {Easley}\ and\ \citenamefont
  {O'hara}(1992)}]{easley1992time}%
  \BibitemOpen
  \bibfield  {author} {\bibinfo {author} {\bibfnamefont {D.}~\bibnamefont
  {Easley}}\ and\ \bibinfo {author} {\bibfnamefont {M.}~\bibnamefont
  {O'hara}},\ }\bibfield  {title} {\bibinfo {title} {Time and the process of
  security price adjustment},\ }\href@noop {} {\bibfield  {journal} {\bibinfo
  {journal} {The Journal of finance}\ }\textbf {\bibinfo {volume} {47}},\
  \bibinfo {pages} {577} (\bibinfo {year} {1992})}\BibitemShut {NoStop}%
\bibitem [{\citenamefont {Fischer}\ and\ \citenamefont
  {Krauss}(2018)}]{fischer2018deep}%
  \BibitemOpen
  \bibfield  {author} {\bibinfo {author} {\bibfnamefont {T.}~\bibnamefont
  {Fischer}}\ and\ \bibinfo {author} {\bibfnamefont {C.}~\bibnamefont
  {Krauss}},\ }\bibfield  {title} {\bibinfo {title} {Deep learning with long
  short-term memory networks for financial market predictions},\ }\href@noop {}
  {\bibfield  {journal} {\bibinfo  {journal} {European journal of operational
  research}\ }\textbf {\bibinfo {volume} {270}},\ \bibinfo {pages} {654}
  (\bibinfo {year} {2018})}\BibitemShut {NoStop}%
\bibitem [{\citenamefont {Varadharajan}\ \emph {et~al.}(2024)\citenamefont
  {Varadharajan}, \citenamefont {Smith}, \citenamefont {Kalla}, \citenamefont
  {Samaah}, \citenamefont {Polimetla},\ and\ \citenamefont
  {Kumar}}]{varadharajan2024stock}%
  \BibitemOpen
  \bibfield  {author} {\bibinfo {author} {\bibfnamefont {V.}~\bibnamefont
  {Varadharajan}}, \bibinfo {author} {\bibfnamefont {N.}~\bibnamefont {Smith}},
  \bibinfo {author} {\bibfnamefont {D.}~\bibnamefont {Kalla}}, \bibinfo
  {author} {\bibfnamefont {F.}~\bibnamefont {Samaah}}, \bibinfo {author}
  {\bibfnamefont {K.}~\bibnamefont {Polimetla}},\ and\ \bibinfo {author}
  {\bibfnamefont {G.~R.}\ \bibnamefont {Kumar}},\ }\bibfield  {title} {\bibinfo
  {title} {Stock closing price and trend prediction with lstm-rnn},\
  }\href@noop {} {\bibfield  {journal} {\bibinfo  {journal} {Journal of
  Artificial Intelligence and Big Data}\ }\textbf {\bibinfo {volume} {4}},\
  \bibinfo {pages} {877} (\bibinfo {year} {2024})}\BibitemShut {NoStop}%
\bibitem [{\citenamefont {Mourya}\ \emph {et~al.}(2025)\citenamefont {Mourya},
  \citenamefont {Leipold},\ and\ \citenamefont
  {Adhikari}}]{mourya2025contextual}%
  \BibitemOpen
  \bibfield  {author} {\bibinfo {author} {\bibfnamefont {S.}~\bibnamefont
  {Mourya}}, \bibinfo {author} {\bibfnamefont {H.}~\bibnamefont {Leipold}},\
  and\ \bibinfo {author} {\bibfnamefont {B.}~\bibnamefont {Adhikari}},\
  }\bibfield  {title} {\bibinfo {title} {Contextual quantum neural networks for
  stock price prediction},\ }\href@noop {} {\bibfield  {journal} {\bibinfo
  {journal} {arXiv preprint arXiv:2503.01884}\ } (\bibinfo {year}
  {2025})}\BibitemShut {NoStop}%
\bibitem [{\citenamefont {Srivastava}\ \emph {et~al.}(2023)\citenamefont
  {Srivastava}, \citenamefont {Belekar}, \citenamefont {Shahakar} \emph
  {et~al.}}]{srivastava2023potential}%
  \BibitemOpen
  \bibfield  {author} {\bibinfo {author} {\bibfnamefont {N.}~\bibnamefont
  {Srivastava}}, \bibinfo {author} {\bibfnamefont {G.}~\bibnamefont {Belekar}},
  \bibinfo {author} {\bibfnamefont {N.}~\bibnamefont {Shahakar}}, \emph
  {et~al.},\ }\bibfield  {title} {\bibinfo {title} {The potential of quantum
  techniques for stock price prediction},\ }in\ \href@noop {} {\emph {\bibinfo
  {booktitle} {2023 IEEE International Conference on Recent Advances in Systems
  Science and Engineering (RASSE)}}}\ (\bibinfo {organization} {IEEE},\
  \bibinfo {year} {2023})\ pp.\ \bibinfo {pages} {1--7}\BibitemShut {NoStop}%
\bibitem [{\citenamefont {Stamatopoulos}\ \emph {et~al.}(2020)\citenamefont
  {Stamatopoulos}, \citenamefont {Egger}, \citenamefont {Sun}, \citenamefont
  {Zoufal}, \citenamefont {Iten}, \citenamefont {Shen},\ and\ \citenamefont
  {Woerner}}]{stamatopoulos2020option}%
  \BibitemOpen
  \bibfield  {author} {\bibinfo {author} {\bibfnamefont {N.}~\bibnamefont
  {Stamatopoulos}}, \bibinfo {author} {\bibfnamefont {D.~J.}\ \bibnamefont
  {Egger}}, \bibinfo {author} {\bibfnamefont {Y.}~\bibnamefont {Sun}}, \bibinfo
  {author} {\bibfnamefont {C.}~\bibnamefont {Zoufal}}, \bibinfo {author}
  {\bibfnamefont {R.}~\bibnamefont {Iten}}, \bibinfo {author} {\bibfnamefont
  {N.}~\bibnamefont {Shen}},\ and\ \bibinfo {author} {\bibfnamefont
  {S.}~\bibnamefont {Woerner}},\ }\bibfield  {title} {\bibinfo {title} {Option
  pricing using quantum computers},\ }\href@noop {} {\bibfield  {journal}
  {\bibinfo  {journal} {Quantum}\ }\textbf {\bibinfo {volume} {4}},\ \bibinfo
  {pages} {291} (\bibinfo {year} {2020})}\BibitemShut {NoStop}%
\bibitem [{\citenamefont {Emmanoulopoulos}\ and\ \citenamefont
  {Dimoska}(2022)}]{emmanoulopoulos2022quantum}%
  \BibitemOpen
  \bibfield  {author} {\bibinfo {author} {\bibfnamefont {D.}~\bibnamefont
  {Emmanoulopoulos}}\ and\ \bibinfo {author} {\bibfnamefont {S.}~\bibnamefont
  {Dimoska}},\ }\bibfield  {title} {\bibinfo {title} {Quantum machine learning
  in finance: Time series forecasting},\ }\href@noop {} {\bibfield  {journal}
  {\bibinfo  {journal} {arXiv preprint arXiv:2202.00599}\ } (\bibinfo {year}
  {2022})}\BibitemShut {NoStop}%
\bibitem [{\citenamefont {Orlandi}\ \emph {et~al.}(2024)\citenamefont
  {Orlandi}, \citenamefont {Barbierato},\ and\ \citenamefont
  {Gatti}}]{orlandi2024enhancing}%
  \BibitemOpen
  \bibfield  {author} {\bibinfo {author} {\bibfnamefont {F.}~\bibnamefont
  {Orlandi}}, \bibinfo {author} {\bibfnamefont {E.}~\bibnamefont
  {Barbierato}},\ and\ \bibinfo {author} {\bibfnamefont {A.}~\bibnamefont
  {Gatti}},\ }\bibfield  {title} {\bibinfo {title} {Enhancing financial time
  series prediction with quantum-enhanced synthetic data generation: A case
  study on the s\&p 500 using a quantum wasserstein generative adversarial
  network approach with a gradient penalty},\ }\href@noop {} {\bibfield
  {journal} {\bibinfo  {journal} {Electronics}\ }\textbf {\bibinfo {volume}
  {13}},\ \bibinfo {pages} {2158} (\bibinfo {year} {2024})}\BibitemShut
  {NoStop}%
\bibitem [{\citenamefont {Mironowicz}\ \emph {et~al.}(2024)\citenamefont
  {Mironowicz}, \citenamefont {Mandarino}, \citenamefont {Yilmaz},
  \citenamefont {Ankenbrand} \emph {et~al.}}]{mironowicz2024applications}%
  \BibitemOpen
  \bibfield  {author} {\bibinfo {author} {\bibfnamefont {P.}~\bibnamefont
  {Mironowicz}}, \bibinfo {author} {\bibfnamefont {A.}~\bibnamefont
  {Mandarino}}, \bibinfo {author} {\bibfnamefont {A.}~\bibnamefont {Yilmaz}},
  \bibinfo {author} {\bibfnamefont {T.}~\bibnamefont {Ankenbrand}}, \emph
  {et~al.},\ }\bibfield  {title} {\bibinfo {title} {Applications of quantum
  machine learning for quantitative finance},\ }\href@noop {} {\bibfield
  {journal} {\bibinfo  {journal} {arXiv preprint arXiv:2405.10119}\ } (\bibinfo
  {year} {2024})}\BibitemShut {NoStop}%
\bibitem [{\citenamefont {Zhou}(2025)}]{zhou2025quantum}%
  \BibitemOpen
  \bibfield  {author} {\bibinfo {author} {\bibfnamefont {J.}~\bibnamefont
  {Zhou}},\ }\bibfield  {title} {\bibinfo {title} {Quantum finance: Exploring
  the implications of quantum computing on financial models},\ }\href@noop {}
  {\bibfield  {journal} {\bibinfo  {journal} {Computational Economics}\ ,\
  \bibinfo {pages} {1}} (\bibinfo {year} {2025})}\BibitemShut {NoStop}%
\bibitem [{\citenamefont {Cohen}\ \emph {et~al.}(2020)\citenamefont {Cohen},
  \citenamefont {Khan},\ and\ \citenamefont {Alexander}}]{cohen2020portfolio}%
  \BibitemOpen
  \bibfield  {author} {\bibinfo {author} {\bibfnamefont {J.}~\bibnamefont
  {Cohen}}, \bibinfo {author} {\bibfnamefont {A.}~\bibnamefont {Khan}},\ and\
  \bibinfo {author} {\bibfnamefont {C.}~\bibnamefont {Alexander}},\ }\bibfield
  {title} {\bibinfo {title} {Portfolio optimization of 60 stocks using
  classical and quantum algorithms},\ }\href@noop {} {\bibfield  {journal}
  {\bibinfo  {journal} {arXiv preprint arXiv:2008.08669}\ } (\bibinfo {year}
  {2020})}\BibitemShut {NoStop}%
\bibitem [{\citenamefont {Fujii}\ and\ \citenamefont
  {Nakajima}(2017)}]{fujii2017harnessing}%
  \BibitemOpen
  \bibfield  {author} {\bibinfo {author} {\bibfnamefont {K.}~\bibnamefont
  {Fujii}}\ and\ \bibinfo {author} {\bibfnamefont {K.}~\bibnamefont
  {Nakajima}},\ }\bibfield  {title} {\bibinfo {title} {Harnessing
  disordered-ensemble quantum dynamics for machine learning},\ }\href@noop {}
  {\bibfield  {journal} {\bibinfo  {journal} {Physical Review Applied}\
  }\textbf {\bibinfo {volume} {8}},\ \bibinfo {pages} {024030} (\bibinfo {year}
  {2017})}\BibitemShut {NoStop}%
\bibitem [{\citenamefont {Rosato}\ \emph {et~al.}(2025)\citenamefont {Rosato},
  \citenamefont {Ceschini}, \citenamefont {Succetti}, \citenamefont {Chen},\
  and\ \citenamefont {Panella}}]{rosato2025study}%
  \BibitemOpen
  \bibfield  {author} {\bibinfo {author} {\bibfnamefont {A.}~\bibnamefont
  {Rosato}}, \bibinfo {author} {\bibfnamefont {A.}~\bibnamefont {Ceschini}},
  \bibinfo {author} {\bibfnamefont {F.}~\bibnamefont {Succetti}}, \bibinfo
  {author} {\bibfnamefont {S.~Y.-C.}\ \bibnamefont {Chen}},\ and\ \bibinfo
  {author} {\bibfnamefont {M.}~\bibnamefont {Panella}},\ }\bibfield  {title}
  {\bibinfo {title} {A study on quantum reservoir recurrent models for
  time-constrained volatile sequence forecasting},\ }in\ \href@noop {} {\emph
  {\bibinfo {booktitle} {2025 International Joint Conference on Neural Networks
  (IJCNN)}}}\ (\bibinfo {organization} {IEEE},\ \bibinfo {year} {2025})\ pp.\
  \bibinfo {pages} {1--8}\BibitemShut {NoStop}%
\bibitem [{\citenamefont {Kobayashi}\ \emph {et~al.}(2024)\citenamefont
  {Kobayashi}, \citenamefont {Fujii},\ and\ \citenamefont
  {Yamamoto}}]{kobayashi2024feedback}%
  \BibitemOpen
  \bibfield  {author} {\bibinfo {author} {\bibfnamefont {K.}~\bibnamefont
  {Kobayashi}}, \bibinfo {author} {\bibfnamefont {K.}~\bibnamefont {Fujii}},\
  and\ \bibinfo {author} {\bibfnamefont {N.}~\bibnamefont {Yamamoto}},\
  }\bibfield  {title} {\bibinfo {title} {Feedback-driven quantum reservoir
  computing for time-series analysis},\ }\href@noop {} {\bibfield  {journal}
  {\bibinfo  {journal} {PRX quantum}\ }\textbf {\bibinfo {volume} {5}},\
  \bibinfo {pages} {040325} (\bibinfo {year} {2024})}\BibitemShut {NoStop}%
\bibitem [{\citenamefont {Ivaki}\ \emph {et~al.}(2025)\citenamefont {Ivaki},
  \citenamefont {Lazarides},\ and\ \citenamefont
  {Ala-Nissila}}]{ivaki2025quantum}%
  \BibitemOpen
  \bibfield  {author} {\bibinfo {author} {\bibfnamefont {M.~N.}\ \bibnamefont
  {Ivaki}}, \bibinfo {author} {\bibfnamefont {A.}~\bibnamefont {Lazarides}},\
  and\ \bibinfo {author} {\bibfnamefont {T.}~\bibnamefont {Ala-Nissila}},\
  }\bibfield  {title} {\bibinfo {title} {Quantum reservoir computing on random
  regular graphs},\ }\href@noop {} {\bibfield  {journal} {\bibinfo  {journal}
  {Physical Review A}\ }\textbf {\bibinfo {volume} {112}},\ \bibinfo {pages}
  {012622} (\bibinfo {year} {2025})}\BibitemShut {NoStop}%
\bibitem [{\citenamefont {Yoder}\ \emph {et~al.}(2025)\citenamefont {Yoder},
  \citenamefont {Schoute}, \citenamefont {Rall}, \citenamefont {Pritchett},
  \citenamefont {Gambetta}, \citenamefont {Cross}, \citenamefont {Carroll},\
  and\ \citenamefont {Beverland}}]{yoder2025tour}%
  \BibitemOpen
  \bibfield  {author} {\bibinfo {author} {\bibfnamefont {T.~J.}\ \bibnamefont
  {Yoder}}, \bibinfo {author} {\bibfnamefont {E.}~\bibnamefont {Schoute}},
  \bibinfo {author} {\bibfnamefont {P.}~\bibnamefont {Rall}}, \bibinfo {author}
  {\bibfnamefont {E.}~\bibnamefont {Pritchett}}, \bibinfo {author}
  {\bibfnamefont {J.~M.}\ \bibnamefont {Gambetta}}, \bibinfo {author}
  {\bibfnamefont {A.~W.}\ \bibnamefont {Cross}}, \bibinfo {author}
  {\bibfnamefont {M.}~\bibnamefont {Carroll}},\ and\ \bibinfo {author}
  {\bibfnamefont {M.~E.}\ \bibnamefont {Beverland}},\ }\bibfield  {title}
  {\bibinfo {title} {Tour de gross: A modular quantum computer based on
  bivariate bicycle codes},\ }\href@noop {} {\bibfield  {journal} {\bibinfo
  {journal} {arXiv preprint arXiv:2506.03094}\ } (\bibinfo {year}
  {2025})}\BibitemShut {NoStop}%
\bibitem [{\citenamefont {Bourassa}\ \emph {et~al.}(2021)\citenamefont
  {Bourassa}, \citenamefont {Alexander}, \citenamefont {Vasmer}, \citenamefont
  {Patil}, \citenamefont {Tzitrin}, \citenamefont {Matsuura}, \citenamefont
  {Su}, \citenamefont {Baragiola}, \citenamefont {Guha}, \citenamefont
  {Dauphinais} \emph {et~al.}}]{bourassa2021blueprint}%
  \BibitemOpen
  \bibfield  {author} {\bibinfo {author} {\bibfnamefont {J.~E.}\ \bibnamefont
  {Bourassa}}, \bibinfo {author} {\bibfnamefont {R.~N.}\ \bibnamefont
  {Alexander}}, \bibinfo {author} {\bibfnamefont {M.}~\bibnamefont {Vasmer}},
  \bibinfo {author} {\bibfnamefont {A.}~\bibnamefont {Patil}}, \bibinfo
  {author} {\bibfnamefont {I.}~\bibnamefont {Tzitrin}}, \bibinfo {author}
  {\bibfnamefont {T.}~\bibnamefont {Matsuura}}, \bibinfo {author}
  {\bibfnamefont {D.}~\bibnamefont {Su}}, \bibinfo {author} {\bibfnamefont
  {B.~Q.}\ \bibnamefont {Baragiola}}, \bibinfo {author} {\bibfnamefont
  {S.}~\bibnamefont {Guha}}, \bibinfo {author} {\bibfnamefont {G.}~\bibnamefont
  {Dauphinais}}, \emph {et~al.},\ }\bibfield  {title} {\bibinfo {title}
  {Blueprint for a scalable photonic fault-tolerant quantum computer},\
  }\href@noop {} {\bibfield  {journal} {\bibinfo  {journal} {Quantum}\ }\textbf
  {\bibinfo {volume} {5}},\ \bibinfo {pages} {392} (\bibinfo {year}
  {2021})}\BibitemShut {NoStop}%
\bibitem [{\citenamefont {Butt}\ \emph {et~al.}(2024)\citenamefont {Butt},
  \citenamefont {Locher}, \citenamefont {Brechtelsbauer}, \citenamefont
  {B{\"u}chler},\ and\ \citenamefont {M{\"u}ller}}]{butt2024measurement}%
  \BibitemOpen
  \bibfield  {author} {\bibinfo {author} {\bibfnamefont {F.}~\bibnamefont
  {Butt}}, \bibinfo {author} {\bibfnamefont {D.~F.}\ \bibnamefont {Locher}},
  \bibinfo {author} {\bibfnamefont {K.}~\bibnamefont {Brechtelsbauer}},
  \bibinfo {author} {\bibfnamefont {H.~P.}\ \bibnamefont {B{\"u}chler}},\ and\
  \bibinfo {author} {\bibfnamefont {M.}~\bibnamefont {M{\"u}ller}},\ }\bibfield
   {title} {\bibinfo {title} {Measurement-free, scalable and fault-tolerant
  universal quantum computing},\ }\href@noop {} {\bibfield  {journal} {\bibinfo
   {journal} {arXiv preprint arXiv:2410.13568}\ } (\bibinfo {year}
  {2024})}\BibitemShut {NoStop}%
\bibitem [{\citenamefont {Jnane}\ \emph {et~al.}(2022)\citenamefont {Jnane},
  \citenamefont {Undseth}, \citenamefont {Cai}, \citenamefont {Benjamin},\ and\
  \citenamefont {Koczor}}]{jnane2022multicore}%
  \BibitemOpen
  \bibfield  {author} {\bibinfo {author} {\bibfnamefont {H.}~\bibnamefont
  {Jnane}}, \bibinfo {author} {\bibfnamefont {B.}~\bibnamefont {Undseth}},
  \bibinfo {author} {\bibfnamefont {Z.}~\bibnamefont {Cai}}, \bibinfo {author}
  {\bibfnamefont {S.~C.}\ \bibnamefont {Benjamin}},\ and\ \bibinfo {author}
  {\bibfnamefont {B.}~\bibnamefont {Koczor}},\ }\bibfield  {title} {\bibinfo
  {title} {Multicore quantum computing},\ }\href@noop {} {\bibfield  {journal}
  {\bibinfo  {journal} {Physical review applied}\ }\textbf {\bibinfo {volume}
  {18}},\ \bibinfo {pages} {044064} (\bibinfo {year} {2022})}\BibitemShut
  {NoStop}%
\bibitem [{\citenamefont {Monroe}\ \emph {et~al.}(2014)\citenamefont {Monroe},
  \citenamefont {Raussendorf}, \citenamefont {Ruthven}, \citenamefont {Brown},
  \citenamefont {Maunz}, \citenamefont {Duan},\ and\ \citenamefont
  {Kim}}]{monroe2014large}%
  \BibitemOpen
  \bibfield  {author} {\bibinfo {author} {\bibfnamefont {C.}~\bibnamefont
  {Monroe}}, \bibinfo {author} {\bibfnamefont {R.}~\bibnamefont {Raussendorf}},
  \bibinfo {author} {\bibfnamefont {A.}~\bibnamefont {Ruthven}}, \bibinfo
  {author} {\bibfnamefont {K.~R.}\ \bibnamefont {Brown}}, \bibinfo {author}
  {\bibfnamefont {P.}~\bibnamefont {Maunz}}, \bibinfo {author} {\bibfnamefont
  {L.-M.}\ \bibnamefont {Duan}},\ and\ \bibinfo {author} {\bibfnamefont
  {J.}~\bibnamefont {Kim}},\ }\bibfield  {title} {\bibinfo {title} {Large-scale
  modular quantum-computer architecture with atomic memory and photonic
  interconnects},\ }\href@noop {} {\bibfield  {journal} {\bibinfo  {journal}
  {Physical Review A}\ }\textbf {\bibinfo {volume} {89}},\ \bibinfo {pages}
  {022317} (\bibinfo {year} {2014})}\BibitemShut {NoStop}%
\bibitem [{\citenamefont {Sang}\ \emph {et~al.}(2025)\citenamefont {Sang},
  \citenamefont {Hour},\ and\ \citenamefont {Han}}]{sang2025toward}%
  \BibitemOpen
  \bibfield  {author} {\bibinfo {author} {\bibfnamefont {S.}~\bibnamefont
  {Sang}}, \bibinfo {author} {\bibfnamefont {L.}~\bibnamefont {Hour}},\ and\
  \bibinfo {author} {\bibfnamefont {Y.}~\bibnamefont {Han}},\ }\bibfield
  {title} {\bibinfo {title} {Toward scalable quantum compilation for modular
  architecture: Qubit mapping and reuse via deep reinforcement learning},\
  }\href@noop {} {\bibfield  {journal} {\bibinfo  {journal} {arXiv preprint
  arXiv:2506.09323}\ } (\bibinfo {year} {2025})}\BibitemShut {NoStop}%
\bibitem [{\citenamefont {Li}\ \emph {et~al.}(2024{\natexlab{b}})\citenamefont
  {Li}, \citenamefont {Santis}, \citenamefont {Harris}, \citenamefont {Chen},
  \citenamefont {Gao}, \citenamefont {Christen}, \citenamefont {Choi},
  \citenamefont {Trusheim}, \citenamefont {Song}, \citenamefont
  {Errando-Herranz} \emph {et~al.}}]{li2024heterogeneous}%
  \BibitemOpen
  \bibfield  {author} {\bibinfo {author} {\bibfnamefont {L.}~\bibnamefont
  {Li}}, \bibinfo {author} {\bibfnamefont {L.~D.}\ \bibnamefont {Santis}},
  \bibinfo {author} {\bibfnamefont {I.~B.}\ \bibnamefont {Harris}}, \bibinfo
  {author} {\bibfnamefont {K.~C.}\ \bibnamefont {Chen}}, \bibinfo {author}
  {\bibfnamefont {Y.}~\bibnamefont {Gao}}, \bibinfo {author} {\bibfnamefont
  {I.}~\bibnamefont {Christen}}, \bibinfo {author} {\bibfnamefont
  {H.}~\bibnamefont {Choi}}, \bibinfo {author} {\bibfnamefont {M.}~\bibnamefont
  {Trusheim}}, \bibinfo {author} {\bibfnamefont {Y.}~\bibnamefont {Song}},
  \bibinfo {author} {\bibfnamefont {C.}~\bibnamefont {Errando-Herranz}}, \emph
  {et~al.},\ }\bibfield  {title} {\bibinfo {title} {Heterogeneous integration
  of spin--photon interfaces with a cmos platform},\ }\href@noop {} {\bibfield
  {journal} {\bibinfo  {journal} {Nature}\ }\textbf {\bibinfo {volume} {630}},\
  \bibinfo {pages} {70} (\bibinfo {year} {2024}{\natexlab{b}})}\BibitemShut
  {NoStop}%
\bibitem [{\citenamefont {IBM}(2025)}]{ibm2025}%
  \BibitemOpen
  \bibfield  {author} {\bibinfo {author} {\bibnamefont {IBM}},\ }\href
  {https://newsroom.ibm.com/2025-06-10-IBM-Sets-the-Course-to-Build-Worlds-First-Large-Scale,-Fault-Tolerant-Quantum-Computer-at-New-IBM-Quantum-Data-Center}
  {\bibinfo {title} {Ibm sets the course to build world's first large-scale,
  fault-tolerant quantum computer at new ibm quantum data center}} (\bibinfo
  {year} {2025}),\ \bibinfo {note} {accessed 2025-07-31}\BibitemShut {NoStop}%
\bibitem [{\citenamefont {Bartee}\ \emph {et~al.}(2025)\citenamefont {Bartee},
  \citenamefont {Gilbert}, \citenamefont {Zuo}, \citenamefont {Das},
  \citenamefont {Tanttu}, \citenamefont {Yang}, \citenamefont
  {Dumoulin~Stuyck}, \citenamefont {Pauka}, \citenamefont {Su}, \citenamefont
  {Lim} \emph {et~al.}}]{bartee2025spin}%
  \BibitemOpen
  \bibfield  {author} {\bibinfo {author} {\bibfnamefont {S.~K.}\ \bibnamefont
  {Bartee}}, \bibinfo {author} {\bibfnamefont {W.}~\bibnamefont {Gilbert}},
  \bibinfo {author} {\bibfnamefont {K.}~\bibnamefont {Zuo}}, \bibinfo {author}
  {\bibfnamefont {K.}~\bibnamefont {Das}}, \bibinfo {author} {\bibfnamefont
  {T.}~\bibnamefont {Tanttu}}, \bibinfo {author} {\bibfnamefont {C.~H.}\
  \bibnamefont {Yang}}, \bibinfo {author} {\bibfnamefont {N.}~\bibnamefont
  {Dumoulin~Stuyck}}, \bibinfo {author} {\bibfnamefont {S.~J.}\ \bibnamefont
  {Pauka}}, \bibinfo {author} {\bibfnamefont {R.~Y.}\ \bibnamefont {Su}},
  \bibinfo {author} {\bibfnamefont {W.~H.}\ \bibnamefont {Lim}}, \emph
  {et~al.},\ }\bibfield  {title} {\bibinfo {title} {Spin-qubit control with a
  milli-kelvin cmos chip},\ }\href@noop {} {\bibfield  {journal} {\bibinfo
  {journal} {Nature}\ ,\ \bibinfo {pages} {1}} (\bibinfo {year}
  {2025})}\BibitemShut {NoStop}%
\bibitem [{\citenamefont {Beverland}\ \emph {et~al.}(2022)\citenamefont
  {Beverland}, \citenamefont {Murali}, \citenamefont {Troyer}, \citenamefont
  {Svore}, \citenamefont {Hoefler}, \citenamefont {Kliuchnikov}, \citenamefont
  {Low}, \citenamefont {Soeken}, \citenamefont {Sundaram},\ and\ \citenamefont
  {Vaschillo}}]{beverland2022assessing}%
  \BibitemOpen
  \bibfield  {author} {\bibinfo {author} {\bibfnamefont {M.~E.}\ \bibnamefont
  {Beverland}}, \bibinfo {author} {\bibfnamefont {P.}~\bibnamefont {Murali}},
  \bibinfo {author} {\bibfnamefont {M.}~\bibnamefont {Troyer}}, \bibinfo
  {author} {\bibfnamefont {K.~M.}\ \bibnamefont {Svore}}, \bibinfo {author}
  {\bibfnamefont {T.}~\bibnamefont {Hoefler}}, \bibinfo {author} {\bibfnamefont
  {V.}~\bibnamefont {Kliuchnikov}}, \bibinfo {author} {\bibfnamefont {G.~H.}\
  \bibnamefont {Low}}, \bibinfo {author} {\bibfnamefont {M.}~\bibnamefont
  {Soeken}}, \bibinfo {author} {\bibfnamefont {A.}~\bibnamefont {Sundaram}},\
  and\ \bibinfo {author} {\bibfnamefont {A.}~\bibnamefont {Vaschillo}},\
  }\bibfield  {title} {\bibinfo {title} {Assessing requirements to scale to
  practical quantum advantage},\ }\href@noop {} {\bibfield  {journal} {\bibinfo
   {journal} {arXiv preprint arXiv:2211.07629}\ } (\bibinfo {year}
  {2022})}\BibitemShut {NoStop}%
\bibitem [{\citenamefont {Andreev}\ \emph {et~al.}(2021)\citenamefont
  {Andreev}, \citenamefont {Balanov}, \citenamefont {Fromhold}, \citenamefont
  {Greenaway}, \citenamefont {Hramov}, \citenamefont {Li}, \citenamefont
  {Makarov},\ and\ \citenamefont {Zagoskin}}]{andreev2021emergence}%
  \BibitemOpen
  \bibfield  {author} {\bibinfo {author} {\bibfnamefont {A.}~\bibnamefont
  {Andreev}}, \bibinfo {author} {\bibfnamefont {A.}~\bibnamefont {Balanov}},
  \bibinfo {author} {\bibfnamefont {T.}~\bibnamefont {Fromhold}}, \bibinfo
  {author} {\bibfnamefont {M.}~\bibnamefont {Greenaway}}, \bibinfo {author}
  {\bibfnamefont {A.}~\bibnamefont {Hramov}}, \bibinfo {author} {\bibfnamefont
  {W.}~\bibnamefont {Li}}, \bibinfo {author} {\bibfnamefont {V.}~\bibnamefont
  {Makarov}},\ and\ \bibinfo {author} {\bibfnamefont {A.}~\bibnamefont
  {Zagoskin}},\ }\bibfield  {title} {\bibinfo {title} {Emergence and control of
  complex behaviors in driven systems of interacting qubits with dissipation},\
  }\href@noop {} {\bibfield  {journal} {\bibinfo  {journal} {npj Quantum
  Information}\ }\textbf {\bibinfo {volume} {7}},\ \bibinfo {pages} {1}
  (\bibinfo {year} {2021})}\BibitemShut {NoStop}%
\bibitem [{\citenamefont {De~Clerk}\ and\ \citenamefont
  {Savel'ev}(2022)}]{de2022investigation}%
  \BibitemOpen
  \bibfield  {author} {\bibinfo {author} {\bibfnamefont {L.}~\bibnamefont
  {De~Clerk}}\ and\ \bibinfo {author} {\bibfnamefont {S.}~\bibnamefont
  {Savel'ev}},\ }\bibfield  {title} {\bibinfo {title} {An investigation of
  higher order moments of empirical financial data and their implications to
  risk},\ }\href@noop {} {\bibfield  {journal} {\bibinfo  {journal} {Heliyon}\
  }\textbf {\bibinfo {volume} {8}} (\bibinfo {year} {2022})}\BibitemShut
  {NoStop}%
\bibitem [{\citenamefont {De~Clerk}\ and\ \citenamefont
  {Savel’ev}(2022)}]{de2022ai}%
  \BibitemOpen
  \bibfield  {author} {\bibinfo {author} {\bibfnamefont {L.}~\bibnamefont
  {De~Clerk}}\ and\ \bibinfo {author} {\bibfnamefont {S.}~\bibnamefont
  {Savel’ev}},\ }\bibfield  {title} {\bibinfo {title} {Ai algorithms for
  fitting garch parameters to empirical financial data},\ }\href@noop {}
  {\bibfield  {journal} {\bibinfo  {journal} {Physica A: Statistical Mechanics
  and its Applications}\ }\textbf {\bibinfo {volume} {603}},\ \bibinfo {pages}
  {127869} (\bibinfo {year} {2022})}\BibitemShut {NoStop}%
\bibitem [{\citenamefont {Zhu}\ \emph {et~al.}(2024)\citenamefont {Zhu},
  \citenamefont {Ehlers}, \citenamefont {Nurdin},\ and\ \citenamefont
  {Soh}}]{zhu2024practical}%
  \BibitemOpen
  \bibfield  {author} {\bibinfo {author} {\bibfnamefont {C.}~\bibnamefont
  {Zhu}}, \bibinfo {author} {\bibfnamefont {P.~J.}\ \bibnamefont {Ehlers}},
  \bibinfo {author} {\bibfnamefont {H.~I.}\ \bibnamefont {Nurdin}},\ and\
  \bibinfo {author} {\bibfnamefont {D.}~\bibnamefont {Soh}},\ }\bibfield
  {title} {\bibinfo {title} {Practical and scalable quantum reservoir
  computing},\ }\href@noop {} {\bibfield  {journal} {\bibinfo  {journal} {arXiv
  preprint arXiv:2405.04799}\ } (\bibinfo {year} {2024})}\BibitemShut {NoStop}%
\bibitem [{\citenamefont {McCaul}\ \emph {et~al.}(2025)\citenamefont {McCaul},
  \citenamefont {Totero~Gongora}, \citenamefont {Otieno}, \citenamefont
  {Savel’ev}, \citenamefont {Zagoskin},\ and\ \citenamefont
  {Balanov}}]{mccaul2025minimalquantumreservoirshamiltonian}%
  \BibitemOpen
  \bibfield  {author} {\bibinfo {author} {\bibfnamefont {G.}~\bibnamefont
  {McCaul}}, \bibinfo {author} {\bibfnamefont {J.~S.}\ \bibnamefont
  {Totero~Gongora}}, \bibinfo {author} {\bibfnamefont {W.}~\bibnamefont
  {Otieno}}, \bibinfo {author} {\bibfnamefont {S.}~\bibnamefont {Savel’ev}},
  \bibinfo {author} {\bibfnamefont {A.}~\bibnamefont {Zagoskin}},\ and\
  \bibinfo {author} {\bibfnamefont {A.~G.}\ \bibnamefont {Balanov}},\
  }\bibfield  {title} {\bibinfo {title} {Minimal quantum reservoirs with
  hamiltonian encoding},\ }\href {https://doi.org/10.1063/5.0282921} {\bibfield
   {journal} {\bibinfo  {journal} {Chaos: An Interdisciplinary Journal of
  Nonlinear Science}\ }\textbf {\bibinfo {volume} {35}},\ \bibinfo {pages}
  {093135} (\bibinfo {year} {2025})}\BibitemShut {NoStop}%
\end{thebibliography}%

\end{document}